\documentclass[acmsmall,screen]{acmart}

\usepackage{tcolorbox}
\usepackage{fontawesome}
\usepackage{float}
\usepackage[normalem]{ulem}
\usepackage{geometry}
\usepackage{multirow,bigstrut}
\usepackage{colortbl}
\usepackage{longtable}
\usepackage{changepage}
\usepackage[printonlyused]{acronym} % Acronyms with \ac
\usepackage{bookmark}
\usepackage{natbib}
\usepackage{placeins}
\usepackage{enumitem}
\usepackage{caption}
\usepackage{tabularx}
\usepackage{ragged2e}

\newcolumntype{P}[1]{>{\raggedright\arraybackslash}p{#1}}
\newcolumntype{B}[1]{>{\justifying\arraybackslash}p{#1}}

\definecolor{darkgreen}{RGB}{31, 148, 51}
\definecolor{darkred}{RGB}{179, 32, 21}

\newenvironment{checklist}
{\begin{itemize}[label=\(\square\), leftmargin=1.5em, noitemsep, topsep=2pt, itemsep=2pt]}
{\end{itemize}}

\tcbset{
  pitfallbox/.style={
    colback=white,
    colframe=black,
    boxrule=0.5pt,
    arc=2pt,
    left=8pt,
    right=8pt,
    top=8pt,
    bottom=8pt,
    fonttitle=\bfseries,
  }
}

\hypersetup{
  pdfinfo={
  pdfproducer={},
  Title={},
  Subject={},
  Author={},
  }
}

\makeatletter
\AtBeginDocument{%
  \renewcommand*{\AC@hyperlink}[2]{%
    \begingroup
      \hypersetup{hidelinks}%
      \hyperlink{#1}{#2}%
    \endgroup
  }%
}
\newcommand{\ExtendedDataFiguresOn}{%
  \gdef\@captype{figure}%
  \renewcommand{\fnum@figure}{Extended Data Fig.~\thefigure}%
}
\newcommand{\ExtendedDataFiguresOff}{%
  \renewcommand{\fnum@figure}{Figure~\thefigure}%
}

\makeatother

\AtBeginDocument{%
  \providecommand\BibTeX{{%
    \normalfont B\kern-0.5em{\scshape i\kern-0.25em b}\kern-0.8em\TeX}}}

\begin{document}

%\DeclareUrlCommand\ULurl@{%
%  \def\UrlFont{\ttfamily\color{blue}}%
%  \def\UrlLeft{\uline\bgroup}%
%  \def\UrlRight{\egroup}}
%\def\ULurl@#1{\hyper@linkurl{\ULurl@@{#1}}{#1}}
%\DeclareRobustCommand*\ULurl{\hyper@normalise\ULurl@}

%%
%% The "title" command has an optional parameter,
%% allowing the author to define a "short title" to be used in page headers.
\title[Current validation practice undermines surgical AI development]{Current validation practice undermines surgical AI development}

%%
%% The "author" command and its associated commands are used to define
%% the authors and their affiliations.
%% Of note is the shared affiliation of the first two authors, and the
%% "authornote" and "authornotemark" commands
%% used to denote shared contribution to the research.

\author{Annika Reinke}
\authornote{\textbf{Corresponding author:} Annika Reinke: a.reinke@dkfz-heidelberg.de}
% \authornote{The complete list of affiliations can be found in Appendix~\ref{app:authors}.}
\affiliation{%
  \institution{German Cancer Research Center (DKFZ) Heidelberg, Division of Intelligent Medical Systems and HI Helmholtz Imaging}
  %\country{Im Neuenheimer Feld 223, 69120 Heidelberg, Germany}
  \country{Heidelberg, Germany}
}

%% Extended core team
\author{Ziying O. Li}
\affiliation{%
  \institution{University of Cambridge}
  \country{Cambridge, UK}
}

\author{Minu D. Tizabi}
\affiliation{%
  \institution{German Cancer Research Center (DKFZ) Heidelberg, Division of Intelligent Medical Systems}
  %\country{Im Neuenheimer Feld 223, 69120 Heidelberg, Germany}
  \country{Heidelberg, Germany}
}
\affiliation{%
  \institution{National Center for Tumor Diseases (NCT), NCT Heidelberg, a partnership between DKFZ and University Medical Center Heidelberg}
  %\country{Im Neuenheimer Feld 460, 69120 Heidelberg, Germany}
  \country{Heidelberg, Germany}
}

\author{Pascaline André}
\affiliation{
    \institution{Sorbonne Université, Institut du Cerveau - Paris Brain Institute - ICM, CNRS, Inria}
    \country{Paris, France}
}
\affiliation{
    \institution{Inserm, AP-HP, Hôpital de la Pitié-Salpêtrière}
    \country{Paris, France}
}

\author{Marcel Knopp}
\affiliation{%
  \institution{German Cancer Research Center (DKFZ) Heidelberg, Division of Intelligent Medical Systems and HI Helmholtz Imaging}
  %\country{Im Neuenheimer Feld 223, 69120 Heidelberg, Germany}
  \country{Heidelberg, Germany}
}
\affiliation{%
  \institution{Faculty of Mathematics and Computer Science, Heidelberg University}
  %\country{Seminarstraße 2, 69117 Heidelberg, Germany}
  \country{Heidelberg, Germany}
}

\author{Mika M. Rother}
\affiliation{%
  \institution{German Cancer Research Center (DKFZ) Heidelberg, Division of Intelligent Medical Systems and HI Helmholtz Imaging}
  %\country{Im Neuenheimer Feld 223, 69120 Heidelberg, Germany}
  \country{Heidelberg, Germany}
}

\author{Ines P. Machado}
\affiliation{
    \institution{Department of Oncology, University of Cambridge}
    \country{Cambridge, UK}
}

%% Alphabetical order
\author{Maria S. Altieri}
\affiliation{
\institution{University of Pennsylvania}
\country{PA, USA}
}

\author{Deepak Alapatt}
\affiliation{
\institution{University of Strasbourg}
\country{Strasbourg, France}
}
\affiliation{
\institution{Scialytics}
\country{Strasbourg, France}
}

\author{Sophia Bano}
\affiliation{
\institution{UCL Hawkes Institute and Department of Computer Science, University College London}
\country{London, United Kingdom}
}

\author{Sebastian Bodenstedt}
\affiliation{
\institution{National Center for Tumor Diseases (NCT), NCT/UCC Dresden, a partnership between DKFZ, Faculty of Medicine and University Hospital Carl Gustav Carus, TUD Dresden University of Technology, and Helmholtz-Zentrum Dresden-Rossendorf (HZDR)}
\country{Dresden, Germany}
}

\author{Oliver Burgert}
\affiliation{
\institution{Reutlingen University}
\country{Reutlingen, Germany}
}

\author{Elvis C.S. Chen}
\affiliation{
\institution{Department of Electrical and Computer Engineering/Department of Medical Biophysics/Robarts Research Institute, Western University}
\country{London, Canada}
}
\affiliation{
\institution{Lawson Health Research Institute}
\country{London, Canada}
}

\author{Justin W. Collins}
\affiliation{
\institution{Division of Surgery \& Interventional Science, University College London}
\country{London, UK}
}
\affiliation{
\institution{Division of Uro-oncology, University College London Hospital}
\country{London, UK}
}

\author{Olivier Colliot}
\affiliation{
\institution{Sorbonne Université, Institut du Cerveau - Paris Brain Institute - ICM, CNRS, Inria}
\country{Paris, France}
}
\affiliation{
\institution{Inserm, AP-HP, Hôpital de la Pitié-Salpêtrière}
\country{Paris, France}
}

\author{Evangelia Christodoulou}
\affiliation{
\institution{German Cancer Research Center (DKFZ) Heidelberg, Division of Intelligent Medical Systems}
\country{Heidelberg, Germany}
}
% \affiliation{
% \institution{National Center for Tumor Diseases (NCT), NCT Heidelberg, a partnership between DKFZ and Heidelberg University Hospital}
% \country{Heidelberg, Germany}
% }

\author{Tobias Czempiel}
\affiliation{
\institution{EnAcuity Ltd.}
\country{London, UK }
}
\affiliation{
\institution{EnAcuity Ltd.}
\country{UCL Hawkes Institute, University College London}
}

\author{Adrito Das}
\affiliation{
\institution{UCL Hawkes Institute, University College London}
\country{London, UK}
}

\author{Reuben Docea}
\affiliation{
\institution{National Center for Tumor Diseases (NCT), NCT/UCC Dresden, a partnership between DKFZ, Faculty of Medicine and University Hospital Carl Gustav Carus, TUD Dresden University of Technology, and Helmholtz-Zentrum Dresden-Rossendorf (HZDR)}
\country{Dresden, Germany}
% Fetscherstraße 74, 01307 Dresden, Germany
}
\affiliation{
\institution{German Cancer Research Center (DKFZ)}
\country{Heidelberg, Germany}
}
\affiliation{
\institution{Faculty of Medicine and University Hospital Carl Gustav Carus, TUD Dresden University of Technology}
\country{Dresden, Germany}
}
\affiliation{
\institution{Helmholtz-Zentrum Dresden-Rossendorf (HZDR)}
\country{Dresden, Germany}
}

\author{Daniel Donoho}
\affiliation{
\institution{Children’s National Hospital}
\country{Washington, D.C., USA}
}

\author{Qi Dou}
\affiliation{
\institution{Department of Computer Science and Engineering, The Chinese University of Hong Kong}
\country{Hong Kong}
}

\author{Jennifer Eckhoff}
\affiliation{
\institution{Department for General, Visceral, Thoracic and Transplant Surgery, University Hospital Cologne}
\country{Cologne, Germany}
}

\author{Sandy Engelhardt}
\affiliation{
\institution{Department of Cardiac Surgery, Heidelberg University Hospital}
\country{Heidelberg, Germany}
}
\affiliation{
\institution{Medical Faculty of Heidelberg University, Heidelberg University}
\country{Heidelberg, Germany}
}
\affiliation{
\institution{DZHK Partnersite Heidelberg-Mannheim}
\country{Heidelberg, Germany}
}

\author{Gabor Fichtinger}
\affiliation{
\institution{School of Computing, Queen's University}
\country{Kingston, Canada}
}

\author{Philipp Fuernstahl}
\affiliation{
\institution{Research in Orthopedic Computer Science Group, Balgrist University Hospital, University of Zurich}
\country{Zurich, Switzerland}
}

\author{Pablo García Kilroy}
\affiliation{
\institution{Verb Surgical Inc.}
\country{Santa Clara, USA}
}

\author{Stamatia Giannarou}
\affiliation{
\institution{Hamlyn Centre for Robotic Surgery, Department of Surgery and Cancer, Imperial College London}
\country{London, UK}
}

\author{Stephen Gilbert}
\affiliation{
\institution{Else Kröner Fresenius Center for Digital Health, TUD Dresden University of Technology}
\country{Dresden, Germany}
}
\affiliation{
\institution{Faculty of Business and Economics, TUD Dresden University of Technology}
\country{Dresden, Germany}
}

\author{Ines Gockel}
\affiliation{
\institution{Section of Esophageal Surgery, Shanghai Chest Hospital}
\country{Shanghai, China}
}

\author{Patrick Godau}
\affiliation{
\institution{German Cancer Research Center (DKFZ) Heidelberg, Division of Intelligent Medical Systems}
\country{Heidelberg, Germany}
}
\affiliation{
\institution{National Center for Tumor Diseases (NCT), NCT Heidelberg, a partnership between DKFZ and University Hospital Heidelberg}
\country{Heidelberg, Germany}
}
\affiliation{
\institution{Faculty of Mathematics and Computer Science, Heidelberg University}
\country{Heidelberg, Germany}
}
\affiliation{
\institution{HIDSS4Health - Helmholtz Information and Data Science School for Health}
\country{Karlsruhe/Heidelberg, Germany}
}

\author{Jan Gödeke}
\affiliation{
\institution{Department of Paediatric Surgery, Dr. von Hauner Children´s Hospital, LMU University Hospital}
\country{Munich, Germany}
% Lindwurmstrasse 4, 80337 Munich, Germany
}

\author{Teodor P. Grantcharov}
\affiliation{
\institution{Stanford University}
\country{Palo Alto, USA}
}

\author{Tamas Haidegger}
\affiliation{
\institution{Obuda University}
\country{Budapest, Hungary}
}
\affiliation{
\institution{Queen's University}
\country{Kingston, Canada}
}
\affiliation{
\institution{Austrian Center for Medical Innovation and Technology (ACMIT) Gmbh}
\country{Austria, Vienna}
}

\author{Alexander Hann}
\affiliation{
\institution{Interventional and Experimental Endoscopy (InExEn), Department of Internal Medicine 2, University Hospital Würzburg}
\country{Würzburg, Germany}
}

\author{Makoto Hashizume}
\affiliation{
\institution{College of NISHINIHON Nursing \& Medical Care}
\country{Fukuoka, Japan}
}
\affiliation{
\institution{Kyushu University}
\country{Fukuoka, Japan}
}

\author{Charles Heitz}
\affiliation{
    \institution{Sorbonne Université, Institut du Cerveau - Paris Brain Institute - ICM, CNRS, Inria}
    \country{Paris, France}
}
\affiliation{
    \institution{Inserm, AP-HP, Hôpital de la Pitié-Salpêtrière}
    \country{Paris, France}
}

\author{Rebecca Hisey}
\affiliation{
    \institution{National Center for Tumor Diseases (NCT), NCT/UCC Dresden, a partnership between DKFZ, Faculty of Medicine and University Hospital Carl Gustav Carus, TUD Dresden University of Technology, and Helmholtz-Zentrum Dresden-Rossendorf (HZDR)}
    \country{Dresden, Germany}
}

\author{Hanna Hoffmann}
\affiliation{
    \institution{National Center for Tumor Diseases (NCT), NCT/UCC Dresden, a partnership between DKFZ, Faculty of Medicine and University Hospital Carl Gustav Carus, TUD Dresden University of Technology, and Helmholtz-Zentrum Dresden-Rossendorf (HZDR)}
    \country{Dresden, Germany}
}

\author{Arnaud Huaulmé}
\affiliation{
\institution{University of Rennes, INSERM, LTSI - UMR 1099, F35000}
\country{Rennes, France}
}

\author{Paul F. Jäger}
\affiliation{
\institution{Google DeepMind}
\country{London, UK}
}

\author{Pierre Jannin}
\affiliation{
\institution{University of Rennes, INSERM, LTSI - UMR 1099, F35000}
\country{Rennes, France}
}

\author{Anthony Jarc}
\affiliation{
\institution{Intuitive}
\country{Sunnyvale, CA, USA}
}

\author{Rohit Jena}
\affiliation{
\institution{Penn Image Computing and Science Laboratory, Department of Radiology, University of Pennsylvania}
\country{Philadelphia, PA, USA}
}

\author{Yueming Jin}
\affiliation{
\institution{Department of Biomedical Engineering, and Department of Electrical and Computer Engineering, National University of Singapore}
\country{Singapore}
}

\author{Leo Joskowicz}
\affiliation{
\institution{School of Computer Science and Engineering, The Hebrew University of Jerusalem}
\country{Jerusalem, Israel}
}

\author{Luc Joyeux}
\affiliation{
\institution{Texas Children’s Center for Translational Fetal Research, Texas Children’s Fetal Center, Department of Obstetrics and Gynecology; Division of Pediatric Surgery, Michael E. DeBakey Department of Surgery; Texas Children’s Hospital and Baylor College of Medicine}
\country{Houston, TX, USA}
}

\author{Max Kirchner}
\affiliation{
    \institution{National Center for Tumor Diseases (NCT), NCT/UCC Dresden, a partnership between DKFZ, Faculty of Medicine and University Hospital Carl Gustav Carus, TUD Dresden University of Technology, and Helmholtz-Zentrum Dresden-Rossendorf (HZDR)}
    \country{Dresden, Germany}
}

\author{Axel Krieger}
\affiliation{
\institution{Department of Mechanical Engineering, Johns Hopkins University}
\country{Baltimore, MD, USA}
% 3400 N. Charles Street, Baltimore, MD 21218, USA
}

\author{Gernot Kronreif}
\affiliation{
\institution{Austrian Center for Medical Innovation and Technology (ACMIT) Gmbh}
\country{Vienna, Austria}
}

\author{Kyle Lam}
\affiliation{
\institution{Department of Surgery and Cancer, Imperial College London}
\country{London, UK}
}

\author{Shlomi Laufer}
\affiliation{
\institution{Faculty of Data and Decision Sciences, Technion - Israel Institute of Technology}
\country{Haifa, Israel}
}

\author{Joël L. Lavanchy}
\affiliation{
\institution{University Digestive Health Care Center -Clarunis}
\country{Basel, Switzerland}
% 4002
}
\affiliation{
\institution{Department of Biomedical Engineering, University of Basel }
\country{Allschwil, Switzerland}
% 4123
}

\author{Gyusung I. Lee}
\affiliation{
\institution{American College of Surgeons}
\country{Chicago, USA}
}

\author{Robert Lim}
\affiliation{
\institution{Department of Surgery, Wake Forest University School of Medicine at Charlotte, Atrium Health Carolinas}
\country{North Carolina, USA}
}

\author{Peng Liu}
\affiliation{
    \institution{National Center for Tumor Diseases (NCT), NCT/UCC Dresden, a partnership between DKFZ, Faculty of Medicine and University Hospital Carl Gustav Carus, TUD Dresden University of Technology, and Helmholtz-Zentrum Dresden-Rossendorf (HZDR)}
    \country{Dresden, Germany}
}

\author{Lucas Luttner}
\affiliation{%
  \institution{German Cancer Research Center (DKFZ) Heidelberg, Division of Intelligent Medical Systems}
  %\country{Im Neuenheimer Feld 223, 69120 Heidelberg, Germany}
  \country{Heidelberg, Germany}
}
\affiliation{%
  \institution{Medical Faculty, Heidelberg University}
  %\country{Seminarstraße 2, 69117 Heidelberg, Germany}
  \country{Heidelberg, Germany}
}

\author{Hani J. Marcus}
\affiliation{
\institution{UCL Queen Square Institute of Neurology, UCL}
\country{London, UK}
}

\author{Pietro Mascagni}
\affiliation{
\institution{Bioimage Analysis Center, Fondazione Policlinico Agostino Gemelli IRCCS}
\country{Rome, Italy}
}
\affiliation{
\institution{Institute of Image-Guided Surgery, IHU-Strasbourg}
\country{Strasbourg, France}
}

\author{Leon Mayer}
\affiliation{%
  \institution{German Cancer Research Center (DKFZ) Heidelberg, Division of Intelligent Medical Systems}
  %\country{Im Neuenheimer Feld 223, 69120 Heidelberg, Germany}
  \country{Heidelberg, Germany}
}
\affiliation{%
  \institution{Medical Faculty, Heidelberg University}
  %\country{Seminarstraße 2, 69117 Heidelberg, Germany}
  \country{Heidelberg, Germany}
}

\author{Ozanan R. Meireles}
\affiliation{
\institution{Massachusetts General Hospital, Harvard Medical School}
\country{Boston, MA, USA}
}

\author{Beat P. Mueller}
\affiliation{
\institution{University Digestive Health Care Center Basel}
\country{Basel, Switzerland}
}

\author{Lars Mündermann}
\affiliation{
\institution{KARL STORZ SE \& Co. KG}
\country{Tuttlingen, Germany}
}

\author{Hirenkumar Nakawala}
\affiliation{
\institution{Independent researcher}
\country{London, UK}
}

\author{Nassir Navab}
\affiliation{
\institution{Chair for Computer Aided Medical Procedures (CAMP), TU Munich}
\country{Munich, Germany}
}

\author{Abdourahmane Ndong}
\affiliation{
\institution{Gaston Berger University}
\country{Saint-Louis, Senegal}
}

\author{Juliane Neumann}
\affiliation{
\institution{University of Leipzig, Innovation Center Computer-Assisted Surgery (ICCAS)}
\country{Leipzig, Germany}
}

\author{Felix Nickel}
\affiliation{
\institution{Department of General, Visceral, and Thoracic Surgery, University Medical Center Hamburg-Eppendorf}
\country{Hamburg, Germany}
% Martinistrasse 52, 20251 Hamburg, Germany
}

\author{Marco Nolden}
\affiliation{
\institution{Institute Division of Medical Image Computing, German Cancer Research Center (DKFZ)}
\country{Heidelberg, Germany}
}
\affiliation{
\institution{Pattern Analysis and Learning Group, Department of Radiation Oncology, Heidelberg University Hospital}
\country{Heidelberg, Germany}
}

\author{Chinedu Nwoye}
\affiliation{
\institution{Intuitive}
\country{Sunnyvale, CA, USA}
}
\affiliation{
\institution{University of Strasbourg}
\country{Strasbourg, France}
}
\affiliation{
\institution{IHU Strasbourg}
\country{Strasbourg, France}
}

\author{Namkee Oh}
\affiliation{
\institution{Department of Surgery, Samsung Medical Center}
\country{Seoul, Republic of Korea}
}

\author{Nicolas Padoy}
\affiliation{
\institution{University of Strasbourg, CNRS, INSERM, ICube, UMR7357}
\country{Strasbourg, France}
}
\affiliation{
\institution{IHU Strasbourg}
\country{Strasbourg, France}
}

\author{Thomas Pausch}
\affiliation{
\institution{Department of General, Visceral and Transplantation Surgery, University Hospital Heidelberg}
\country{Heidelberg, Germany}
}
\affiliation{
\institution{Institute of Medical Informatics, Heidelberg University}
\country{Heidelberg, Germany}
}
\affiliation{
\institution{ISSO Partnership Between DKFZ and University Medical Center Heidelberg, National Center for Tumor Diseases}
\country{Heidelberg, Germany}
}

\author{Micha Pfeiffer}
\affiliation{
\institution{National Center for Tumor Diseases (NCT), NCT/UCC Dresden, a partnership between DKFZ, Faculty of Medicine and University Hospital Carl Gustav Carus, TUD Dresden University of Technology, and Helmholtz-Zentrum Dresden-Rossendorf (HZDR)}
\country{Dresden, Germany}
}
\affiliation{
\institution{Helmholtz-Zentrum Dresden-Rossendorf (HZDR)}
\country{Dresden, Germany}
}
\affiliation{
\institution{University Hospital Carl Gustav Carus, Technische Universität Dresden}
\country{Dresden, Germany}
}

\author{Tim Rädsch}
\affiliation{%
  \institution{German Cancer Research Center (DKFZ) Heidelberg, Division of Intelligent Medical Systems and HI Helmholtz Imaging}
  %\country{Im Neuenheimer Feld 223, 69120 Heidelberg, Germany}
  \country{Heidelberg, Germany}
}

\author{Hongliang Ren}
\affiliation{
\institution{Department of Electronic Engineering, The Chinese University of Hong Kong (CUHK)}
\country{Hong Kong}
}

\author{Nicola Rieke}
\affiliation{
\institution{NVIDIA}
\country{Munich, Germany}
}

\author{Dominik Rivoir}
\affiliation{
\institution{National Center for Tumor Diseases (NCT), NCT/UCC Dresden, a partnership between DKFZ, Faculty of Medicine and University Hospital Carl Gustav Carus, TUD Dresden University of Technology, and Helmholtz-Zentrum Dresden-Rossendorf (HZDR)}
\country{Dresden, Germany}
}
\affiliation{
\institution{Centre for Tactile Internet with Human-in-the-Loop (CeTI), TUD Dresden University of Technology}
\country{Dresden, Germany}
}

\author{Duygu Sarikaya}
\affiliation{
\institution{School of Computer Science, University of Leeds}
\country{Leeds, UK}
}

\author{Samuel Schmidgall}
\affiliation{
\institution{Department of Electrical and Computer Engineering, Johns Hopkins University}
\country{Baltimore, MD, USA}
% 3400 N. Charles Street, Baltimore, MD 21218, USA
}

\author{Matthias Seibold}
\affiliation{
\institution{Research in Orthopedic Computer Science, Balgrist University Hospital}
\country{Zurich, Switzerland}
}

\author{Silvia Seidlitz}
\affiliation{
\institution{German Cancer Research Center (DKFZ) Heidelberg, Division of Intelligent Medical Systems}
\country{Heidelberg, Germany}
}
\affiliation{
\institution{Helmholtz Information and Data Science School for Health}
\country{Heidelberg/Karlsruhe, Germany}
}
\affiliation{
\institution{Faculty of Mathematics and Computer Science, Heidelberg University}
\country{Heidelberg, Germany}
}
\affiliation{
\institution{National Center for Tumor Diseases (NCT), NCT Heidelberg, a partnership between DKFZ and University Medical Center Heidelberg}
\country{Heidelberg, Germany}
}

\author{Alexander Seitel}
\affiliation{
\institution{German Cancer Research Center (DKFZ) Heidelberg, Division of Intelligent Medical Systems}
\country{Heidelberg, Germany}
}

\author{Lalith Sharan}
\affiliation{
\institution{University of Strasbourg, CNRS, INSERM, ICube, UMR7357}
\country{Strasbourg, France}
}
\affiliation{
\institution{IHU Strasbourg}
\country{Strasbourg, France}
}

\author{Jeffrey H. Siewerdsen}
\affiliation{
\institution{The University of Texas MD Anderson Cancer Center}
\country{Houston, USA}
}

\author{Vinkle Srivastav}
\affiliation{
\institution{University of Strasbourg, CNRS, INSERM, ICube, UMR7357}
\country{Strasbourg, France}
}
\affiliation{
\institution{IHU Strasbourg}
\country{Strasbourg, France}
}
\affiliation{
\institution{Department of Data Science and AI, Wadhwani School of Data Science and AI (WSAI), Indian Institute of Technology (IIT) Madras}
\country{Chennai, India}
% Department of Data Science and AI, Wadhwani School of Data Science and AI (WSAI), Indian Institute of Technology (IIT) Madras, Chennai, 600036, India.
}

\author{Raphael Sznitman}
\affiliation{
\institution{University of Bern}
\country{Bern, Switzerland}
}

\author{Russell Taylor}
\affiliation{
\institution{Johns Hopkins University}
\country{Baltimore, MD, USA}
% 3400 N. Charles Street, Baltimore, MD 21218, USA
}

\author{Thuy N. Tran}
\affiliation{
\institution{German Cancer Research Center (DKFZ) Heidelberg, Division of Intelligent Medical Systems}
\country{Heidelberg, Germany}
}

\author{Matthias Unberath}
\affiliation{
\institution{Johns Hopkins University}
\country{Baltimore, MD, USA}
% 3400 N. Charles Street, Baltimore, MD 21218, USA
}

\author{Fons van der Sommen}
\affiliation{
\institution{Eindhoven University of Technology}
\country{Eindhoven, The Netherlands}
}

\author{Martin Wagner}
\affiliation{
\institution{Department of Visceral, Thoracic and Vascular Surgery, Faculty of Medicine and University Hospital Carl Gustav Carus, TUD Dresden University of Technology}
\country{Dresden, Germany}
% Fetscherstraße 74, 01307 Dresden, Germany
}
\affiliation{
\institution{Centre for the Tactile Internet with Human‑in‑the‑Loop (CeTI), TUD Dresden University of Technology}
\country{Dresden, Germany}
}

\author{Amine Yamlahi}
\affiliation{%
  \institution{German Cancer Research Center (DKFZ) Heidelberg, Division of Intelligent Medical Systems}
  %\country{Im Neuenheimer Feld 223, 69120 Heidelberg, Germany}
  \country{Heidelberg, Germany}
}
\affiliation{%
  \institution{National Center for Tumor Diseases (NCT), NCT Heidelberg, a partnership between DKFZ and University Medical Center Heidelberg}
  %\country{Im Neuenheimer Feld 460, 69120 Heidelberg, Germany}
  \country{Heidelberg, Germany}
}

\author{Shaohua K. Zhou}
\affiliation{
\institution{University of Science and Technology of China (USTC)}
\country{Hefei, Anhui, China}
}

\author{Aneeq Zia}
\affiliation{
\institution{Intuitive}
\country{Sunnyvale, CA, USA}
}

%% Extended core senior authors
\author{Amin Madani}
\affiliation{%
  \institution{Surgical Artificial Intelligence Research Academy, University Health Network}
  \country{Toronto, Canada}
}
\affiliation{%
  \institution{Department of Surgery, University of Toronto}
  \country{Toronto, Canada}
}

\author{Danail Stoyanov}
\affiliation{%
  \institution{University College London}
  \country{London, UK}
}
\affiliation{%
  \institution{Medtronic}
  \country{London, UK}
}

\author{Stefanie Speidel}
\affiliation{%
  \institution{National Center for Tumor Diseases (NCT), NCT/UCC Dresden, a partnership between DKFZ, Faculty of Medicine and University Hospital Carl Gustav Carus, TUD Dresden University of Technology, and Helmholtz-Zentrum Dresden-Rossendorf (HZDR)}
  %\country{Fiedlerstraße 19, 01307 Dresden, Germany}
  \country{Dresden, Germany}
}
\affiliation{%
  \institution{Centre for Tactile Internet with Human-in-the-Loop (CeTI), TUD Dresden University of Technology}
  %\country{Georg-Schumann-Str. 11, 01069 Dresden, Germany}
  \country{Dresden, Germany}
}

%% Senior authors
\author{Daniel A. Hashimoto}
\authornote{\textbf{These authors jointly supervised this work:} Daniel A. Hashimoto, Fiona R. Kolbinger, Lena Maier-Hein}
\affiliation{%
  \institution{Departments of Surgery, Computer and Information Science, University of Pennsylvania}
  %\country{3400 Spruce St, Philadelphia, PA 19104, USA}
  \country{PA, USA}
}

\author{Fiona R. Kolbinger}
\authornotemark[2]
\affiliation{%
  \institution{Weldon School of Biomedical Engineering, Purdue University}
  %\country{206 S Martin Jischke Dr, West Lafayette, IN, 47907, USA}
  \country{IN, USA}
}
\affiliation{%
  \institution{Department of Visceral, Thoracic and Vascular Surgery, University Hospital and Faculty of Medicine Carl Gustav Carus, TUD Dresden University of Technology}
  %\country{Fetscherstr. 74, 01037 Dresden, Germany}
  \country{Dresden, Germany}
}

\author{Lena Maier-Hein}
\authornotemark[2]
\affiliation{%
  \institution{German Cancer Research Center (DKFZ) Heidelberg, Division of Intelligent Medical Systems and HI Helmholtz Imaging}
  %\country{Im Neuenheimer Feld 223, 69120 Heidelberg, Germany}
  \country{Heidelberg, Germany}
}
\affiliation{%
  \institution{Faculty of Mathematics and Computer Science and Medical Faculty, Heidelberg University}
  %\country{Seminarstraße 2, 69117 Heidelberg, Germany}
  \country{Heidelberg, Germany}
}
\affiliation{%
  \institution{National Center for Tumor Diseases (NCT), NCT Heidelberg, a partnership between DKFZ and University Medical Center Heidelberg}
  %\country{Im Neuenheimer Feld 460, 69120 Heidelberg, Germany}
  \country{Heidelberg, Germany}
}

\renewcommand{\shortauthors}{Reinke et al.}

%%
%% The abstract is a short summary of the work to be presented in the
%% article.
\begin{abstract}
%\newpage
\textbf{Abstract:} 
Surgical data science (SDS) is rapidly advancing, yet clinical adoption of artificial intelligence (AI) in surgery remains limited, with inadequate validation as an important contributing factor. Existing validation practices often neglect the temporal and hierarchical structure of intraoperative videos, yielding misleading or clinically irrelevant results. We introduce a comprehensive catalogue of validation pitfalls in AI-based surgical video analysis, derived from a multi-stage Delphi process with 92 international experts. Pitfalls span three categories: (1) data, (2) metric selection/configuration, and (3) aggregation and reporting. A systematic review of surgical AI papers reveals that these pitfalls are widespread. Experiments on surgical video datasets show that ignoring temporal and hierarchical data structures can understate uncertainty, obscure critical failure modes, and alter algorithm rankings. To address these shortcomings, we provide consensus-based best practices compiled. Together, this work provides an evidence-based framework for rigorous validation of surgical video analysis algorithms, guiding benchmarking, reporting, regulatory review, and clinical translation.

\end{abstract}

%%
%% Keywords. The author(s) should pick words that accurately describe
%% the work being presented. Separate the keywords with commas.

%%
%% This command processes the author and affiliation and title
%% information and builds the first part of the formatted document.
\maketitle
\setlength{\parskip}{0.5em}

\section{Introduction} \label{sec:main}
Surgical data science (SDS) was formally introduced in 2017 as a distinct field at the intersection of surgery, data science, and artificial intelligence (AI) \cite{maier2017surgical}. The highly interdisciplinary domain leverages data acquisition, analysis, and modeling to enhance surgical decision-making, execution, training, and patient outcomes throughout the surgical care pathway. Since then, SDS has shown remarkable growth in AI-based publications \cite{li2025artificial} (e.g., \cite{hashimoto2019computer, madani2022artificial, kiyasseh2023vision, khan2024artificial}). However, clinical translation of SDS methods remains limited. For example, while over 1,000 AI-enabled medical devices have received authorization from the US Food and Drug Administration (FDA) since 2017, only six AI-enabled devices have been specifically approved by the FDA General and Plastic Surgery Devices panel, and only 17 for the field of gastroenterology-urology \cite{FDA} as of 2025. 

Multiple barriers limit the clinical translation of SDS methods. These mainly include regulatory and economic hurdles, workflow and hardware integration, and clinical utility. Although barriers are manifold, a key cross-cutting concern can be observed in the lack of robust and rigorous validation of AI algorithms \cite{burger2024unmet, madai2021artificial, chouffani2024not, gong2025knowledge, bernstam2022artificial, olaye2023gap, carstens2025artificial}. Failures in more mature domains such as radiology and clinical decision support illustrate how inadequate validation can contribute to unsuccessful real-world implementation. For instance, IBM Watson for Oncology was validated on narrow, curated cases from a single institution, limiting generalizability and undermining clinical safety and economic viability \cite{strickland2019ibm}. Google Health’s diabetic retinopathy system achieved high retrospective accuracy, yet deployment revealed unvalidated workflow constraints (e.g., image acquisition conditions and throughput), reducing usability without improving clinical benefit \cite{beede2020human}. The Epic Sepsis Model, deployed on proprietary internal validation, degraded substantially under independent external validation, raising concerns about patient safety and regulatory oversight \cite{wong2021external}. Closer to surgery, computer-aided polyp detection systems were validated on improved detection metrics, yet a randomized trial in community-based practice showed no improvement in clinically relevant adenoma detection, while increasing irrelevant findings and procedure time \cite{wei2022boxpolyp, berzin2023navigating}.

These examples illustrate that validation limited to retrospective accuracy or narrow datasets fails to capture clinical utility, robustness, generalizability, and workflow impact that determine real-world adoption. Policy reports and reviews likewise identify insufficient validation as a recurring challenge to AI deployment in healthcare \cite{europeancommission2025}, and a \textit{Nature Medicine} opinion piece described a "benchmarking crisis" in biomedical machine learning, highlighting the lack of standardized validation as a bottleneck impeding progress and patient benefit \cite{mahmood2025benchmarking}. Validation methodology has also been addressed in the computer vision and video analysis literature for specific tasks and benchmarks, including video analysis, indexing,  retrieval, and video object segmentation (e.g., \cite{awad2021trecvid, perazzi2016benchmark, joly2007argos}). However, these works focus on improving validation of individual tasks or benchmarks rather than systematically identifying recurring validation pitfalls that generalize across tasks.

\begin{figure}[h]
    \centering
    \includegraphics[width=1\linewidth]{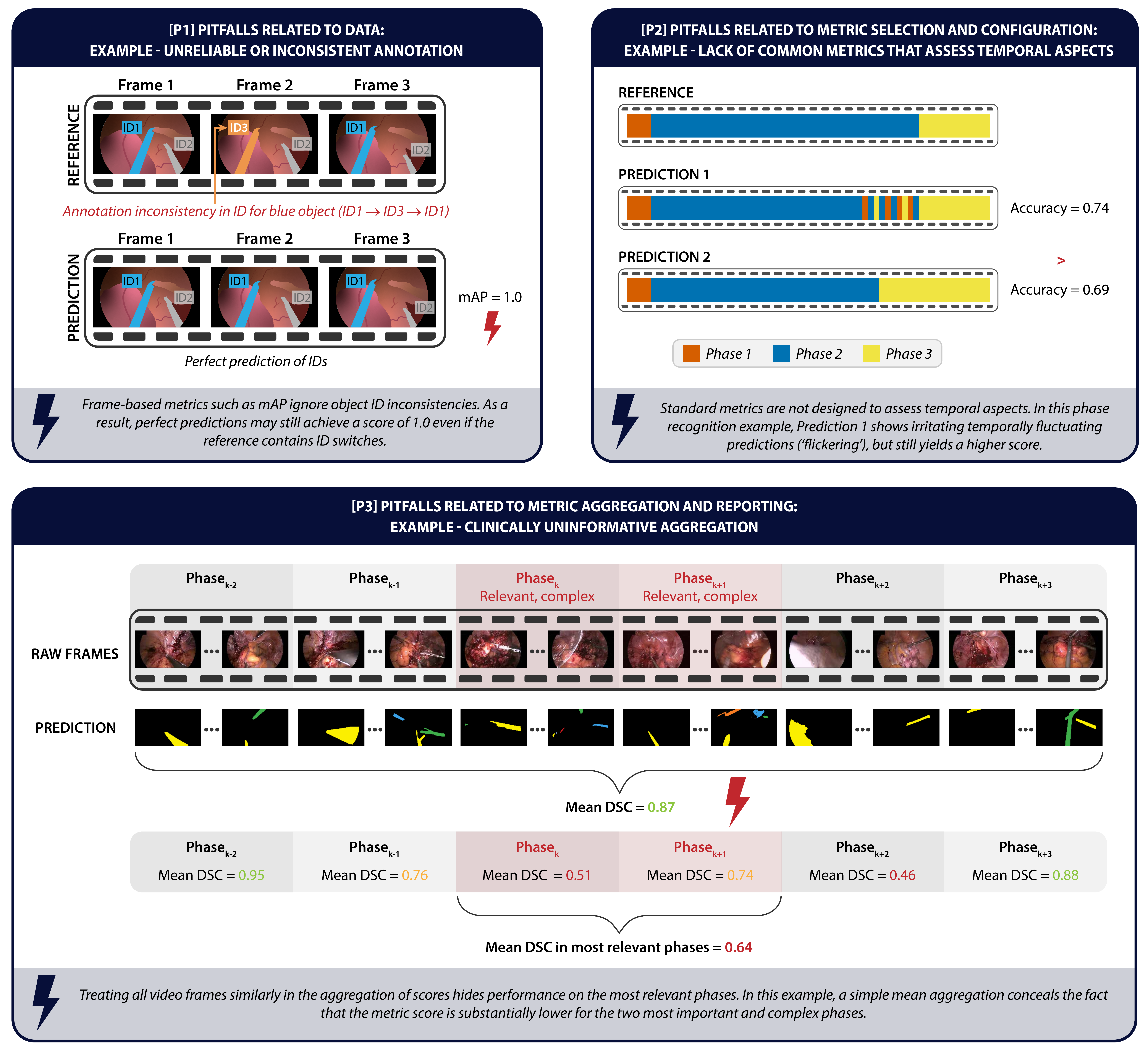}
    \caption{\textbf{Examples of validation pitfalls in surgical video analysis related to data, metric selection \& configuration, and metric aggregation \& reporting.} (a) Inconsistent object identifiers (IDs) in the reference can invalidate performance assessment. Frame-based metrics such as mean Average Precision (mAP), which do not assess object identity, may not reveal annotation errors. Conversely, tracking-based metrics would penalize correct predictions because errors originate from the reference annotation. (b) Lack of common metrics that assess temporal aspects: Standard metrics such as Accuracy were not designed to assess temporal aspects. In this example, \textit{Prediction 1} shows rapid temporal fluctuations ('flickering') in the phase predictions, i.e., predictions that alternate rapidly between correct and incorrect phases across consecutive frames in a task with expected temporal continuity, but still yields a higher Accuracy compared to \textit{Prediction 2} with a temporally more consistent result. (c) Clinically uninformative aggregation: In this example, aggregating the Dice similarity coefficient (DSC) for instrument segmentation with a simple mean over all frames conceals the fact that the DSC is substantially lower for the two most important and complex phases of the procedure. Shown frames were selected from the Heidelberg colorectal data set for surgical data science in the sensor operating room (HeiCo) that were used for the Robust Medical Instrument Segmentation challenge (RobustMIS) data \cite{maier2021heidelberg, ross2021comparative}.}
    \label{fig:fig1}
\end{figure}

Validation outcomes strongly depend on how clinical objectives are translated into quantitative assessment, in particular through the choice of performance metrics. 
Previous work has shown that inappropriate metrics can compromise the validity in AI-driven medical image analysis \cite{carstens2025artificial, reinke2024understanding, funke2023metrics, kofler2023we, vaassen2020evaluation}. For example, metrics that do not reflect the underlying biomedical research question can severely undermine the validity of validation outcomes \cite{reinke2021common, reinke2024understanding}. To address this problem, the global \textit{Metrics Reloaded} initiative introduced metric-related pitfalls and a problem-aware metric recommendation framework for classification, detection, and segmentation \cite{maier2024metrics}.

While this work has gained much support in the research community, it was designed for image-based problems. Unlike radiology or digital pathology, which are image-centered, AI-enabled optimization of intraoperative surgical behaviors is largely dependent on the analysis of spatiotemporal (video) data of the surgical field. Figure~\ref{fig:fig1} exemplifies three consequences of neglecting temporal aspects during validation:
\begin{enumerate}[label=(\alph*)]
    \item annotation inconsistencies across frames can result in misleading metric values, even if predictions are correct,
    \item commonly used metrics may fail to capture temporal stability and therefore not distinguish between temporally stable and rapidly fluctuating ('flickering') predictions in tasks where temporal continuity is expected, and
    \item simple, un-weighted aggregation over video frames may obscure poor performance during clinically critical phases.
\end{enumerate}

Despite their importance, our findings reveal that temporal relations are rarely considered in practice. This becomes particularly apparent during result aggregation, where videos are typically split into frames treated as independent images, thus ignoring temporal continuity and structural dependencies. However, surgical videos contain highly redundant, strongly correlated adjacent frames. Na\"ive aggregation across such frames violates the assumption of independent and identically distributed (i.i.d.) samples, which underlies many statistical analyses including confidence interval (CI) estimation and significance testing. Consequently, performance estimates can be biased and misleading, particularly when redundant frames dominate over clinically critical but less frequent moments.

The present work aims to advance more rigorous and clinically grounded validation of surgical AI. It began with a workshop at the 2023 annual meeting of the Society of American Gastrointestinal and Endoscopic Surgeons (SAGES), which laid the foundation for translating the \textit{Metrics Reloaded} initiative into the video-centric context of surgery. Over the following three years, we conducted a large-scale Delphi process with more than 90 experts to systematically identify and refine validation pitfalls specific to surgical video analysis.

Specifically, this work makes the following key contributions to the safe clinical adoption of surgical AI:
\begin{itemize}
    \item \textbf{Consensus-based catalog of validation pitfalls:} We introduce a comprehensive, structured single point of access for pitfalls in the validation of surgical AI. The catalogue is organized along the validation pipeline (data, metric selection and configuration, metric aggregation and reporting) and links each pitfall to potential consequences and real-world risks. It resulted from a combined approach including a literature review, agentic internet research, and a consensus-driven expert process.
    \item \textbf{Evidence for high occurrence of pitfalls:} A systematic literature review provides empirical evidence that these pitfalls frequently occur in current surgical AI studies, including widespread reliance on frame-wise validation and naive aggregation strategies that ignore the temporal and hierarchical structure of surgical video data.
    \item \textbf{Experimental demonstration of pitfall impact:} Using surgical data, we experimentally quantify how these pitfalls distort performance assessment, mask critical failure modes, and affect uncertainty estimates and algorithm rankings.
    \item\textbf{Practical, pitfall-specific best practices:} Building on the identified pitfalls and supporting evidence, we derive expert-consensus best practices to support more robust validation practice and reporting in surgical AI based on a multi-stage Delphi process.
\end{itemize}

%------------------------------------
\section{Results}\label{sec:results}
Over the past three years, we conducted a structured process involving 92 experts from surgery, computer vision, and data science across 68 institutions, ensuring broad perspectives across both surgical practice and technical disciplines. The initiative began with a hypothesis-generating workshop with dedicated focus group discussions at the SAGES 2023 meeting before evolving into a multi-stage Delphi process. This iterative process yielded a consensus-based catalog of validation pitfalls for surgical video analysis, supported by a systematic review demonstrating their prevalence and experiments quantifying their impact.

\subsection{A multi-stage, multi-stakeholder Delphi process revealed numerous pitfalls in surgical AI validation}\label{subsec:results-pitfalls}
To systematically identify common validation flaws in surgical AI, we combined empirical evidence with structured expert consensus. Our method comprised three pillars: (1) literature review in PubMed and Google Scholar, (2) agentic internet search tools, and (3) a six-stage Delphi process involving SDS experts and clinicians (see Methods (Section~\ref{sec:methods})). This approach enabled us to compile, refine, and validate a comprehensive list of pitfalls that may compromise surgical AI validation.

\begin{figure}[h]
    \centering
    \includegraphics[width=1\linewidth]{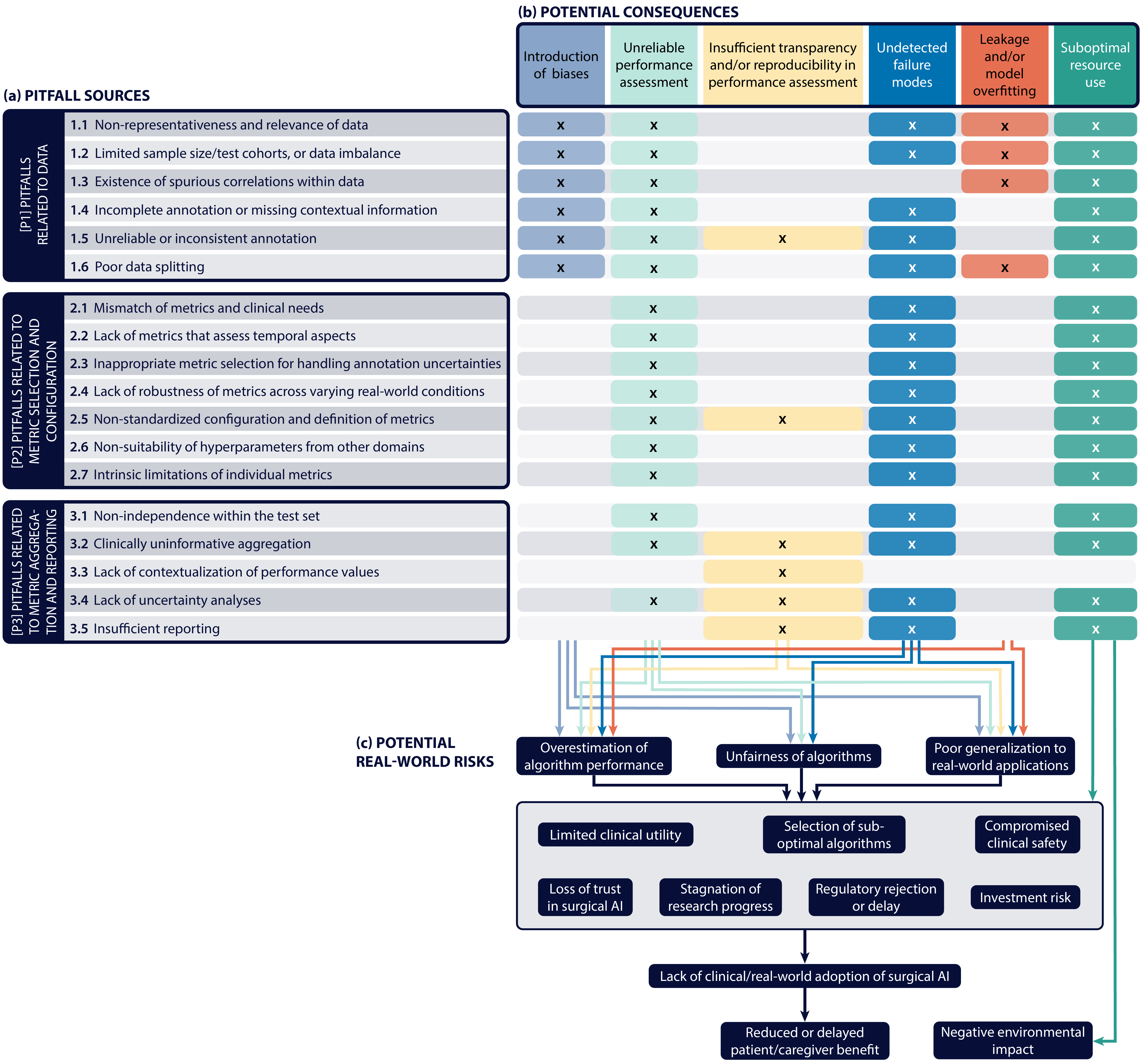}
    \caption{\textbf{Pitfalls related to validation of surgical AI may have severe consequences and real-world risks.} (a) Overview of pitfalls collected in a multi-stage Delphi process involving over 90 experts. Pitfalls were classified into "pitfalls related to data [P1]", "metric selection and configuration [P2]", and "metric aggregation and reporting [P3]". (b) Connections between pitfalls and potential consequences. A colored box marked with an "x" indicates that a pitfall may potentially lead to that consequence. (c) Connections between consequences and potential real-world risks. Lines indicate a "potentially leads to" connection between consequences and risks. Descriptions for each pitfall as well as consequences and risks can be found in Supplementary Tables~\ref{tab:pitfall-descr}, \ref{tab:consequences}, and \ref{tab:risks}.}
    \label{fig:fig2}
\end{figure}

Figure~\ref{fig:fig2} summarizes the pitfall catalog into three categories: [P1] pitfalls related to data, [P2] pitfalls related to metric selection and configuration, and [P3] pitfalls related to metric aggregation and reporting. Each category represents a distinct level at which validation can be compromised. Each pitfall was mapped to specific potential consequences, such as biases, unreliable performance assessment, or undetected failure modes, and to associated real-world risks (e.g., regulatory delay, compromised surgical safety; Figure~\ref{fig:fig2}). Concrete definitions are provided in Supplementary Table~\ref{tab:pitfall-descr}. Below, we outline the identified pitfalls and highlight why they are particularly critical in surgical AI validation. While some issues are known from other machine learning domains, surgery adds unique layers of complexity.

\paragraph{[P1] Pitfalls related to data} 
Flaws in how data are acquired, curated, or partitioned.
\begin{itemize}
    \item \textit{P1.1: Non-representativeness and low relevance of data:} The lack of representative data is a familiar challenge in machine learning, where diversity is key for reliable validation. In surgery, however, this problem is amplified by diverse sources of variability, including variations across hospitals, operating room (OR) setups, surgical technique, variations in data quality (Extended Data Figure~\ref{fig:extended-fig1}a), pronounced geographical imbalance of available datasets (Extended Data Figure~\ref{fig:extended-fig1}b), and heterogeneous surgical teams and experience levels.
    \item \textit{P1.2: Limited sample size/test cohorts, or data imbalance:} Sample size limitations are widely discussed in medical imaging AI (e.g., \cite{christodoulou2025false}), but surgical data is particularly hard to collect. Many procedures are not routinely recorded, and rare, but safety-critical events may appear only occasionally. Single videos can last hours and require extensive annotation \cite{carstens2025artificial}. Moreover, the resulting datasets often exhibit class imbalance and small, heterogeneous test cohorts, which can lead to unstable and unreliable performance estimates (Extended Data Figure~\ref{fig:extended-fig2}).
    \item \textit{P1.3: Existence of spurious correlations within data:} Vision models often exploit accidental cues, but surgical datasets introduce additional confounders such as specific scopes, surgical team compositions, or OR layouts (Extended Data Figure~\ref{fig:extended-fig3}a).
    \item \textit{P1.4: Incomplete annotation or missing contextual information:} Incomplete labels reduce validity generally, but surgical videos heavily depend on temporal context, which is compromised by annotating only a fraction of frames (Extended Data Figure~\ref{fig:extended-fig3}b).
    \item \textit{P1.5: Unreliable or inconsistent annotation:} High inter-rater variability is common in medical imaging \cite{kolbinger2025appendix300, joskowicz2019inter, funke2023metrics}, yet surgical tasks are especially ambiguous (e.g., phase transitions, fine tool-tissue interactions). Even trained raters frequently disagree and maintaining consistency across hour-long videos is particularly challenging (Figure~\ref{fig:fig1}a).
    \item \textit{P1.6: Poor data splitting:} Data leakage is a common problem in machine learning. Surgical videos are highly redundant, i.e., one patient generates thousands of dependent frames. Without strict separation between data subsets, results may measure memorization rather than true generalization to unseen procedures (Extended Data Figure~\ref{fig:extended-fig4}).
\end{itemize}

\paragraph{[P2] Pitfalls related to metric selection and configuration}
Flaws in the choice or setup of performance metrics.
\begin{itemize}
    \item \textit{P2.1: Mismatch of metrics and clinical needs:} Choosing metrics that reflect clinical objectives is important for every research field \cite{maier2024metrics}. However, the gap between existing metrics and needs is particularly wide for surgical applications. Systems must operate in real time, keep latency within safe limits, and produce outputs that remain temporally stable across rapidly changing scenes. Existing measures rarely capture them (Extended Data Figure~\ref{fig:extended-fig5}), and, for many clinically relevant aspects, no established metric may yet exist.
    \item \textit{P2.2: Lack of common metrics that assess temporal aspects:} Temporal reasoning is crucial in surgery. However, standard metrics operate on a frame level or by simply aggregating frames without considering temporal dynamics (Figure~\ref{fig:fig1}b). Consequently, errors during safety-critical phases, or instability over time, may remain hidden.
    \item \textit{P2.3: Inappropriate metric selection for handling annotation uncertainties:} Ambiguous labels occur across domains, but phase boundaries or subtle tool-tissue contacts, among others, make ambiguity even more complex in surgical videos. In surgical reality, human spatiotemporal understanding is often associated with considerable inter-rater variability and inherent uncertainty \cite{kolbinger2025appendix300}. Metrics assuming confident labels can either exaggerate or underestimate errors (Extended Data Figure~\ref{fig:extended-fig6}).
    \item \textit{P2.4: Lack of metric robustness across varying real-world conditions:} Metrics should be consistent across real-world conditions. In surgery, even well-defined measures may behave inconsistently when OR conditions vary – for example, when lighting changes, smoke, or blood partially obscure the field of view, the camera moves, or objects change in size or move in and out of view. Such factors can distort point estimates, despite stable model behaviors (Extended Data Figure~\ref{fig:extended-fig7}).
    \item \textit{P2.5: Non-standardized configuration and definition of metrics:} In many AI applications, inconsistent thresholds or averaging rules reduce comparability. In surgery, even slight differences in how a metric is configured, such as overlap thresholds or smoothing windows, can obscure failures in short, safety-critical steps (e.g., vessel clipping) or make studies with the same metric incomparable (Extended Data Figure~\ref{fig:extended-fig8}a).
    \item \textit{P2.6: Non-suitability of hyperparameters from unrelated domains:} Translating hyperparameters from generic vision tasks is common practice to support standardization. However, in surgical videos, object sizes, motion speed, and safety requirements differ; often, a coarser threshold is already sufficient to track where instruments or anatomy are located within the scene, while overly strict settings may conceal whether an algorithm can follow events robustly over time (Extended Data Figure~\ref{fig:extended-fig8}b).
    \item \textit{P2.7: Intrinsic limitations of individual metrics:} Every metric comes with limitations. Translating standard metrics to surgical video analysis introduces additional challenges. Single scores may overlook brief but high-risk errors, fail to capture stability across time, or ignore how mistakes propagate through multi-step procedures.
\end{itemize}

\paragraph{[P3] Pitfalls related to metric aggregation and reporting}
Flaws in summarizing and presenting results.
\begin{itemize}
    \item \textit{P3.1: Non-independence within the test set:} Correlated samples are a known concern in performance validation. In surgical video analysis, however, thousands of adjacent frames or several clips from the same patient may appear in the test set, inflating apparent confidence and masking how a system behaves on genuinely new procedures (Figure~\ref{fig:fig4}).
    \item \textit{P3.2: Clinically uninformative aggregation:} Aggregating scores is common practice, but simple averaging over all frames can hide poor performance during high-risk phases (Figure~\ref{fig:fig1}c) or overweight patients with longer procedures, obscuring performance on shorter, potentially riskier cases. The lack of stratification by clinically relevant conditions further conceals failure modes that may only appear under specific challenges or surgical contexts (Figure~\ref{fig:fig5}).
    \item \textit{P3.3: Lack of contextualization of performance values:} Point estimates without context are problematic in any field, but even more so in surgical AI, as they can be highly misleading (Figure 6 and Extended Data Figure~\ref{fig:extended-fig9}a). For example, a high Accuracy may mainly reflect routine phases, while errors cluster in moments of adverse events.
    \item \textit{P3.4: Lack of uncertainty reporting:} Uncertainty estimates are often neglected in AI validation. For surgical systems, missing information on confidence or calibration limits the clinicians’ ability to decide when model outputs can be trusted during an operation  (Extended Data Figure~\ref{fig:extended-fig9}b). This effect is even more critical if hierarchical data structures are not considered (see P3.1).
    \item \textit{P3.5: Insufficient reporting:} Sparse or incomplete reporting undermines reproducibility everywhere, yet for surgical applications, the consequences are immediate. Without clear descriptions of data sources, inclusion criteria, metric definitions, and aggregation methods (Figure~\ref{fig:fig7}), it is impossible to judge whether results cover critical steps or rare complications.
\end{itemize}

\subsection{Validation flaws are widespread in common practice}\label{subsec:results-literature}
\begin{figure}[h]
    \centering
    \includegraphics[width=0.6\linewidth]{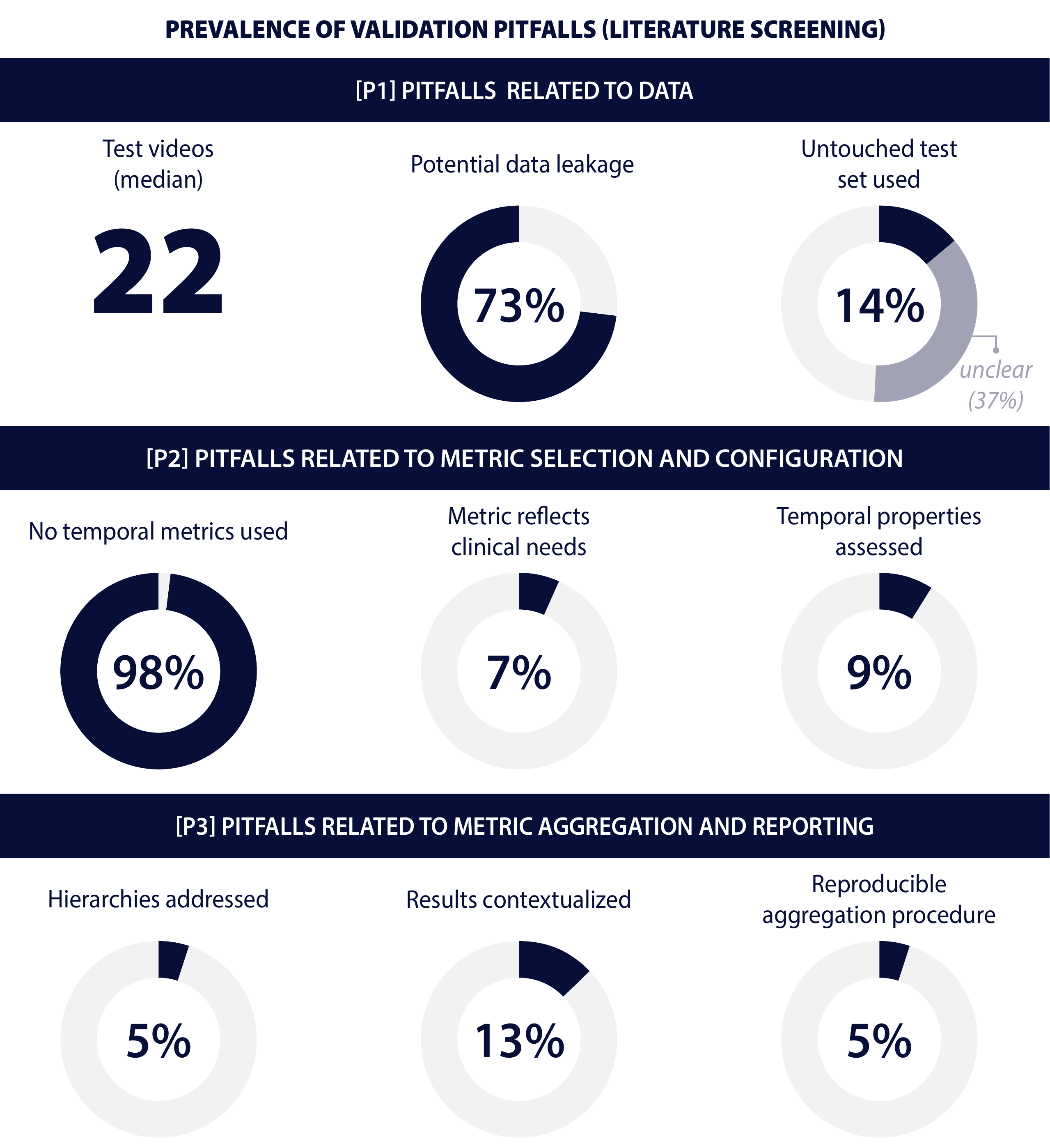}
    \caption{\textbf{Validation and reporting flaws are widespread in common practice.} Selected key insights from a literature screening of 2023 Medical Image Computing and Computer Assisted Intervention (MICCAI) conference surgical data science papers (n = 46) demonstrate that validation and reporting flaws are widespread across all three pitfall categories: [P1] data, [P2] metric selection and configuration, and [P3] aggregation and reporting.}
    \label{fig:fig3}
\end{figure}

While pitfalls can occur in any study, their prevalence in state-of-the-art surgical AI publications remained unclear. To address this, we systematically screened all papers at the 2023 Medical Image Computing and Computer Assisted Intervention (MICCAI) conference, the leading international venue for the field and a representative sample of current practice. Key results are summarized in Figure~\ref{fig:fig3} and \ref{suppl-screening}.

Of all papers meeting the inclusion criteria (n = 46), 74\% used surgical video data. The screening revealed shortcomings across datasets, metrics, aggregation, and reporting.

\textbf{Surgical data cohorts were typically small and fragmented.} The median dataset contained 37 training, 10 validation, and 22 test videos (minimum: 6, 2, and 2 videos, respectively). 79\% of datasets were only used once, with 38 distinct datasets across papers. Only 47\% explicitly reported an untouched test set, while this was unclear in 37\% of the papers.

\textbf{Temporal and modality-specific considerations were largely missing.} 77\% of papers did not assess properties specific to temporal data, and only a single paper used a temporal consistency metric (Extended Data Figure~\ref{fig:extended-fig10}).

\textbf{Metric use was heterogeneous and rarely justified.} Across all papers, 41 metrics were used only once (Extended Data Figure~\ref{fig:extended-fig10}). The most commonly used metric was Accuracy. Only 30\% properly justified their metric choice, and 20\% relied on popularity alone. In addition, in 80\% of papers, it was unclear whether clinical relevance had been considered when selecting metrics.

\textbf{Aggregation practices were insufficient.} Aggregation procedures were unclear or not described at all in 66\% of papers. Among studies involving hierarchical structures, (e.g., patient-level), only 5\% explicitly accounted for their dependencies. 80\% did not contextualize performance values, for example against human baseline or clinical thresholds.

\textbf{Reporting was incomplete and rarely reproducible.} 59\% of papers did not (fully) report dataset sizes. Only one paper reported CIs, and one reported prediction intervals. Notably, 98\% of papers did not report inter-rater variability. Ethical, legal, and social aspects were largely absent; 78\% lacked ethical reporting, 89\% ignored fairness or biases, and 91\% omitted social, legal, or governance considerations. Ultimately, only one paper reported sufficient detail to enable reproducibility; all others lacked relevant details in one or several aspects.

To assess the robustness of our findings over time, we conducted an additional targeted screening of MICCAI 2025 SDS papers (n = 89) using a simplified screening protocol (Methods Section~\ref{sec:methods}). Overall, results were largely consistent. Differences were generally small, with most values differing by only several percentage points, and no systematic trend towards improved validation practices. A comparison of the 2023 and 2025 screenings is provided in Supplementary Table~\ref{tab:2023-2025-comp}.

Notable deviations were only observed for a few parameters. The proportion of papers with unclear reporting of an untouched test set increased substantially (37\% to 74\%), and the proportion of metrics used only once also increased markedly. Other characteristics, including dataset sizes and general reporting patterns, remained broadly comparable.

\subsection{Experiments demonstrate consequences of pitfalls using real-world data}\label{subsec:results-experiments}
To move beyond theoretical examples, we experimentally investigated the consequences of selected pitfalls using representative surgical datasets. As surgical videos are typically long, temporally structured, and safety-critical, the manner in which results are aggregated and reported can strongly influence the visibility and interpretation of algorithm weaknesses. Given the video- and time-sensitive nature of surgical video data, and the limited empirical evidence in the literature, we focused our experiments on pitfalls concerning metric aggregation and reporting [P3]. All experimental procedures are described in the Methods (Section~\ref{sec:methods}).

\subsubsection*{Dependent test samples inflate confidence}\label{subsec:exp-p31}
Surgical data are inherently hierarchical. Frames from a single video are not independent, as they share a patient and are influenced by the performing surgeon, the hospital, or the used surgical tools. Ignoring this dependence can distort the confidence intervals (CIs) that characterize model uncertainty and that are explicitly required for assessing model reliability and clinical readiness in medical imaging AI, including under U.S. Food and Drug Administration (FDA) regulatory frameworks \cite{food2019recommended, us2007statistical, 21cfr8922060, 21cfr8922070, christodoulou2024confidence}. Yet as shown above, only a fraction of studies addresses hierarchical data, and little empirical evidence exists on how this affects CIs under realistic dependencies. 

\begin{figure}[h]
    \centering
    \includegraphics[width=1\linewidth]{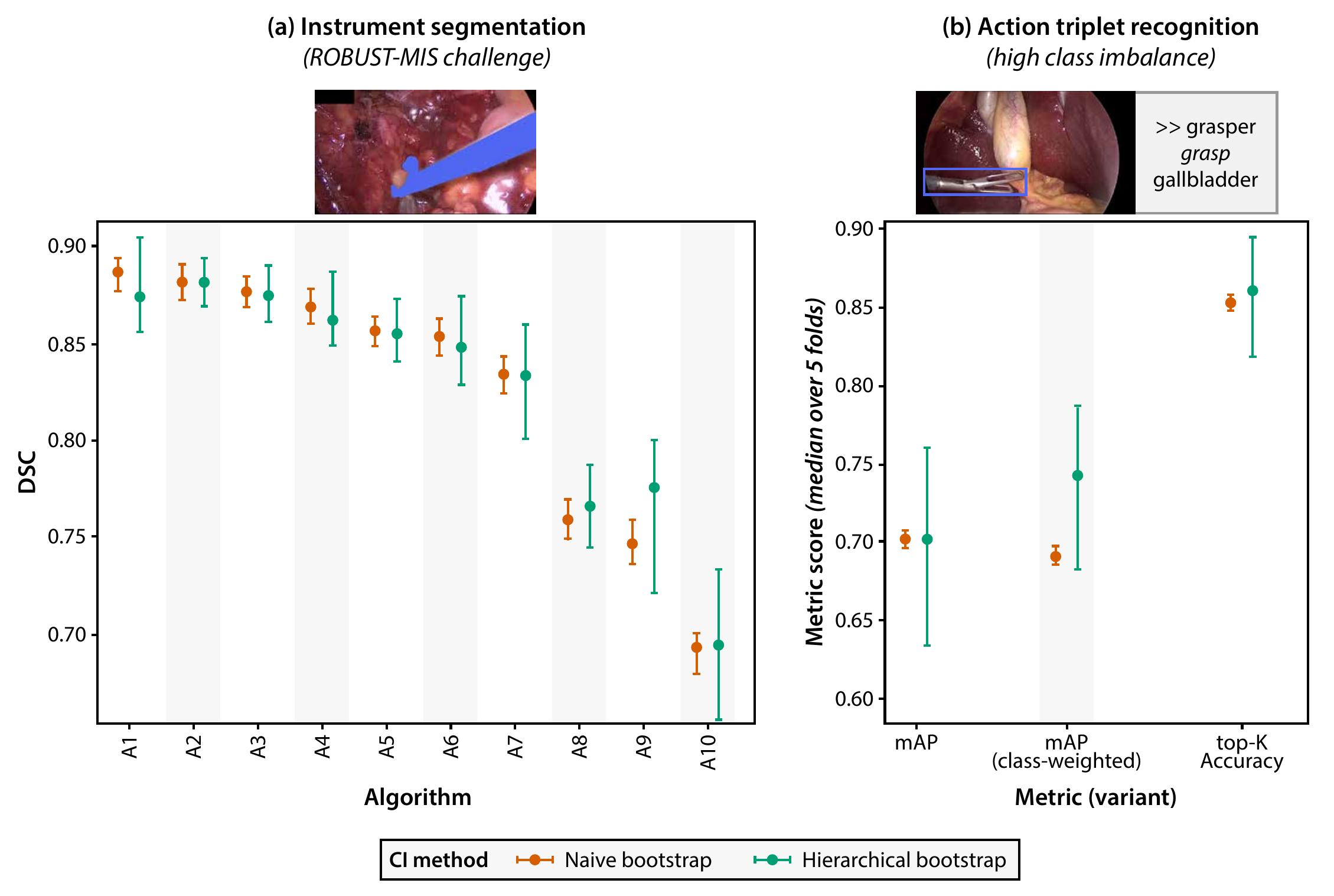}
    \caption{\textbf{Common practice leads to large underestimation of confidence intervals.} Experimental evidence for two representative tasks ((a) binary instrument segmentation (example frame taken from the Heidelberg colorectal data set for surgical data science in the sensor operating room (HeiCo) that was used for the Robust Medical Instrument Segmentation challenge (RobustMIS) data \cite{maier2021heidelberg, ross2021comparative})) and (b) action triplet recognition (example frame taken from the Cholectriplet2021 data \cite{nwoye2023cholectriplet2021})). Confidence intervals (CIs) were computed either per na\"ive bootstrap, assuming all samples were independent (orange), or with a hierarchical bootstrap that accounts for the inherent hierarchical data structure (green), introduced by the dependencies between frames originating from the same video (i.e., patient case). The na\"ive approach only yields narrow CIs and underestimates uncertainty, whereas the hierarchical bootstrap produces wider, more reliable CIs. Points indicate the mean DSC performance, error bars represent 95\% CIs obtained from 1,000 bootstrap iterations. Na\"ive bootstrap resampled individual frames, whereas hierarchical bootstrap resampled videos and subsequently frames within videos. The analyses were based on a total of n = 2,231 frames from ten videos in (a) and n = 3,946 frames from 45 videos over five folds in (b). Abbreviations: Dice similarity coefficient (DSC), mean Average Precision (mAP).}
    \label{fig:fig4}
\end{figure}

To determine this impact, we analyzed (1) the Robust Medical Instrument Segmentation (RobustMIS) 2019 challenge dataset \cite{ross2021comparative} for binary instrument segmentation, which is a large-scale benchmark dataset with multi-team predictions, and (2) the CholecTriplet dataset \cite{nwoye2023cholectriplet2021} for surgical action triplet recognition, which is one of the most frequently used SDS datasets \cite{carstens2025artificial}. For both tasks, we compared CIs derived from a na\"ive bootstrap approach, which does not account for hierarchical dependencies, against CIs from a hierarchical bootstrap (Figure~\ref{fig:fig4}). 

Accounting for hierarchical data structure led to substantially wider CIs, which reflects the additional variance introduced at the video (i.e., patient) level that is ignored when assuming independence across all samples (na\"ive approach). Concretely, for binary segmentation, the CI widths increased by a median factor of more than 2 for both Dice similarity coefficient (DSC; 2.3x wider) and Normalized surface Dice (NSD; 2.1x wider). For surgical action recognition, CIs were 13.3x wider for mean Average Precision (mAP), 10.9x wider for weighted mAP, and 7.2x wider for top-5 Accuracy. These findings demonstrate that ignoring data dependencies can substantially understate model uncertainty, potentially giving a false sense of algorithm reliability in surgical settings.

\newpage
\subsubsection*{Averages hide critical failures}\label{subsec:exp-p32}
Many studies in surgical AI summarize results as a single overall score, averaging performance across all frames or cases. While this practice is convenient, there is little evidence on how such aggregation may conceal errors under conditions where reliability is most critical for patient safety. In surgery, visual and technical challenges, such as smoke or rapid tool motion, can strongly affect algorithm robustness, yet these factors are rarely analyzed in validation reports.

\begin{figure}[h]
    \centering
    \includegraphics[width=1\linewidth]{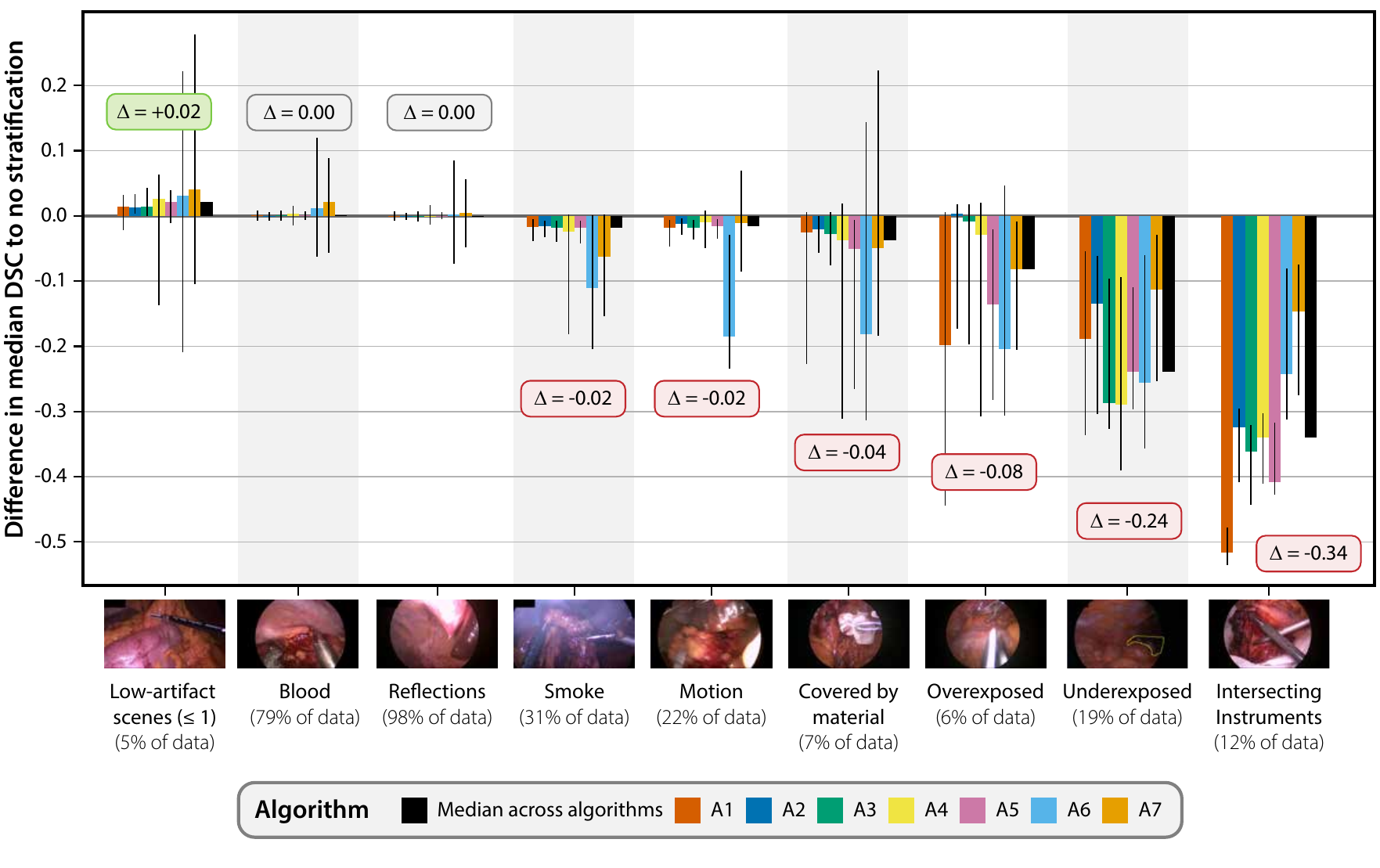}
    \caption{\textbf{Lack of stratification of performance values hides performance drops for relevant, potentially confounding image properties.} The bar plot shows the difference in median instance Dice similarity score (DSC) for the task of surgical instrument instance segmentation between stratified and unstratified validation across algorithms (A1-A7) as well as their median performance (black bar). Hierarchical 95\% confidence intervals (error bars; 1,000 bootstrap replicates) quantify the uncertainty of the estimated median performance differences for each sub dataset and were obtained by resampling surgeries and subsequently frames within surgeries with replacement. Sample sizes for the respective imaging-property subsets were: low-artifact scenes (n = 119), blood (n = 1,765 frames), reflections (n = 2,190), smoke (n = 702), motion (n = 497), covered by material (n = 158), overexposed (n = 126), underexposed (n = 429), intersecting instruments (n = 260), total (n = 2,231). The performance varies substantially across different challenging conditions such as motion or underexposure. Here, algorithms show substantial drops in DSC for cases with potentially confounding imaging properties. The median delta in performance is provided per image property ($\Delta$). For this example, the results of the seven algorithms (A1 - A7) from the multi-instance segmentation task of the Robust Medical Instrument Segmentation (RobustMIS) \cite{ross2021comparative} were used. Shown example frames were selected from the Heidelberg colorectal data set for surgical data science in the sensor operating room (HeiCo) that were used for the Robust Medical Instrument Segmentation challenge (RobustMIS) data \cite{maier2021heidelberg, ross2021comparative}.}
    \label{fig:fig5}
\end{figure}

To shed light on this problem, we compared global results with stratified analysis on multi-instance instrument segmentation results from the RobustMIS challenge \cite{ross2021comparative}, using metadata describing various relevant, potentially confounding image properties \cite{ross2023beyond} (Figure~\ref{fig:fig5}). While aggregated DSC scores suggested stable performance across algorithms, stratification by clinically relevant conditions revealed considerable performance drops. For instance, the median DSC decreased by 0.34 (up to 0.52 for one algorithm) in frames with intersecting instruments, and smaller but clear declines appeared for smoke and motion artefacts. Purely reporting globally aggregated values can therefore be highly misleading, whereas stratification exposes failure cases in safety-critical situations.

\newpage
\subsubsection*{Missing context hides temporal errors}
Temporal grounding is becoming increasingly relevant for surgical video understanding, particularly with the emergence of long-context vision-language models. In temporal grounding tasks, predictions are often considered correct if they fall within a predefined temporal tolerance around the reference timestamp (e.g., \cite{deliege2021soccernet}). While this enables concise performance reporting, it provides little information about the magnitude of temporal errors beyond the acceptance threshold. However, whether a prediction misses the reference by a few seconds or by several minutes can make a substantial difference for its practical usefulness.

\begin{figure}[h]
    \centering
    \includegraphics[width=1\linewidth]{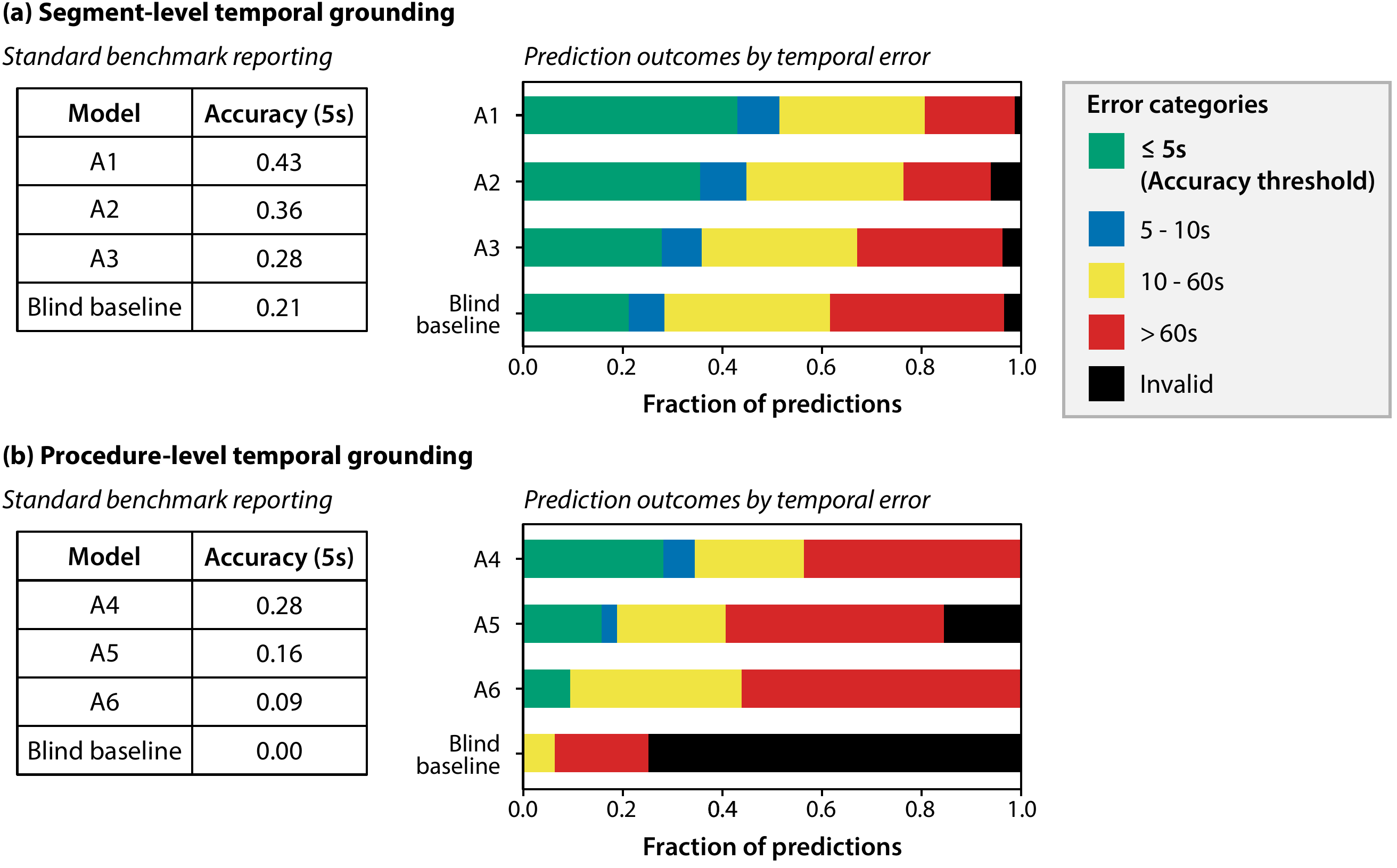}
    \caption{\textbf{Threshold-based reporting provides limited information about temporal prediction errors.} Standard reporting using Accuracy at a 5s threshold is compared with temporal error distributions for the temporal grounding task on the HeiCo-FOCUS benchmark \cite{Heicofocus} using the (a) 'Segment' and (b) 'Procedure' tracks. While Accuracy reports whether predictions fall within the predefined acceptance threshold, temporal error distributions additionally quantify the magnitude of temporal deviations and invalid predictions. The dataset contained 377 question-answer pairs for the 'Segment' track and 32 question-answer pairs for the 'Procedure' track.}
    \label{fig:fig6}
\end{figure}

To investigate this limitation, we analyzed temporal grounding results from our recently proposed HeiCo-FOCUS benchmark \cite{Heicofocus} (Figure~\ref{fig:fig6}), which evaluated state-of-the-art vision-language models on short- and long-context surgical video understanding, with an additional blind baseline for comparison. Beyond reporting whether predictions fell within a predefined acceptance threshold, we analyzed temporal deviations between predicted and reference timestamps to distinguish small from substantial localization errors.

The analysis revealed additional context of model behavior that was not apparent from the Accuracy scores alone. Across models, the temporal error distributions showed whether prediction errors were predominantly close to the acceptance threshold or represented substantial temporal deviations, while also revealing the frequency of invalid predictions. This additional context provides a more informative characterization of prediction errors than a binary threshold-based classification.

\newpage
\subsubsection*{Aggregation choices can flip the winner}\label{subsec:exp-p35}
Surgical video analysis involves several hierarchy levels and the way results are aggregated across them can substantially affect reported performance. Yet, as shown in Section~\ref{subsec:results-literature}, the majority of studies do not explain the exact aggregation procedure, leaving readers unable to judge whether rankings or scores reflect clinically meaningful behavior. Here, we systematically assessed how different aggregation strategies influenced conclusions.

\begin{figure}[h]
    \centering
    \includegraphics[width=0.8\linewidth]{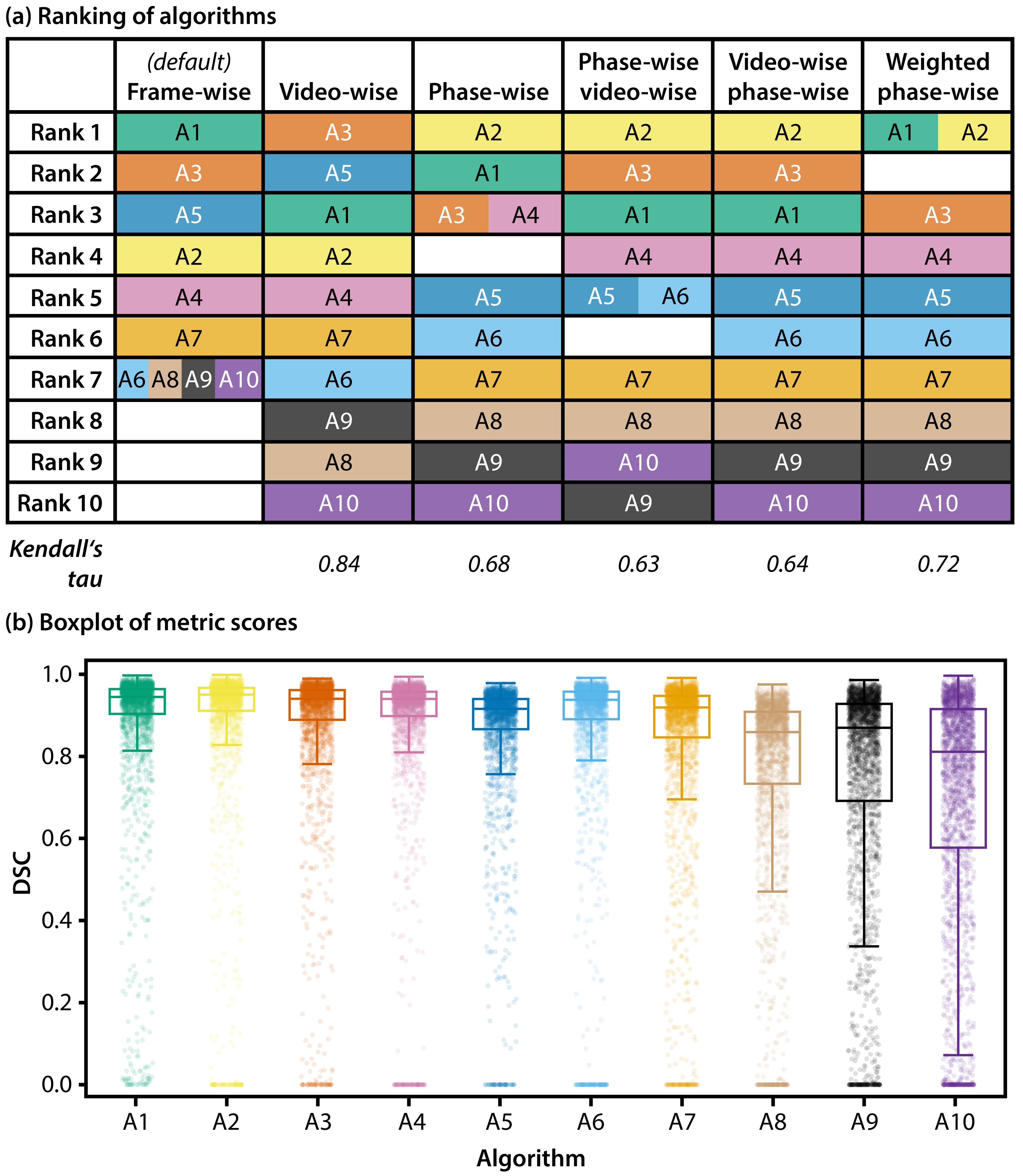}
    \caption{\textbf{Different validation strategies lead to varying algorithm rankings.} (a) Different aggregation strategies such as over all frames (frame-wise aggregation), over videos (video-wise aggregation), or over phases (phase-wise aggregation) produce different rankings. Kendall’s tau is shown in comparison to the default rankings (frame-wise). Similarly to the original challenge, we used the 5\% percentile as aggregation operator to reflect worst-case performance. For this example, the results of the ten algorithms (A1 - A10) from the binary segmentation task of the Robust Medical Instrument Segmentation (RobustMIS) \cite{ross2021comparative} were used. (b) Corresponding boxplots of per-frame metric scores for the same algorithms. Box centres indicate the median, boxes the interquartile range (IQR; 25th-75th percentiles), whiskers the most extreme values within 1.5$\times$ the IQR, and light dots the individual per-frame DSC scores per algorithm. The analysis included a total of n = 2,231 frames from ten surgical procedures.}
    \label{fig:fig7}
\end{figure}

We analyzed results from the binary segmentation task of the RobustMIS challenge \cite{ross2021comparative} and applied six different aggregation strategies: frame-wise, video-wise, phase-wise, phase-wise video-wise, video-wise phase-wise, and weighted phase-wise (see Methods (Section~\ref{sec:methods}) for detailed descriptions of each strategy). We then compared the results for each strategy with the default frame-wise aggregation (Figure~\ref{fig:fig7}). Similarly to the original challenge, we used the 5\% percentile as the aggregation operator to reflect worst-case performance.

The median Kendall’s tau correlation \cite{kendall1938new} across the different rankings compared to the default ranking was 0.68, indicating high variance in the leaderboards. The original winner changed in 80\% of rankings, with a median absolute rank change of 1 and a maximum change of 3. Negative rank shifts occurred in 58\% of cases, positive in 28\%, and no change in 14\%. As shown in Figure~\ref{fig:fig7}b, when performance differences between algorithms were modest, even small changes in the aggregation led to substantial ranking shifts, questioning the reliability of the winner. These findings show that minor reporting omissions such as unspecified aggregation can substantially affect ranking conclusions and potentially influence which algorithms are prioritized for clinical translation.

\newpage
\subsection*{Practical best practices to mitigate validation pitfalls}\label{subsec:res-best-practices}
Building on the identified pitfalls and their empirical consequences, the Delphi consortium derived practical, pitfall-specific best practices, reflecting expert consensus across surgical data science. These best practices are designed to support more robust validation practice in surgical video AI by distinguishing between measures to avoid the occurrence of specific pitfalls as well as strategies to appropriately deal with them when these cannot be fully avoided, for example due to data, annotation, or workflow constraints. The best practices are intended to provide a comprehensive, structured overview of methodologically sound validation approaches identified through the Delphi process, acknowledging that their implementation depends on the specific study context and available resources.

\begin{figure}[h]
    \centering
    \includegraphics[width=1\linewidth]{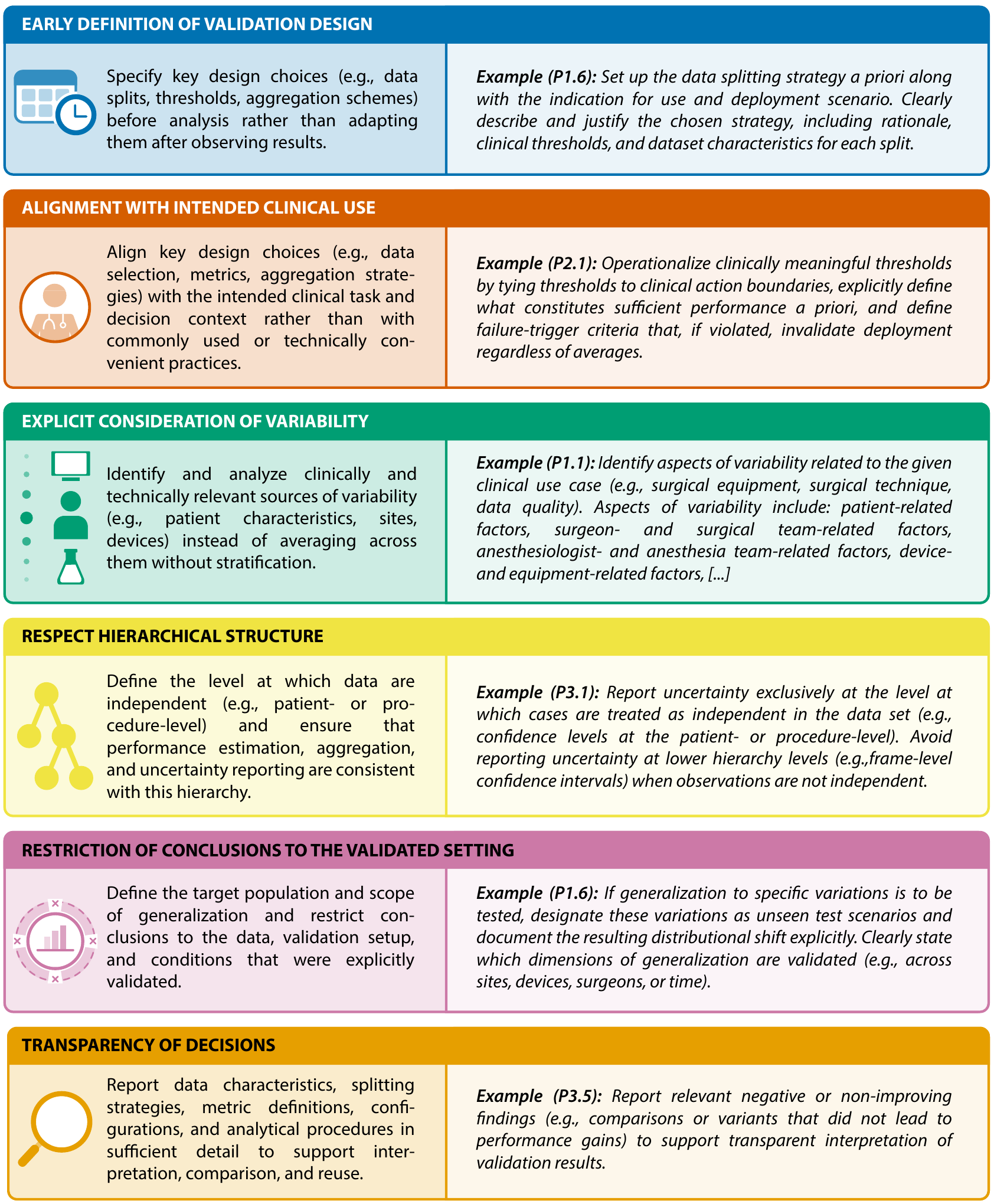}
    \caption{\textbf{Overarching themes across pitfall-specific best practices.} The themes provide a condensed summary of central cross-cutting patterns identified across the best practices developed by the Delphi consortium. For each theme, an illustrative example is provided. The full list of best practices is given in \ref{suppl-best-practices}.}
    \label{fig:fig8}
\end{figure}

The full list of best practices includes 124 items and is provided in \ref{suppl-best-practices}. Across best practices, recurring structural patterns were identified, including early specification of validation design choices, alignment with intended clinical use, explicit consideration of variability, respect for hierarchical data structure, restriction of conclusions to the validated setting, and transparent reporting of decisions (Figure~\ref{fig:fig8}).

\newpage
\section{Discussion}\label{sec:discussion}
Our work provides a comprehensive, expert-driven taxonomy of validation pitfalls in surgical AI, supported by empirical evidence and experiments. By linking methodological flaws to surgical risks, our framework highlights the need for rigorous validation to ensure safe and effective AI deployment. Our findings demonstrate that common validation practices frequently ignore the temporal and hierarchical structure of surgical data, leading to overconfident or clinically irrelevant conclusions. Through a Delphi process with experts from surgery, machine learning, biostatistics, and regulatory affairs, we identified and contextualized 18 critical pitfalls. This multi-stakeholder approach ensured that the collected pitfalls are technically sound and clinically meaningful, making them actionable for algorithm developers through regulatory reviewers. The main contributions of this work are:
\begin{itemize}
    \item We provide a comprehensive single point of access of validation pitfalls in surgical AI, making surgical data scientists and clinicians aware of common limitations.
    \item We quantify their prevalence in SDS literature.
    \item We experimentally demonstrate how selected pitfalls affect reported performance, uncertainty estimates, and algorithm rankings.
    \item We provide consensus-based best practices for improved validation practice.
\end{itemize}
Some of these pitfalls have been discussed in computer vision and machine learning literature, but those insights are scattered across domain-agnostic, technically dense publications and are rarely tailored to surgical data science. Our contribution is to combine expert consensus, empirical literature screening, and experiments into a structured point of access.

The collected pitfalls span all stages of the validation process. For example, at the data level, geographical imbalance of surgical and endoscopic datasets (Extended Data Figure~\ref{fig:extended-fig1}b) shows how limited geographic diversity can restrict representativeness, as many public benchmarks were acquired from only a few regions. Recent initiatives, such as the Critical View of Safety Challenges 2024 and 2025 \cite{alapatt2025sages} improve global inclusion, yet benchmark datasets still predominantly reflect surgical practice patterns from Western regions. Beyond geographical and institutional biases, non-representativeness in surgical video data may also arise from operator-related heterogeneity. In particular, videos recorded in academic or training centers may involve multiple operators with different experience levels (e.g., trainees and attending surgeons), introducing additional variance if operator roles are not annotated or reported. Misalignment between metrics and clinical objectives remains common, as metrics tailored to surgeon-specific needs or temporal aspects are often lacking. Consequently, studies frequently rely on frame-based scores such as Accuracy or DSC, which fail to capture the temporal and hierarchical complexity of surgical workflows and can obscure clinically relevant weaknesses. At the aggregation level, na\"ive frame-wise aggregation across temporally dependent data can mislead performance estimates, conceal critical failure modes and alter algorithm rankings.

We show that the identified pitfalls are widespread across the surgical AI literature. Our systematic screening revealed frequent issues such as unstratified aggregation, lack of uncertainty reporting, unsuitable metric selection, and poor documentation of validation procedures. Despite temporality being inherent to intraoperative surgical workflows and temporal metrics exist, they were rarely used in practice. Although surgical video AI tasks involve complex, structured data with multiple classes, temporal dependencies, and hierarchical structure, validation frequently remains limited to simple Accuracy metrics. As shown by our experiments, na\"ive frame-wise validation can mask critical failure modes and substantially underestimated uncertainty. Together, these findings underscore the need for more rigorous, problem-tailored validation in surgical AI.

Building on our findings, we derived practical, pitfall-specific best practices through Delphi consensus. They encompass measures to avoid specific pitfalls and strategies to address unavoidable limitations transparently. Their applicability depends on the study context, such as available data, annotation resources, and intended use. Accordingly, their relevance varies across method development stages, from exploratory studies to benchmarking and evaluations supporting claims of generalizability or clinical readiness. Importantly, individual high-quality surgical AI studies already demonstrate selected aspects of these best practices. Examples include geographically and procedurally diverse multi-center cohorts with independent external validation (e.g., \cite{lavanchy2023preserving, protserov2024development, goodman2024analyzing}), application-specific performance metrics (e.g., \cite{zia2019novel, dergachyova2016automatic}), and improved validation through appropriate dataset splitting, stratification, and aggregation (e.g., \cite{kostiuchik2024surgical, ross2023beyond}). Our collection is intended to support informed selection of best practices for a given use case rather than an exhaustive checklist or compliance framework, while encouraging transparent documentation of limitations when constraints prevent full implementation. The suggested best practices could support writing and reviewing scientific articles in SDS.

Beyond technical adherence to best practices, clinical translation requires alignment with evolving regulatory frameworks and healthcare stakeholder needs, complicated by a fundamental difference between diagnostic and interventional AI. While diagnostic AI often produces a single, static prediction (e.g., "lesion present"), interventional AI operates in a dynamic, time-dependent environment where the causal link between model output at one timepoint (e.g., a "cut" vs. "no-cut" recommendation) and complication is difficult to isolate. This fuzzy error attribution poses hurdles for traditional validation paradigms and regulatory bodies, which seek a direct mapping of model performance to patient safety. Demonstrating clinical readiness will therefore require validation that moves beyond aggregate metrics toward implementation science approaches accounting for the temporal dynamics of surgical workflow and the interaction between AI guidance and the surgeon’s intraoperative decision-making.

Our work comes with several limitations. While our taxonomy targets surgical video analysis, its generalizability to other temporally structured domains (e.g., cardiology) requires further study. Although our Delphi expert consortium included 92 international experts and covered diverse expertise, it may not have captured the full diversity of surgical subfields and regulatory perspectives. Participation varied across Delphi rounds, raising the potential for biases in the weighting and selection of pitfalls. Our systematic literature review covered only 46 articles. Although the sample size was limited, MICCAI is the leading conference for medical image analysis and computer-assisted interventions and is therefore representative for the field. To assess whether our findings extend beyond a single conference year, we performed an additional targeted screening of 89 MICCAI 2025 SDS papers. This analysis was conducted by a single rater and focused on the key pitfalls reported in the Results rather than the full set of parameters. While less comprehensive than our primary screening, it yielded similar trends across the evaluated aspects, supporting the robustness of our main findings. Our experiments only tackled selected pitfalls. While pitfalls such as data leakage have been demonstrated in the broader machine learning community, we focused on aspects that are especially critical in surgery: aggregation under temporal and hierarchical structures. Other pitfalls are illustrated in smaller analyses in Extended Data Figures~\ref{fig:extended-fig1}-\ref{fig:extended-fig9}. 

Several open research directions emerge from our pitfall taxonomy. Future work should move beyond surrogate metrics toward validation that reflects clinical benefit and patient outcomes. This includes defining sufficient performance within specific clinical contexts and establishing comparability across heterogeneous tasks. Achieving this goal requires close collaboration between AI developers and clinicians. Such multidisciplinary input is essential for interpreting metrics in relation to workflow impact, safety, and patient outcomes. Clear validation phases integrating governance, stakeholder input, and standardized reporting need to be defined and integrated into clinical trial design. Post-deployment monitoring, addressing catastrophic-failure risk, and enabling effective human-AI collaboration in the OR will be equally crucial. From a technical and adoption perspective, progress depends on harmonizing label ontologies and annotation protocols, enabling privacy-preserving multimodal validation, and assessing behavioural consistency across samples and software versions. Finally, embedding clinician priorities, workflow impact, and real-time safety mechanisms should become integral validation goals.

In summary, while \textit{Metrics Reloaded} \cite{maier2024metrics} provided metric recommendations for image-based validation, our framework extends this foundation to pitfalls stemming from the temporal and hierarchical complexity of surgical video analysis. We envision this work as a catalyst for improved validation practice and future benchmarking. By raising awareness of widespread pitfalls, we aim to promote more robust, interpretable, and clinically grounded validation. By systematically mapping validation pitfalls to their consequences and providing practical best practices, this work can support integration of validation quality criteria into clinical trial design, regulatory review, and publication guidelines, such as the development or refinement of reporting standards for medical video analysis (e.g., TRIPOD-AI \cite{collins2021protocol}, DECIDE-AI \cite{vasey2022reporting}, or future domain-specific extensions). Going forward, our consortium will focus on translating these pitfalls into surgery-specific metric and aggregation recommendations, further advancing the reliability and clinical readiness of surgical AI models.

\section{Methods}
\label{sec:methods}

\subsection{Terminology}
Throughout this manuscript, we use the term "surgical" in a procedural sense, consistent with the definition of SDS as encompassing "all clinical disciplines in which patient care requires intervention to manipulate anatomical structures with a diagnostic, prognostic, or therapeutic goal, such as surgery, interventional radiology, radiotherapy, and interventional gastroenterology" \cite{maier2017surgical}. This includes closely related interventional domains such as endoscopy, which share video-centric data, workflows, and validation challenges.

\subsection{Ethics}
This study complied with all applicable ethical regulations and the General Data Protection Regulation (GDPR). The analyses were conducted exclusively on fully anonymized video and frame datasets that did not contain identifiable personal data; therefore, formal ethics committee approval was not required. The Delphi survey likewise did not require formal ethics approval. Participation was voluntary. Before participating, all participants were informed about the purpose of the study, the collection and storage of their data, and their right to withdraw from the study and request deletion of their data at any time. Providing personal information (e.g., participant name) was optional. Participation required informed consent to the processing of personal data.

\newpage
\subsection{Identification of validation pitfalls and best practices through a multi-stage Delphi process and complementary searches}
The pitfalls presented in this work were derived through a combination of approaches, centered on a multi-stage, consensus-driven Delphi process conducted by an international, multidisciplinary panel of experts. A Delphi process is a structured consensus-building approach in which experts provide input individually – typically through questionnaires – followed by rounds of controlled feedback and refinement \cite{brown1968delphi}. This methodology is widely recognized in medicine as a way to establish best practices in areas in which the available evidence is limited, inconsistent, or missing \cite{nasa2021delphi}. 

Our Delphi panel initially included 60 international experts from the SDS initiative. To broaden the diversity of the expertise, the consortium was gradually expanded to 92 members across 99 institutions, reflecting both technical and clinical backgrounds. The expert panel was composed of 30\% clinical, 75\% technical, and 5\% shared expertise. 12\% of experts were from industry. The majority of experts were affiliated in Europe (70\%; mostly Germany (36\%) and United Kingdom (12\%)) and North America (25\%; mostly United States of America (20\%)), followed by Asia (7\%) and Africa (2\%).

This initiative started in March 2023 with a scoping survey identifying the most critical problems in validating surgical AI and use cases lacking suitable metrics or showing discrepancies between metrics and clinical needs (participation rate: 33\%).  Building on the survey, we held a kickoff workshop at the SAGES annual meeting in Montréal, Canada (41\% in-person participation), which refined the project scope, set priorities, and agreed to focus the initiative on surgical video understanding, reflecting shared priorities across clinical and technical stakeholders. The core team then performed targeted literature searches to compile candidate pitfalls, while a joint retreat of three research groups at DFKZ, NCT, and UCL refined preliminary pitfalls from practical experience and interdisciplinary perspectives.

A subsequent systematic search across PubMed, Google Scholar, and a general Google search used the following search string: ("surgical data science" OR "surgical artificial intelligence" OR "surgical AI" OR "surgical scene understanding" OR "surgical video analysis") AND ("validation" OR "evaluation" OR "metric") AND ("pitfall" OR "limitation" OR "caveat" OR "drawback" OR "shortcoming" OR "weakness" OR "flaw" OR "disadvantage"). PubMed returned no relevant results, while Google Scholar yielded 704 hits and the general search 94 results (30 non-peer-reviewed). These searches yielded relevant studies but no comprehensive or structured collection of validation pitfalls.

Based on the first survey, workshop, retreat, and literature review, the core team established a preliminary catalog of pitfalls, including their categorization, which served as the starting point for subsequent refinement through the Delphi process. To reduce blind spots, we complemented this with agentic internet research systems, including the deep research tools from OpenAI (o3-based), Google (Gemini 2.5 Pro-based), and Perplexity Pro, to help identify overlooked pitfalls. Additional suggestions were validated by the consortium for relevance and correctness.

In total, we conducted six Delphi rounds (participation rates: 64\%, 70\%, 48\%, 53\%, 46\%, 60\%). Round 1 confirmed the overall project scope, while round 2 refined the pitfall categorization and list, identified missing pitfalls, and collected supporting references. Round 3 focused on linking pitfalls to consequences and risks, as well as an initial collection of best practices, and round 4 sought final consensus on the pitfall catalog and pitfall categorization (agreement: 98\%), as well as optional feedback on figures, experiments (agreement $\geq$90\%), and open research questions. The core team refined the best practices from round 3, which were finalized in rounds 5 and 6, where only items with agreement above 80\% were kept. In the final round, agreement for the list of best practices was above 90\% for each pitfall.

\subsection{Systematic review for prevalence of pitfalls}
From all MICCAI 2023 papers (n = 730), we identified those papers related to SDS (n = 51). From those, five articles were excluded because these did not deal with deep learning-related methods, therefore, several questions did not apply (remaining n = 46). Each paper was screened by two independent screeners, from a total of twelve screeners. Afterwards, a third senior screener compared the results and resolved conflicts. In total, three senior screeners joined in this last step, the papers were divided among them. The screening covered more general aspects such as task or surgery type, but specifically focused on evidence for the identified pitfalls. In line with the Delphi process and after identifying the final list of pitfalls, a follow-up screening, following the same process, was conducted to ensure evidence for all pitfalls. 

To assess the robustness of our findings over time, we additionally performed a targeted screening of MICCAI 2025 SDS papers (n = 89). In contrast to the primary analysis, this screening was conducted by a single rater and focused on the subset of pitfalls reported in the Results section. The purpose was not to repeat the full screening procedure, but to evaluate whether the main findings hold across a more recent set of publications.

\subsection{Experimental Design}
\subsubsection*{Data}
We based our experiments on two widely used datasets in surgical video analysis \cite{carstens2025artificial} that have been used in international challenges, that are highly cited, and cover two key tasks: instrument segmentation and action recognition. 

The RobustMIS challenge 2019 \cite{ross2021comparative} consists of videos from 30 laparoscopic colorectal surgeries, covering rectal resection, proctocolectomy, and sigmoid resection, with 10 videos per surgery type. For the challenge, the data from rectal resection and proctocolectomy surgeries were used for training and internal testing (stages 1 and 2), while sigmoid resection cases (stage 3) were reserved for assessing generalization to an unseen surgery type. For our experiments, we restricted the analysis to binary and multi-instance segmentation in stage 3 (sigmoid resection). For binary segmentation, ten algorithms were involved, for multi-instance segmentation, seven algorithms participated. The challenge metrics included the DSC, which measures the overlap between prediction and reference, and NSD, which assesses boundary accuracy \cite{maier2024metrics} together with their multi-instance variants (MI\_DSC and MI\_NSD). We had access to the frame-level metric scores submitted by all participating teams, which allowed us to analyze the impact of validation choices across a diverse set of real-world algorithms. All results were used in anonymized form to ensure confidentiality.

In addition to the segmentation masks, we utilized structured meta-annotations for the RobustMIS data, describing the presence of common visual artifacts \cite{ross2023beyond}. These meta annotations indicate whether challenges such as blood, smoke, or motion blur are present per frame and instrument. 

We further used the CholecTriplet challenge data \cite{nwoye2023cholectriplet2021}, which is based on 45 laparoscopic cholecystectomy videos (CholecT45). The task involves recognition of surgical action triplets, with annotations for 100 triplet classes, combining instrument, verb, and target. For our experiments, we used a Swin-Base Transformer trained using multitask learning, incorporating instrument, verb, and target and soft-labels generated using a multi-teacher approach \cite{yamlahi2023self}. Model validation was performed using 5-fold cross-validation, following the official CholecT45 setup. 

In addition, we used the first version of the recent HeiCo-FOCUS benchmark \cite{Heicofocus} for long-context surgical video understanding, comprising 30 colorectal surgical procedures. HeiCo-FOCUS contains multiple validation tracks targeting different long-context reasoning capabilities. For this work, we restricted the analysis to the temporal grounding, in which models are required to identify the temporal location of events within surgical videos. The ‘Segment’ track, comprising 377 question-answer pairs, contains tasks on short-term understanding within video segments of up to 5 minutes, whereas the ‘Procedure’ track, comprising 32 question-answer pairs, contains tasks on long-context understanding over complete surgical procedures of up to 296 minutes.

\subsubsection*{Experiment 1: Dependent test samples inflate confidence}
This experiment investigated how ignoring data dependencies in temporally structured surgical video data can lead to substantially underestimated model uncertainty. To ensure relevance across both low-level and high-level prediction tasks, we focused on two widely used benchmark tasks, instrument segmentation (RobustMIS) and surgical action recognition (CholecT45). Both datasets exhibit a hierarchical structure, with multiple correlated frames per patient. 

For the binary segmentation task, we used the results of the ten algorithms of the binary segmentation task of the RobustMIS challenge. The same metrics as in the original challenge were applied, namely the DSC and the NSD. In this dataset, one hierarchical level was considered, namely the patient (i.e., the video; n = 10).

For the surgical action triplet recognition task, results were derived from the CholecT45 dataset using one algorithm (Swin-Base Transformer; see above). We calculated the mAP \cite{maier2024metrics} as done in the challenge. Given the class imbalance across 100 triplet classes, we additionally computed a class-weighted mAP, and included top-5 Accuracy. Metrics were calculated for each cross-validation fold. Again, we considered the patient (i.e., the video; n = 45) as the relevant level of hierarchy.

For both tasks, CIs were estimated using two resampling strategies: the standard bootstrap and the hierarchical bootstrap. In the standard (na\"ive) bootstrap, we performed resampling with replacement of all frames 1,000 times without considering the hierarchical structure \cite{efron1994introduction}. For each resample, we calculated the mean metric for each bootstrap sample, and obtained the empirical quantiles from the bootstrap distribution to calculate 95\% CIs. For the surgical action triplet recognition task, resampling was applied across all frame-level predictions within each cross-validation fold.

In contrast, the hierarchical bootstrap explicitly accounted for dependencies at each hierarchy level \cite{saravanan2020application}. We first resampled the videos, followed by resampling the individual frames within each sampled video. The mean metric was then computed across all resampled frames and videos. This process was repeated 1,000 times, and the empirical quantiles of the metric means were used to estimate the CIs. For the surgical triplet recognition task, this procedure was applied separately within each cross-validation fold, with metrics averaged per video and per fold before CI estimation.

For both CI methods, we calculated CI widths for each algorithm and metric as well as the ratio between hierarchical and na\"ive CI widths.

\subsubsection*{Experiment 2: Averages hide critical failures}
This experiment investigated whether global (non-stratified) aggregation of metric scores can conceal algorithm weaknesses under challenging image conditions. To enable stratified analysis across clinically relevant image characteristics, we focused on the RobustMIS dataset, for which we had access to structured metadata on visual artifacts \cite{ross2023beyond}. While the original study \cite{ross2023beyond} employed these annotations to analyze model robustness across visual conditions, our analysis focused on how global aggregation can obscure property-dependent performance differences that are critical for assessing validation reliability. Specifically, we analyzed the multi-instance segmentation task of the RobustMIS challenge, using MI\_DSC scores from the seven participating algorithms.

For each algorithm, the median MI\_DSC across all frames was calculated (non-stratified baseline). The median was chosen instead of the mean to reduce sensitivity to outliers. Stratified performance was then calculated by restricting the analysis to frames containing individual artifact types. The following properties were considered: blood, reflections, smoke, motion, overexposed, underexposed, intersecting instruments, and low-artifact scenes (i.e., scenes with one or only fewer annotated properties). Because frames could contain multiple artifacts, these subsets were not mutually exclusive.

For each algorithm, we calculated the median MI\_DSC for the full dataset (non-stratified baseline) and within each artifact-specific subset. We then computed the difference between the stratified and non-stratified medians per algorithm. To summarize performance changes across algorithms, we further computed the median of these algorithm-level scores per artifact type and reported the difference between the stratified and non-stratified medians.

To estimate the uncertainty of these differences, we applied hierarchical bootstrapping with 1,000 iterations, considering one hierarchy level, namely the patient (i.e., the video). For each bootstrap iteration, the difference in median MI\_DSC between the stratified and baseline conditions was computed, and CIs were derived from the empirical quantiles.

\subsubsection*{Experiment 3: Missing context hides temporal errors}
This experiment investigated how contextualizing prediction errors beyond predefined acceptance thresholds affects the interpretation of model performance. For this experiment, we focused on the temporal grounding tasks from the HeiCo-FOCUS benchmark \cite{Heicofocus} for long-context surgical video understanding.

As the threshold-based benchmark metric, we used Accuracy, which quantifies the proportion of predictions whose temporal localization error does not exceed 5 seconds. To provide additional context, we further quantified the absolute temporal deviation between predicted and reference timestamps and grouped prediction errors into increasing levels of temporal deviation ($\leq$5s, 5-10s, 10-60s, >60s), while treating invalid predictions separately.

For the ‘Segment’ track, we evaluated GPT-5.5, Gemini 3.1 Pro, and Qwen 3.6 Plus. For the ‘Procedure’ track, we evaluated Gemini 3 Flash, Gemini 3 Flash Lite, and Gemini 3.1 Pro. GPT-5.5 was used as a blind baseline. It was validated without visual input, receiving only the textual prompt and question. This baseline quantifies the extent to which temporal grounding performance can be attributed to textual priors rather than visual information.

\subsubsection*{Experiment 4: Aggregation choices can flip the winner}
This experiment investigated how different aggregation strategies affect algorithm rankings, given that aggregation schemes are rarely reported in practice. We focused on the data from the RobustMIS challenge, as we had access to frame-level performance scores from all participating algorithms, enabling systematic comparison across aggregation strategies. Specifically, we used the DSC scores of the ten participants of the binary segmentation task to simulate alternative ranking outcomes.

Six different aggregation strategies were calculated: 
\begin{enumerate}
    \item \textbf{Frame-wise aggregation:} Performance was aggregated equally over all frames of all videos, irrespective of procedure or phase. This approach served as the default, reflecting common practice and the original challenge.
    \item \textbf{Video-wise aggregation: } Performance was aggregated equally over all frames within each video, irrespective of the surgical phase. The resulting video-level scores were then aggregated.
    \item \textbf{Phase-wise aggregation:} Performance was aggregated equally over all frames within each surgical phase, irrespective of the video. The resulting phase-level scores were then aggregated.
    \item \textbf{Phase-wise video-wise aggregation:} Performance was first aggregated for each phase within each video. Rankings were then aggregated per phase over those rankings, and finally aggregated into one final ranking.
    \item \textbf{Video-wise phase-wise aggregation:} Performance was first aggregated for each phase within each video. Rankings were then aggregated per video over those rankings, and finally aggregated into one final ranking.
    \item \textbf{Weighted phase-wise aggregation:} Performance was aggregated for each phase, and the resulting phase-level scores were then combined into a global score using clinician-defined weights reflecting the clinical relevance and complexity of each phase. For simplicity, phases of low, intermediate, and high relevance or complexity were assigned weights of 1 (phases 0, 4, 5, 6, 9, 12), 2 (phases 2 and 10), and 3 (phases 1 and 8) \cite{maier2021heidelberg}, respectively.
\end{enumerate}

Across all aggregation strategies, the aggregation operator can vary, e.g., mean, median, or other percentiles. In line with the original challenge, we used the 5th percentile to reflect worst-case performance. Agreement between the default (frame-wise) ranking scheme and alternative aggregation strategies was assessed using Kendall’s tau \cite{kendall1938new} to quantify the similarity between two orderings (1: perfect agreement; 0: no association; -1: complete disagreement).

\newpage
\ExtendedDataFiguresOn
\setcounter{figure}{0}
\section{Extended Data}
\label{sec:extended-data}

\begin{figure}[H]
    \centering
    \includegraphics[width=0.7\linewidth]{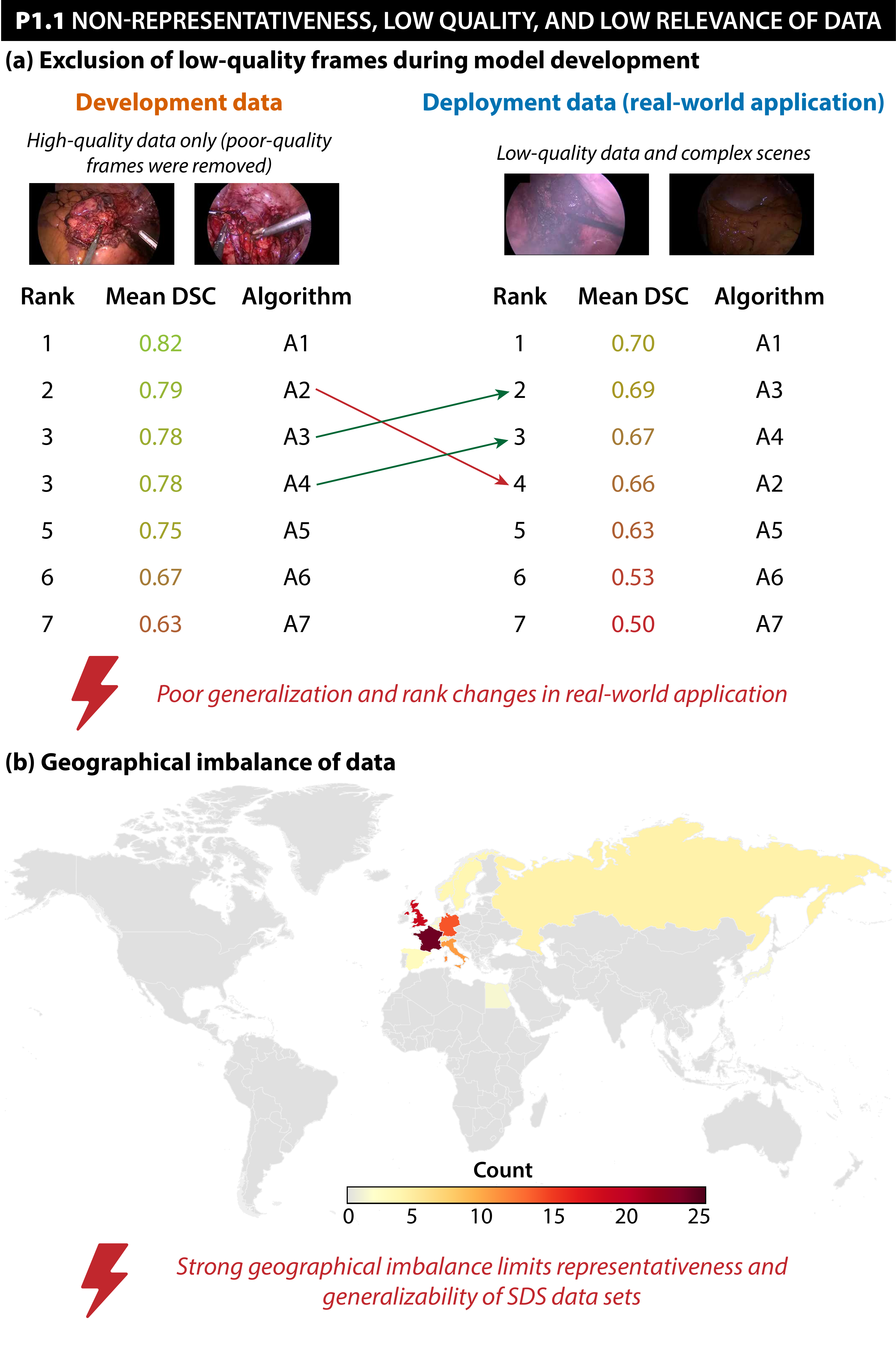}
    \caption{\textbf{P1.1 -- Non-representativeness, low quality, and low relevance of data. }(a) Example of excluding low-quality frames during model development, which can lead to overestimation of algorithm robustness and limited generalization in real-world settings. In this example, algorithms trained on data with such frames omitted perform considerably worse regarding their Dice Similarity Coefficient (DSC) when tested on data containing challenging conditions (results based on data from the Robust Medical Instrument Segmentation (RobustMIS) challenge 2019 \cite{ross2021comparative}). In this case, the performance gap even leads to changes in the relative ranking of algorithms. Mean DSC scores are color-coded (green: high scores; orange: low scores). (b) Example of geographical imbalance of data, highlighting limited representativeness of surgical data science (SDS) datasets. The map shows the geographical distribution of the datasets used in biomedical image analysis challenges involving surgical or endoscopic data conducted between 2018 and 2023 (n = 65 tasks across 14 challenges). Most data originate from a few Western European countries, particularly France, the United Kingdom, and Germany, whereas large parts of the world remain unrepresented.}
    \label{fig:extended-fig1}
\end{figure}

\begin{figure}[H]
    \centering
    \includegraphics[width=0.9\linewidth]{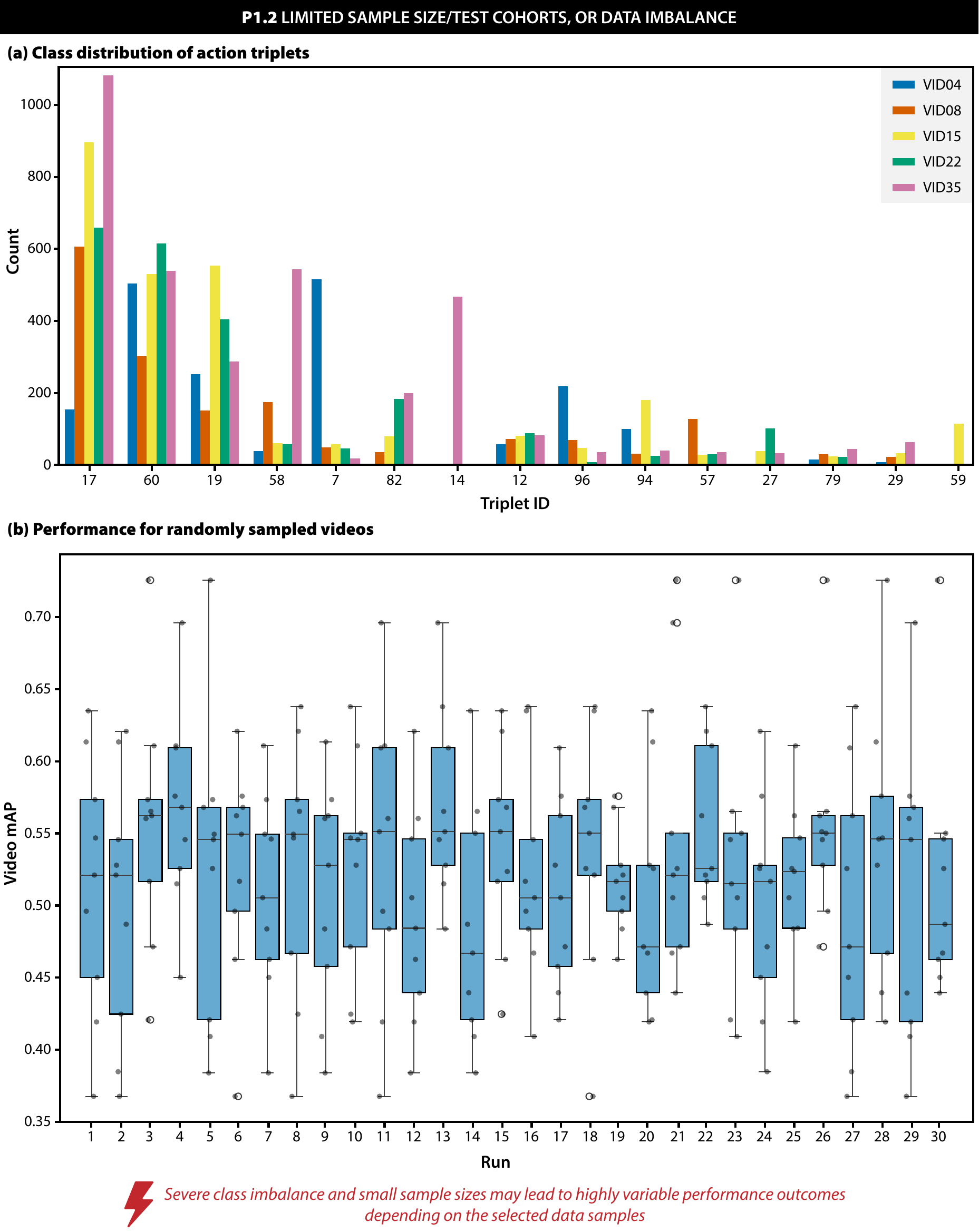}
    \caption{\textbf{P1.2 -- Limited sample size/test cohorts, or data imbalance.} The figure demonstrates the impact of class imbalance and limited test set size on performance stability for the task of surgical action triplet recognition in cholecystectomy (here: CholecT45 \cite{nwoye2023cholectriplet2021}). (a) The distribution of triplet classes across five randomly selected videos, sorted by the total number of occurrences per class (here: top 15 triplet classes), is highly imbalanced, with some classes frequent and many rare or absent, highlighting the risk of inconsistent class coverage. (b) To reflect the data split of CholecT45 (nine videos per fold), we randomly sampled nine videos for 30 runs. In each run, a swin-based transformer model trained on the original CholecT45 training data with multi-task learning and soft labels derived via a multi-teacher strategy was validated on the sampled videos \cite{yamlahi2023self}. The boxplots show high variability in per-video mean Average Precision (mAP) across runs, demonstrating the unstable and unreliable nature of class-averaged metrics such as mAP under the joint influence of limited test cohort size and heterogeneous class coverage. Box centres indicate the median, boxes the interquartile range (IQR; 25th-75th percentiles), whiskers the most extreme values within 1.5$\times$ the IQR, and light dots the individual video mAP values.}
    \label{fig:extended-fig2}
\end{figure}

\begin{figure}[H]
    \centering
    \includegraphics[width=0.8\linewidth]{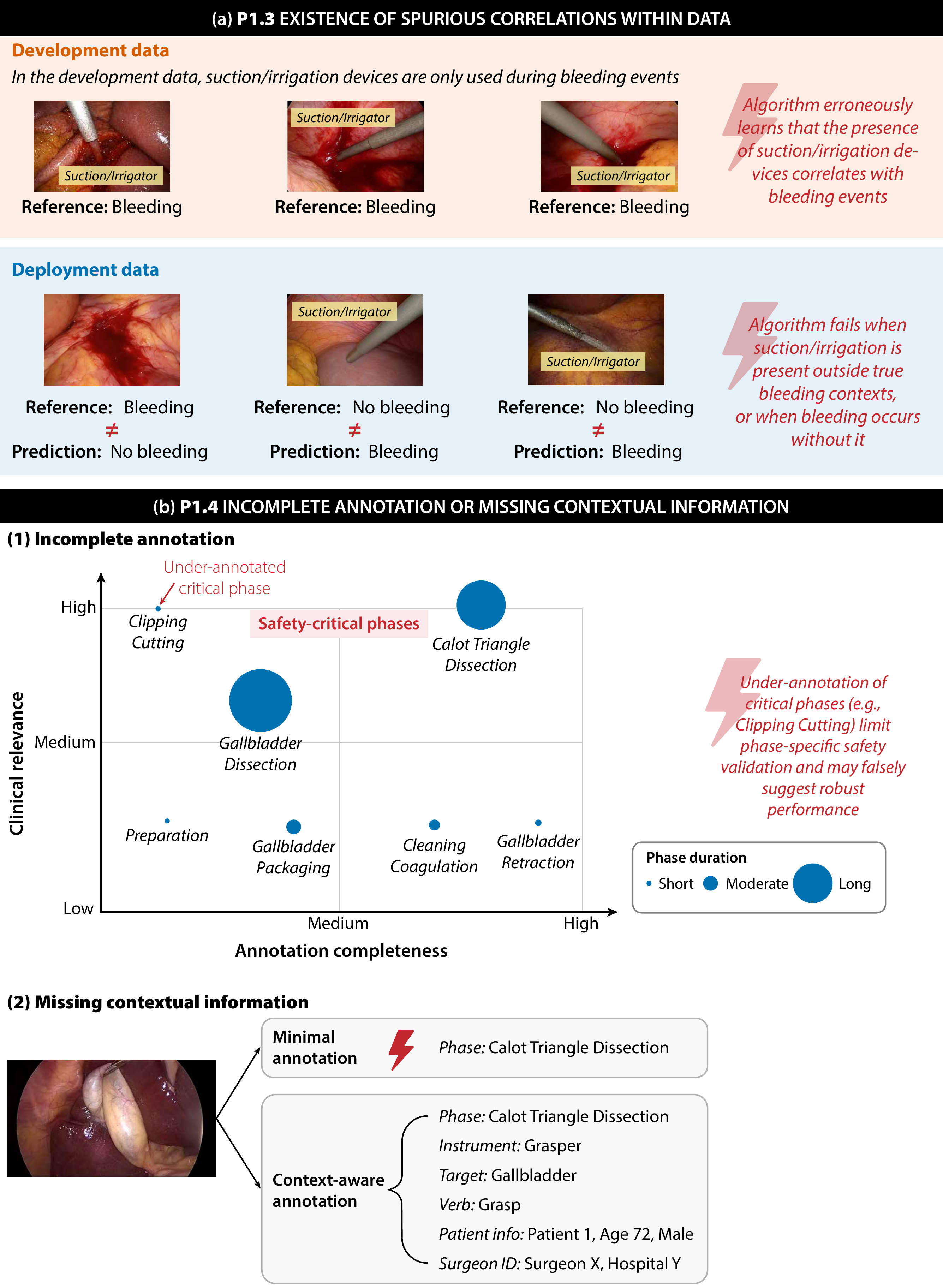}
    \caption{\small\textbf{(a) P1.3 -- Existence of spurious correlations within data.} In this example, a bleeding detection algorithm is trained on a biased development dataset where suction/irrigation devices are only present during bleeding. As a result, the algorithm mistakenly learns to associate the presence of these tools with bleeding events, rather than detecting blood itself. During deployment, the algorithm fails when suction is used outside of bleeding contexts or when blood appears without suction. This illustrates how spurious correlations in biased datasets can undermine clinically important tasks such as bleeding detection. \textbf{(b) P1.4 -- Incomplete annotation or missing contextual information.} Validation may be compromised, especially in safety-critical surgical phases, if annotations are incomplete or lack contextual detail needed to assess model performance. (1) Annotation completeness refers to the extent to which all relevant phases, events, or entities are labeled in sufficient detail for the intended task, but does not necessarily align with clinical relevance. In this example, the Clipping Cutting phase, despite being highly safety-relevant, is poorly annotated, limiting robust validation. In contrast, longer and less critical phases are better annotated, skewing performance estimates. Phase duration is represented by bubble size. (2) Even when frames are annotated, missing contextual information such as anatomical region or metadata can compromise interpretability and validity. In this example, the same surgical frame (from CholecT45 \cite{nwoye2023cholectriplet2021}) is annotated minimally (top; phases only), while the bottom row illustrates a context-aware annotation including semantic and procedural details enabling more meaningful and clinically robust validation.}
    \label{fig:extended-fig3}
\end{figure}

\begin{figure}[H]
    \centering
    \includegraphics[width=0.8\linewidth]{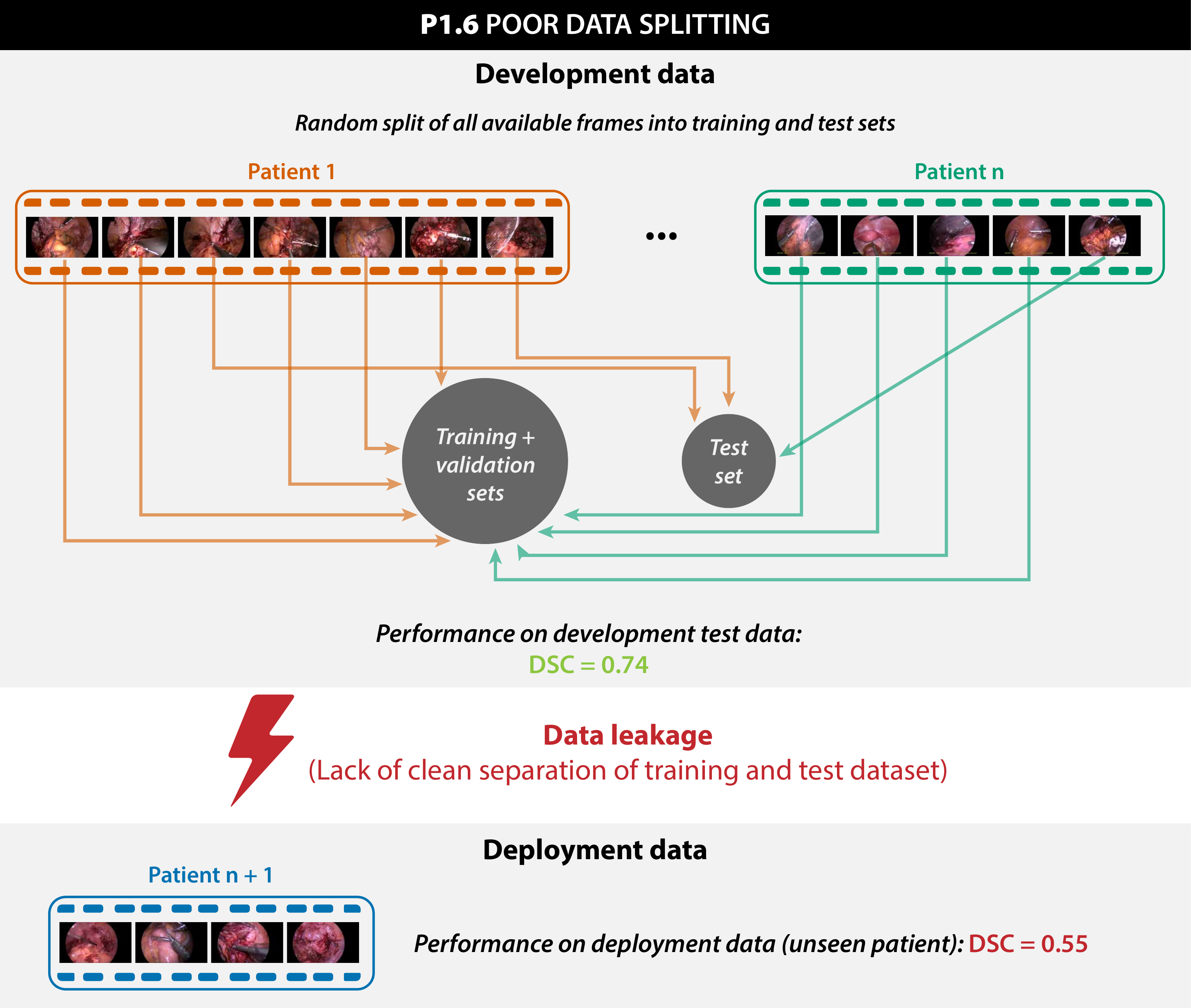}
    \caption{\textbf{P1.6 -- Poor data splitting.} In this example, a random split of frames across training and test sets over all patients leads to data leakage, as images from the same patient appear in both sets. This results in overly optimistic performance on the development test data (Dice Similarity Coefficient (DSC) of 0.74) but substantially lower performance on unseen deployment data (DSC = 0.55). Results are based on the Robust Medical Instrument Segmentation (RobustMIS) challenge 2019 \cite{ross2021comparative} binary segmentation data and a U-Net implementation, comparing training where frames from every patient are split 60/20/20 training/validation/testing with training where patients are used wholly for training, validation, or testing in a 60/20/20 split.}
    \label{fig:extended-fig4}
\end{figure}

\begin{figure}[H]
    \centering
    \includegraphics[width=0.8\linewidth]{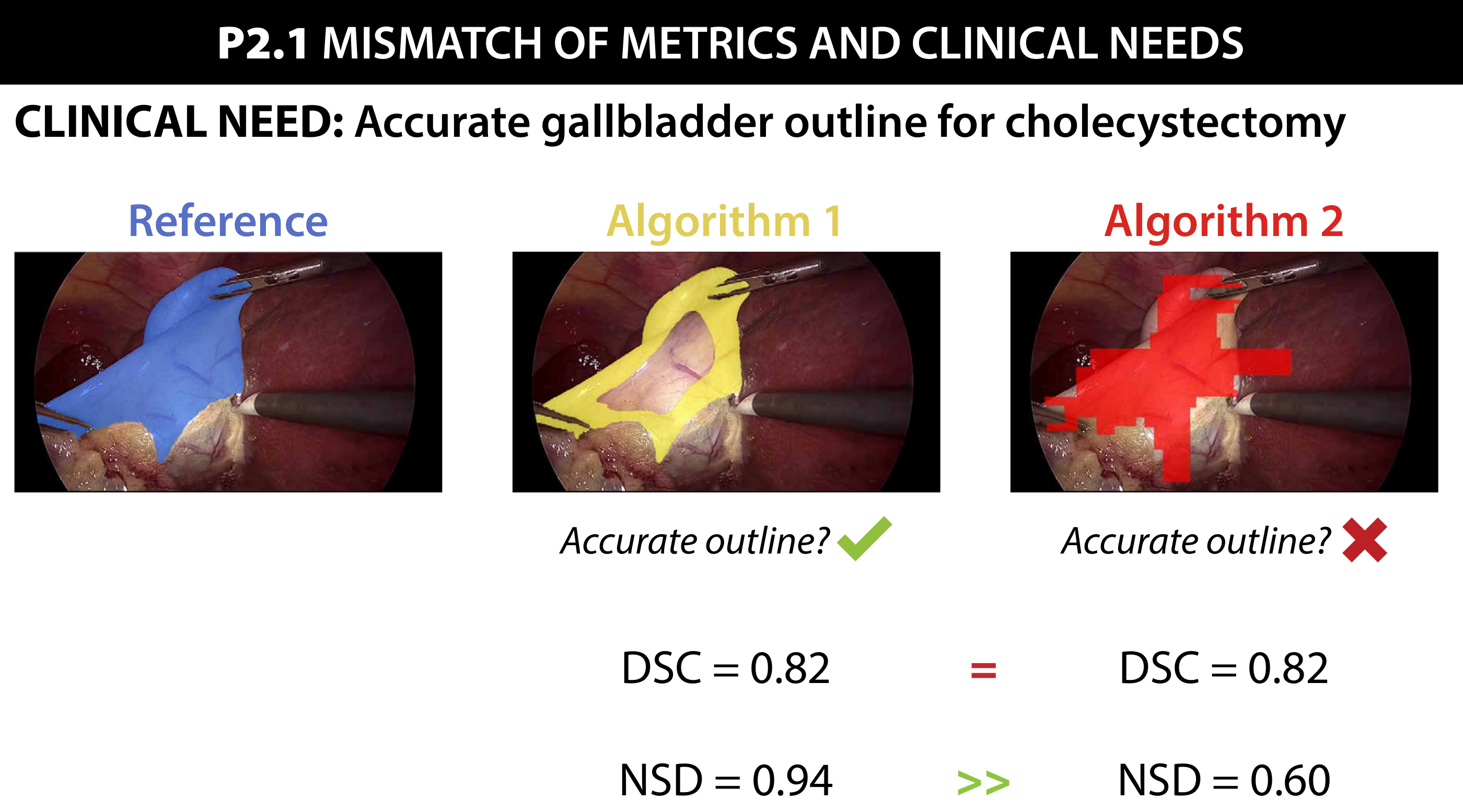}
    \caption{\textbf{P2.1 -- Mismatch of metrics and clinical needs.} Despite both algorithms achieving the same Dice Similarity Coefficient (DSC = 0.82), \textit{Algorithm 1} accurately captures the gallbladder’s outline, while \textit{Algorithm 2} produces a poorly shaped segmentation. The Normalized Surface Distance (NSD) better reflects the clinically relevant boundary accuracy required for this use case. Images from CholecSeg8k \cite{hong2020cholecseg8k}. Note: In clinical practice, certain boundaries (e.g., gallbladder-liver interface) may be more critical than others -- a nuance which is captured by neither DSC nor NSD.}
    \label{fig:extended-fig5}
\end{figure}

\begin{figure}[H]
    \centering
    \includegraphics[width=0.8\linewidth]{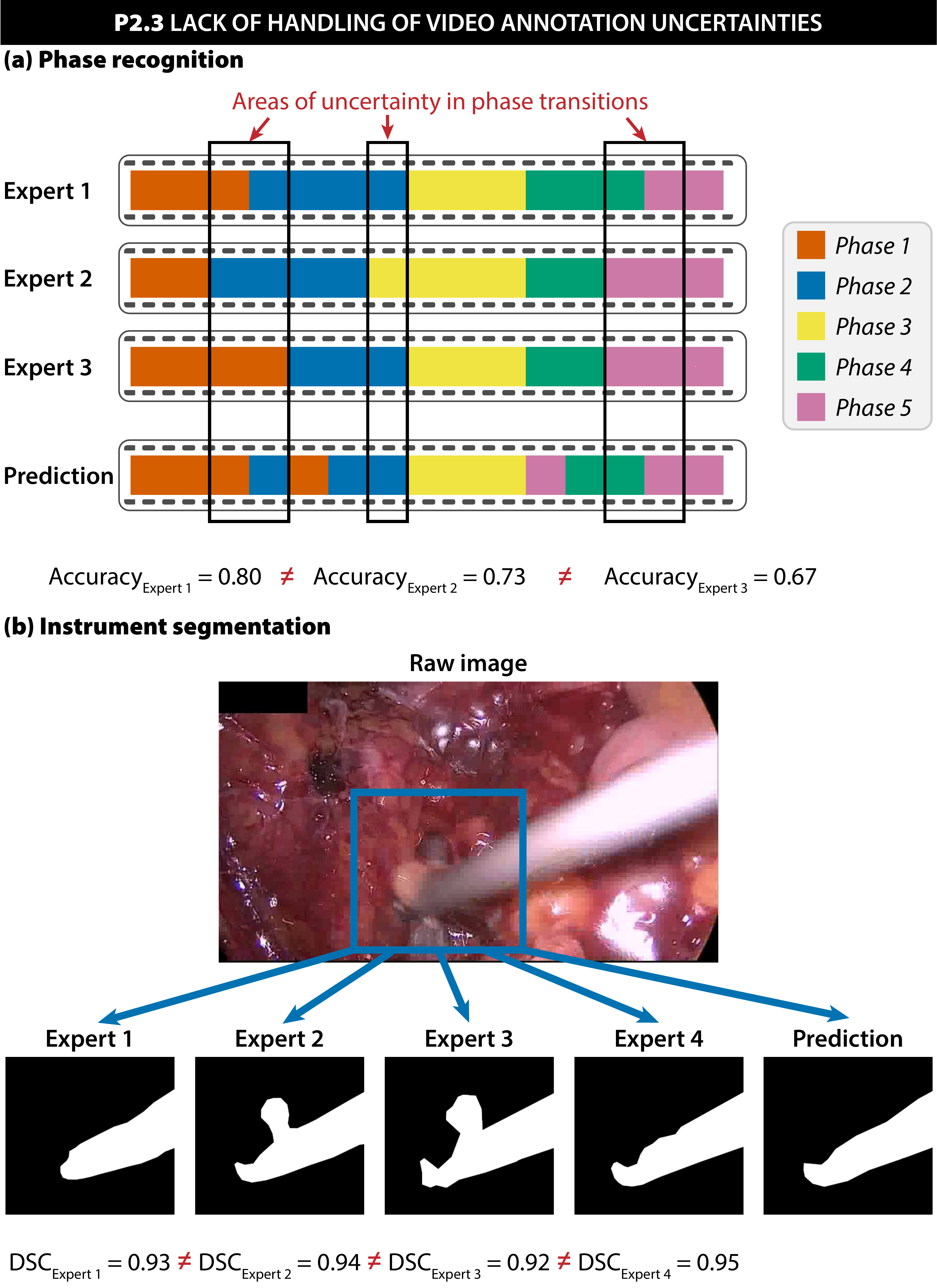}
    \caption{\textbf{P2.3 -- Lack of handling of video annotation uncertainties. }In this example, inconsistencies between expert raters (a) in phase annotations occur at transition points and (b) in instrument segmentation masks. These areas of uncertainty lead to changes in performance scores (here: Accuracy and Dice Similarity Coefficient (DSC)) depending on which expert is considered as the reference. DSC values were computed for the full image, not the zoomed region. Images and annotations in (b) are from inter-rater variability analysis of the Robust Medical Instrument Segmentation (RobustMIS) challenge \cite{ross2021comparative}.}
    \label{fig:extended-fig6}
\end{figure}

\begin{figure}[H]
    \centering
    \includegraphics[width=1\linewidth]{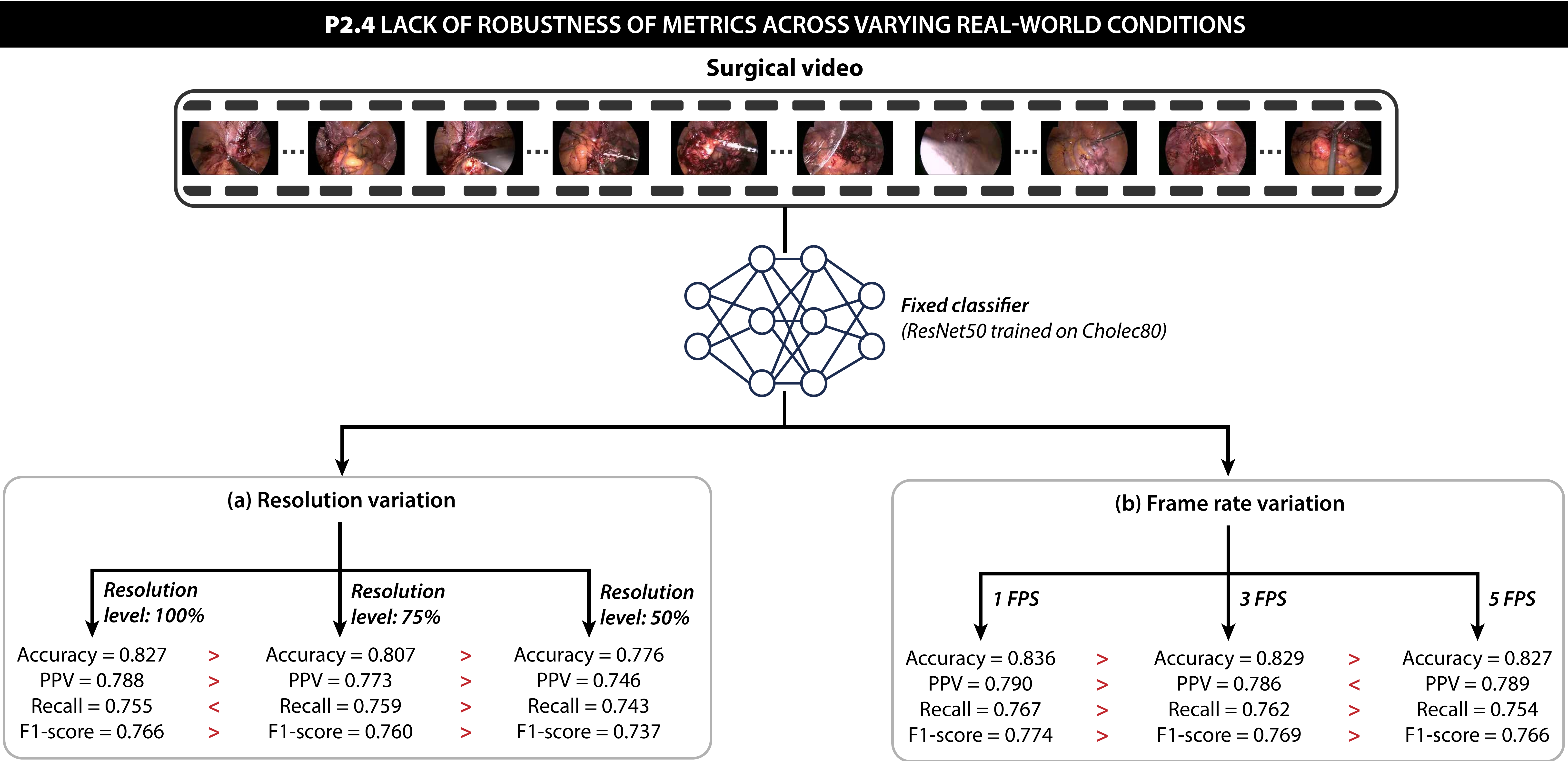}
    \caption{\textbf{P2.4 -- Lack of robustness of metrics across varying real-world conditions (here: image resolution (a) and frame rate (b)).} In this example, a fixed classifier (ResNet50) is validated on surgical video data (Cholec80 \cite{twinanda2016endonet}) under varying real-world conditions: (a) Different spatial resolutions (100\%, 75\%, 50\%) and (b) sampled at 1, 3, and 5 frames per second (FPS), simulating typical variability in real-world acquisition or compression conditions. Despite using identical model weights and validating on the same underlying procedure, performance metrics (here: Accuracy, Positive Predictive Value (PPV), Recall, F1-score) vary noticeably with resolution and frame rate. These variations arise not from changes in the model or task, but from the metric's sensitivity to input resolution, illustrating a lack of robustness in metrics under plausible real-world conditions. }
    \label{fig:extended-fig7}
\end{figure}

\begin{figure}[H]
    \centering
    \includegraphics[width=0.7\linewidth]{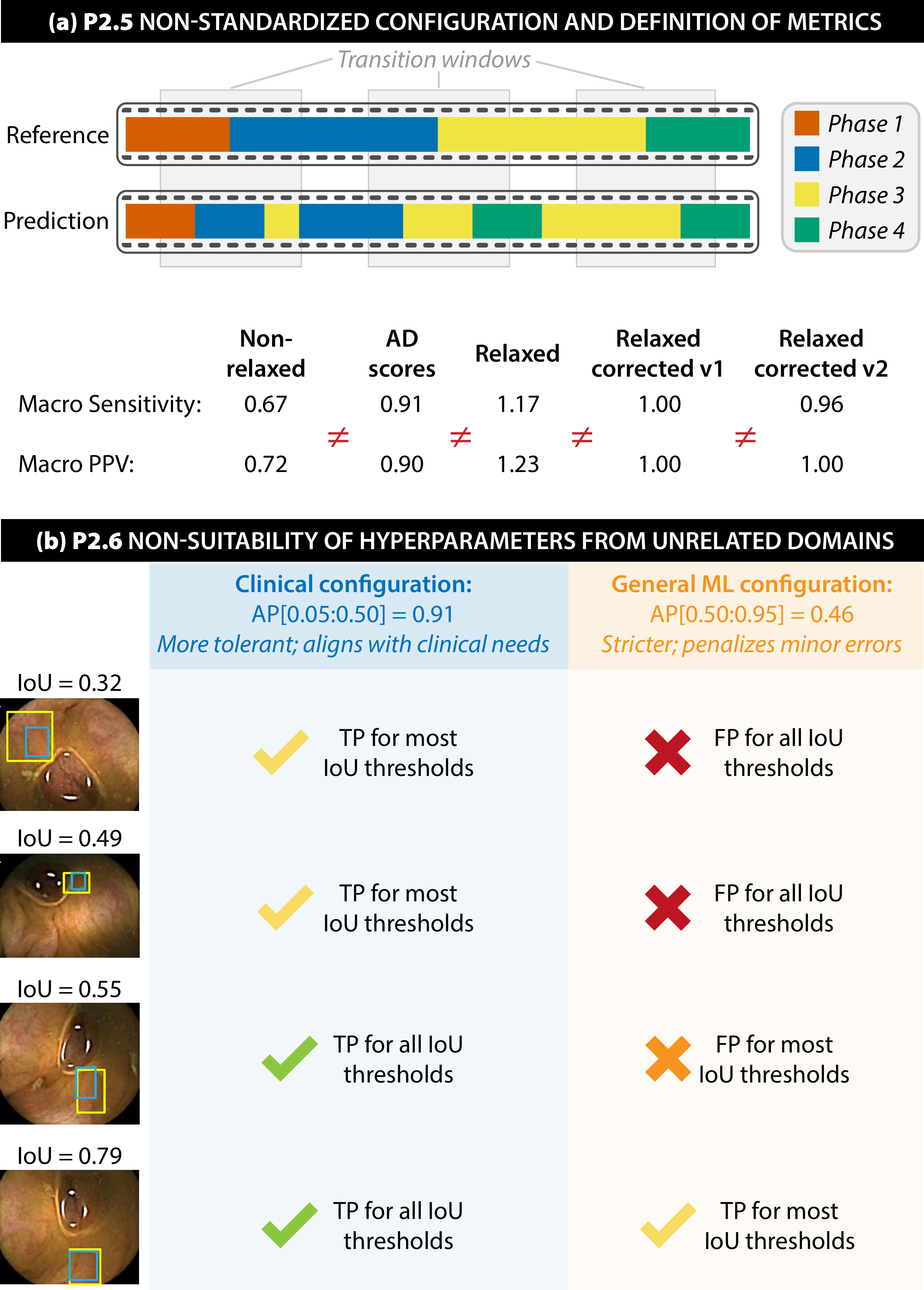}
    \caption{\textbf{(a) P2.5 -- Non-standardized configuration and definition of metrics. }In this example, relaxed metrics, which allow for a less strict definition of True Positives (TPs), are applied in phase transition areas where expert annotations often show inconsistencies. The non-relaxed macro scores, i.e., unweighted mean across classes, are compared to four different variants of relaxed metrics (application-dependent (AD) scores, relaxed metrics for the Cholec80 dataset, and two corrected versions v1 (cutting values at 1.00) and v2 (adapting the denominator of the definition); see \cite{funke2023metrics}), resulting in substantial differences in Sensitivity and Positive Predictive Value (PPV) values. \textbf{(b) P2.6 -- Non-suitability of hyperparameters from unrelated domains.} The choice of detection thresholds (here: Intersection over Union (IoU)) critically impacts the reported detection performance (here: polyp detection during capsule endoscopy). This example compares a clinical threshold configuration (blue box; Average Precision (AP)[0.05:0.50]) and a standard machine learning (ML) configuration (orange box; AP[0.50:0.95]) on the same images. The stricter ML configuration penalizes several detections as False Positives (FP) that would be acceptable in a clinical setting (True Positives (TP)), leading to a much lower overall AP. Note that the phrasing "for most IoU thresholds" refers to a detection being counted as a TP/FP across the majority of thresholds used in the AP calculation.}
    \label{fig:extended-fig8}
\end{figure}

\begin{figure}[H]
    \centering
    \includegraphics[width=0.7\linewidth]{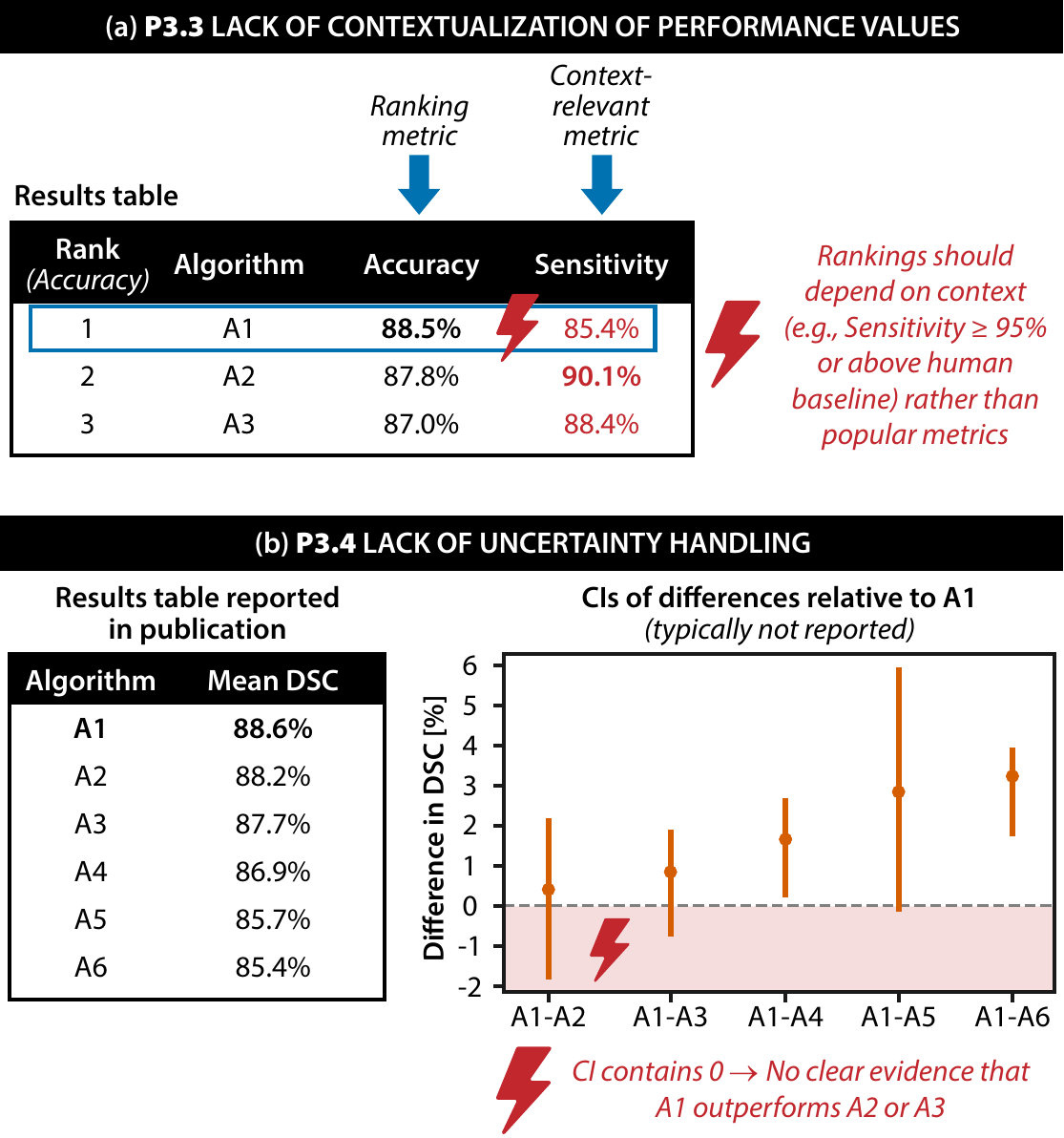}
    \caption{\textbf{(a) P3.3 -- Lack of contextualization of performance values.} In this example, the specific clinical use case prioritizes high Sensitivity, as missing a positive event could have severe consequences. Here, algorithm A3 is ranked highest based on Accuracy but fails to meet the required clinical Sensitivity threshold ($\geq$ 95\%). The lack of contextualization of the performance values conceals the fact that none of the models meet both clinical thresholds. \textbf{(b) P3.4 -- Lack of uncertainty handling.} Performance rankings based solely on point estimates can be misleading without reporting uncertainty. In this example, algorithm A1 is considered the best due to its highest mean Dice Similarity Coefficient (DSC) score. However, confidence intervals (CIs) of the pairwise differences show that for several competitors there is no clear evidence that A1 outperforms them, since the CIs include 0. Circles indicate mean differences, and vertical lines indicate hierarchical 95\% bootstrap CIs of the pairwise differences (1,000 bootstrap iterations). The analysis included n = 2,231 frames from 10 videos. Results are based on RobustMIS 2019 \cite{ross2021comparative} binary segmentation top 6 algorithms.}
    \label{fig:extended-fig9}
\end{figure}

\begin{figure}[H]
    \centering
    \includegraphics[width=1\linewidth]{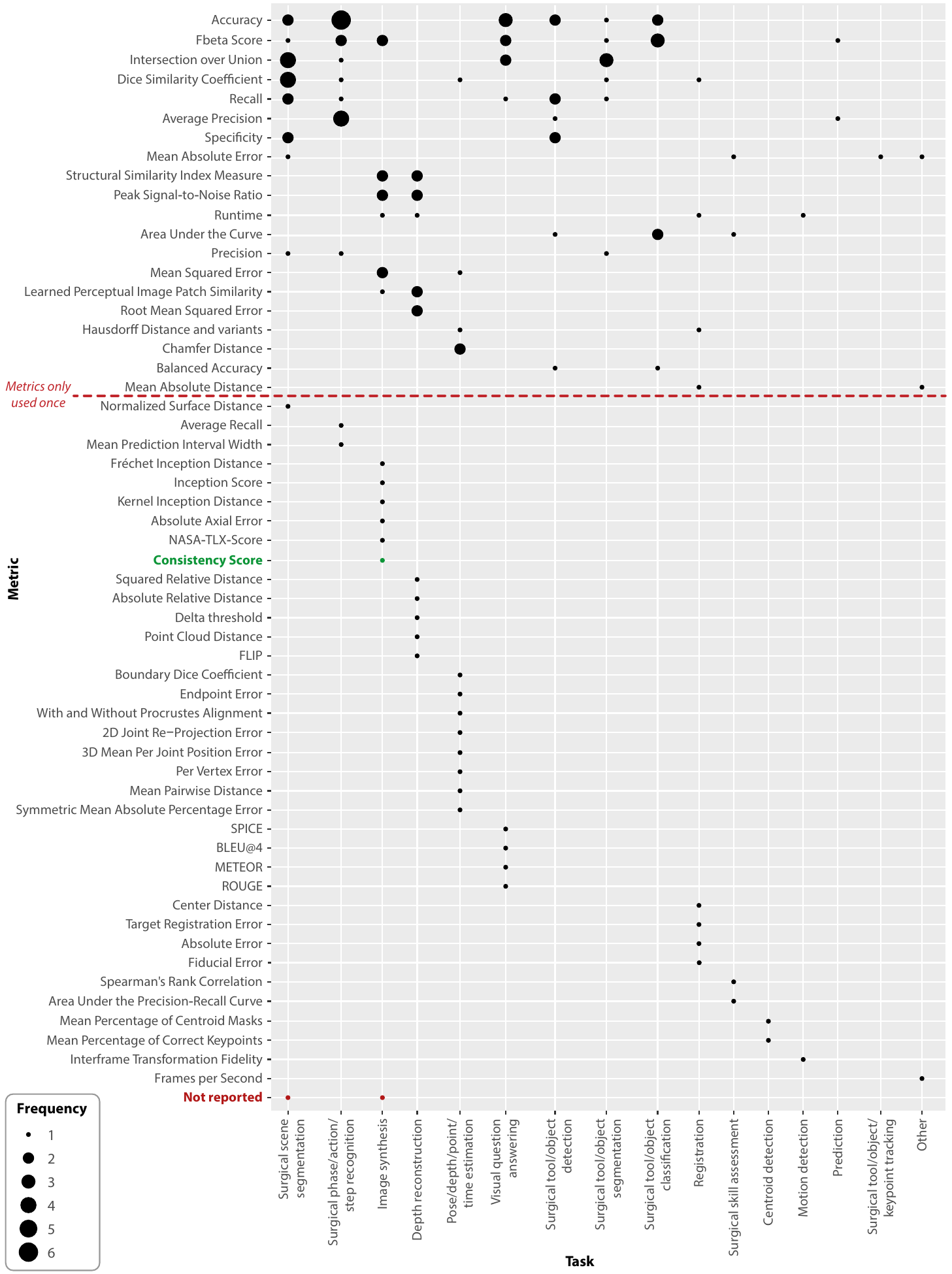}
    \caption{Scatterplot of used metrics in the 2023 Medical Image Computing and Computer Assisted Intervention (MICCAI) conference surgical data science (SDS) papers. Here, the metrics are shown for various SDS tasks. Both tasks and metrics are sorted by frequency of usage (metrics: top to bottom; tasks: left to right). The size of the blobs corresponds to the frequency of metric occurrence. According to the screening, only a single paper reported a temporal metric, the Consistency Score (green).}
    \label{fig:extended-fig10}
\end{figure}

\ExtendedDataFiguresOff

\section{Data availability}
The experiments in this study were conducted using publicly available datasets, namely the RobustMIS 2019 challenge data \cite{ross2021comparative}, CholecT45 data \cite{nwoye2023cholectriplet2021}, and the first version of the HeiCo-FOCUS data \cite{Heicofocus}. Individual challenge results (i.e., metric scores per algorithm) are not publicly available, as participant-level consent for data sharing was not obtained.

\section{Code availability}
Source code for the three experiments is available at \href{https://github.com/IMSY-DKFZ/metrics-reloaded-surgery-pitfalls}{https://github.com/IMSY-DKFZ/metrics-reloaded-surgery-pitfalls}.

\section{Acknowledgements}
We would like to thank Johannes Bender (NCT, Germany), Patrick Beyersdorffer (Reutlingen University, Germany), Isabel Funke (NCT, Germany), Susu Hu (NCT, Germany), Alexander Jenke (NCT, Germany), Denise Junger (Reutlingen University, Germany), Piotr Kalinowsky (DKFZ, Germany), Keno März (NCT, Germany), Dominik Michael (DKFZ, Germany), Wenyao Xi (UCL, UK), Jinjing Xu (NCT, Germany), and Mona Zeinodin (UCL, UK) for their input and discussions on the framework and pitfalls. We would further like to thank Kevin Cleary (Children's National, USA) and Ajit Sachdeva (American College of Surgeons, USA) for participating in the SAGES 2023 workshop.

\section{Funding statement}
A.R.: This work was initiated by the Helmholtz Association of German Research Centers in the scope of the Helmholtz Imaging Incubator (HI). L.M.-H.: This project was supported by the European Research Council (ERC) under the European Union’s Horizon 2020 research and innovation program (NEURAL SPICING, 101002198), the National Center for Tumor Diseases (NCT), Heidelberg’s Surgical Oncology Program, the German Cancer Research Center (DKFZ). E.C.: This publication was further supported through state funds approved by the State Parliament of Baden-Württemberg for the Innovation Campus Health + Life Science Alliance Heidelberg Mannheim. O.C.: The research leading to these results has received funding from the French government under management of Agence Nationale de la Recherche as part of the “France 2030” program (reference ANR-23-IACL-0008, project PRAIRIE-PSAI), as part of the "Investissements d'avenir" program (reference ANR-19-P3IA-0001, project PRAIRIE 3IA Institute and reference ANR-10-IAIHU-06, project Agence Nationale de la Recherche-10-IA Institut Hospitalo-Universitaire-6) and from the European Union’s Horizon Europe Framework Programme (grant number 101136607, project CLARA). R.D.: This project has received funding from the European Union's Horizon Europe research and innovation programme under grant agreement No 101092646. D.D. is funded by NIH K23 EB034110. Q.D.: This work described in this paper was partially supported by a grant from the ANR/RGC Joint Research Scheme sponsored by the Research Grants Council of the Hong Kong Special Administrative Region, China and the French National Research Agency (Project No. A-CUHK402/23). G.F. is supported by Canada Research Chair in Computer-Assisted Surgery. S.Gia. is supported by the Royal Society URF\ R\ 201014. T.H. is a Consolidator Researcher, receiving financial support from the Distinguished Researcher program of Obuda University. His work has been partially supported by ACMIT (Austrian Center for Medical Innovation and Technology), which is funded within the scope of the COMET (Competence Centers for Excellent Technologies) program of the Austrian Government. T.H. is partially supported by Project 2024-1.2.3-HU-RIZONT-00069, implemented with the support provided by the Ministry of Culture and Innovation of Hungary from the National Research, Development and Innovation Fund, financed under the 2024-1.2.3-HU-RIZONT funding scheme. D.A.H. is supported by a grant from the American Surgical Association Foundation. F.R.K. receives support from the German Cancer Research Center (CoBot 2.0), the Joachim Herz Foundation (Add-On Fellowship for Interdisciplinary Life Science), the Central Indiana Corporate Partnership AnalytiXIN Initiative, the Evan and Sue Ann Werling Pancreatic Cancer Research Fund, and the Indiana Clinical and Translational Sciences Institute (EPAR4157) funded, in part, by Grant Number UM1TR004402 from the National Institutes of Health, National Center for Advancing Translational Sciences, Clinical and Translational Sciences Award. The content is solely the responsibility of the authors and does not necessarily represent the official views of the National Institutes of Health. J.L.L. received funding by the Swiss National Science Foundation (P5R5PM 21766), Novartis Foundation for medical-biological Research (\#23C162), and the Vontobel Foundation (0867/2024). H.J.M. is supported by the NIHR UCLH/UCL Biomedical Research Centre. N.P.: This work has received funding from the European Union (ERC, CompSURG, 101088553, PI: Nicolas Padoy). Views and opinions expressed are, however, those of the authors only and do not necessarily reflect those of the European Union or the European Research Council. Neither the European Union nor the granting authority can be held responsible for them. This work has also been supported by French state funds managed within the Plan Investissements d’Avenir by the ANR under reference ANR- 10-IAHU-02 (IHU Strasbourg). H.R.: Hong Kong Research Grants Council Collaborative Research Fund under Grant CRF-C4026-21G. S.S. is supported by the German Research Foundation (DFG, Deutsche Forschungsgemeinschaft) as part of Germany’s Excellence Strategy -- EXC 2050/1 -- Project ID 390696704 -- Cluster of Excellence "Centre for Tactile Internet with Human-in-the-Loop" (CeTI) of Dresden University of Technology. D.S. is supported by the Royal Academy of Engineering Chair in Emerging Technologies. M.W.: This work has been funded by the German Research Foundation (DFG, Deutsche Forschungsgemeinschaft) as part of Germany's Excellence Strategy—EXC 2050/1—Project ID 390696704—Cluster of Excellence "Centre for Tactile Internet with Human-in-the-Loop" (CeTI) of TUD Dresden University of Technology and by the German Federal Ministry of Research, Technology, and Space within the “Surgical AI Hub Germany” project (grant number BMBF 02K223A110). 

\section{Author Contributions Statement}
A.R. initiated and led the study, was a member of the Delphi core team, wrote and reviewed the manuscript, prepared and evaluated all surveys, suggested pitfalls and categorization, performed experiments, served as a conflict resolver for the literature screening, participated in the SAGES workshop, and designed all figures. O.L. suggested pitfalls and categorization, supported survey evaluation and experiments, and reviewed the manuscript. M.D.T. suggested pitfalls and categorization, wrote and reviewed the manuscript. P.A. performed experiments and reviewed the manuscript. M.Kn. performed experiments, wrote and reviewed the manuscript. M.M.R. performed experiments and reviewed the manuscript. I.P.M. was a member of the Delphi expert panel, performed experiments, and reviewed the manuscript. M.S.A. was a member of the Delphi expert panel, participated in the SAGES workshop, and reviewed the manuscript. D.A. was a member of the Delphi expert panel and reviewed the manuscript. S.Ba. was a member of the Delphi expert panel and reviewed the manuscript. S.Bo. was a member of the Delphi expert panel and reviewed the manuscript. O.B. was a member of the Delphi expert panel and reviewed the manuscript. E.C.S.C. was a member of the Delphi expert panel and reviewed the manuscript. J.W.C. was a member of the Delphi expert panel, participated in the SAGES workshop, and reviewed the manuscript. O.C. was a member of the Delphi expert panel and reviewed the manuscript. E.C. was a member of the Delphi expert panel, served as a conflict resolver for the literature screening, and reviewed the manuscript. T.C. was a member of the Delphi expert panel, participated in the SAGES workshop, and reviewed the manuscript. A.D. performed the literature screening and reviewed the manuscript. R.D. was a member of the Delphi expert panel and reviewed the manuscript. D.D. was a member of the Delphi expert panel, participated in the SAGES workshop, and reviewed the manuscript. Q.D. was a member of the Delphi expert panel and reviewed the manuscript. J.E. was a member of the Delphi expert panel, participated in the SAGES workshop, and reviewed the manuscript. S.E. was a member of the Delphi expert panel and reviewed the manuscript. G.F. was a member of the Delphi expert panel and reviewed the manuscript. P.F. was a member of the Delphi expert panel and reviewed the manuscript. P.G.K. was a member of the Delphi expert panel, participated in the SAGES workshop, and reviewed the manuscript. S.Gia. was a member of the Delphi expert panel and reviewed the manuscript. S. Gil. was a member of the Delphi expert panel and reviewed the manuscript. I.G. was a member of the Delphi expert panel and reviewed the manuscript. P.G. was a member of the Delphi expert panel and reviewed the manuscript. J.G. was a member of the Delphi expert panel, participated in the SAGES workshop, and reviewed the manuscript. T.P.G. was a member of the Delphi expert panel and reviewed the manuscript. T.H. was a member of the Delphi expert panel, participated in the SAGES workshop, and reviewed the manuscript. A.Hu. was a member of the Delphi expert panel and reviewed the manuscript. M.H. was a member of the Delphi expert panel and reviewed the manuscript. C.H. performed the literature screening and reviewed the manuscript. R.H. was a member of the Delphi expert panel and reviewed the manuscript. H.H. performed the literature screening and reviewed the manuscript. A.Ha. was a member of the Delphi expert panel and reviewed the manuscript. P.F.J. was a member of the Delphi expert panel and reviewed the manuscript. P.J. was a member of the Delphi expert panel and reviewed the manuscript. A.J. was a member of the Delphi expert panel, participated in the SAGES workshop, and reviewed the manuscript. R.J. was a member of the Delphi expert panel and reviewed the manuscript. Y.J. was a member of the Delphi expert panel and reviewed the manuscript. L.Jos. was a member of the Delphi expert panel and reviewed the manuscript. L.Joy. was a member of the Delphi expert panel and reviewed the manuscript. M.Ki. performed the literature screening and reviewed the manuscript. A.K. was a member of the Delphi expert panel and reviewed the manuscript. G.K. was a member of the Delphi expert panel, participated in the SAGES workshop, and reviewed the manuscript. K.L. was a member of the Delphi expert panel and reviewed the manuscript. S.L. was a member of the Delphi expert panel, participated in the SAGES workshop, and reviewed the manuscript. J.L.L. was a member of the Delphi expert panel and reviewed the manuscript. G.I.L. was a member of the Delphi expert panel, participated in the SAGES workshop, and reviewed the manuscript. R.L. was a member of the Delphi expert panel, participated in the SAGES workshop, and reviewed the manuscript. P.L. performed the literature screening and reviewed the manuscript. L.L. supported experiments and reviewed the manuscript. H.J.M. was a member of the Delphi expert panel and reviewed the manuscript. P.M. was a member of the Delphi expert panel, participated in the SAGES workshop, and reviewed the manuscript. L.Ma. supported experiments and reviewed the manuscript. O.R.M. was a member of the Delphi expert panel, participated in the SAGES workshop, and reviewed the manuscript. B.P.M. was a member of the Delphi expert panel and reviewed the manuscript. L.Mü. was a member of the Delphi expert panel and reviewed the manuscript. H.N. was a member of the Delphi expert panel and reviewed the manuscript. N.N. was a member of the Delphi expert panel and reviewed the manuscript. A.N. was a member of the Delphi expert panel and reviewed the manuscript. J.N. was a member of the Delphi expert panel and reviewed the manuscript. F.N. was a member of the Delphi expert panel and reviewed the manuscript. M.N. was a member of the Delphi expert panel and reviewed the manuscript. C.N. was a member of the Delphi expert panel and reviewed the manuscript. N.O. was a member of the Delphi expert panel and reviewed the manuscript. N.P. was a member of the Delphi expert panel, participated in the SAGES workshop, and reviewed the manuscript. T.P. was a member of the Delphi expert panel and reviewed the manuscript. M.P. was a member of the Delphi expert panel and reviewed the manuscript. T.R. performed the literature screening and reviewed the manuscript. H.R. was a member of the Delphi expert panel and reviewed the manuscript. N.R. was a member of the Delphi expert panel and reviewed the manuscript. D.R. was a member of the Delphi expert panel and reviewed the manuscript. D.Sa. was a member of the Delphi expert panel and reviewed the manuscript. S.Sc. was a member of the Delphi expert panel and reviewed the manuscript. M.S. was a member of the Delphi expert panel and reviewed the manuscript. S.Se. was a member of the Delphi expert panel and reviewed the manuscript. A.S. was a member of the Delphi expert panel and reviewed the manuscript. L.S. was a member of the Delphi expert panel and reviewed the manuscript. J.H.S. was a member of the Delphi expert panel and reviewed the manuscript. V.S. was a member of the Delphi expert panel and reviewed the manuscript. R.S. was a member of the Delphi expert panel and reviewed the manuscript. R.T. was a member of the Delphi expert panel and reviewed the manuscript. T.N.T. was a member of the Delphi expert panel and reviewed the manuscript. M.U. was a member of the Delphi expert panel and reviewed the manuscript. F.v.d.S. was a member of the Delphi expert panel and reviewed the manuscript. M.W. was a member of the Delphi expert panel and reviewed the manuscript. A.Y. was a member of the Delphi expert panel and reviewed the manuscript. S.K.Z. was a member of the Delphi expert panel and reviewed the manuscript. A.Z. was a member of the Delphi expert panel and reviewed the manuscript. A.M. initiated the study, was a member of the Delphi core team, participated in the SAGES workshop, and reviewed the manuscript. D.St. initiated the study, was a member of the Delphi core team, participated in the SAGES workshop, and reviewed the manuscript. S.Sp.. initiated the study, was a member of the Delphi core team, participated in the SAGES workshop, and reviewed the manuscript. D.A.H. initiated and led the study, was a member of the Delphi core team, participated in the SAGES workshop, and wrote and reviewed the manuscript. F.R.K. initiated and led the study, was a member of the Delphi core team, participated in the SAGES workshop, and wrote and reviewed the manuscript. L.M.-H. initiated and led the study, was a member of the Delphi core team, participated in the SAGES workshop, and wrote and reviewed the manuscript.

\section{Competing Interests Statement}
M.S.A. is a speaker of Medtronic and advisor of Johnson \& Johnson. D.A. is a shareholder and has a leadership role in Scialytics. N.W.C.: Nil. O.C. reports having received consulting fees from Therapanacea (2022-2024). OC reports that other principal investigators affiliated to the team which he co-leads have received grants (paid to the institution) from Sanofi (2020-2022) and Biogen (2022-2023). OC reports that his spouse was an employee of myBrainTechnologies (2015-2023) and is an employee of DiamPark. G.F. declares the following conflict of interest: Johns Hopkins University, Harvard Brigham and Women's Hospital. S.G. has or has had consulting relationships with AMBOSS SE, Deutsche Gesellschaft für Internationale Zusammenarbeit (GIZ) GmbH, Flo Ltd, FORUM Institut für Management GmbH, High-Tech Gründerfonds Management GmbH, Lindus Health Ltd, Thymia Ltd, Una Health GmbH, the Saudi Arabia Food and Drug Authority (SFDA) funded through the United Nations Development Programme (UNDP) and Ada Health GmbH, and he holds share options in Ada Health GmbH.". T.P.G. is the Founder of SST. D.A.H. is a consultant for Medtronic. A.J. is an employee of Intuitive. F.R.K. declares advisory roles for Radical Healthcare, USA; and the Surgical Data Science Collective, USA, and has received research funding from Novartis. L.Mü. is an employee at KARL STORZ SE \& Co. KG. H.J.M. is employed by and holds shares in Panda Surgical Limited. P.M. is a co-founder and shareholder of Scialytics. H.N. is affiliated with Proximie Ltd, London, UK. N.P. is co-founder and own shares in Scialytics SAS. N.R. is an employee at NVIDIA. M.M.R. is now working at SAP Fioneer. D.S.: Medtronic, Odin Vision, Panda Surgical, EnAcuity, Helico Medical, Uncovr, Vope Medical. M.W. has received honoraria from KARL STORZ SE \& Co. KG. The remaining authors declare no competing interests.

\renewcommand{\bibsection}{\section{References}}
\bibliography{sample-base}

%=========================================================
% Supplementary Notes
%=========================================================
\clearpage
\appendix

% Reset counters and define "SN" style numbering
\setcounter{section}{0}
\setcounter{figure}{0}
\setcounter{table}{0}
\renewcommand{\figurename}{Supplementary Fig.}
\renewcommand{\tablename}{Supplementary Table}

\renewcommand{\thefigure}{\arabic{figure}}
\renewcommand{\thetable}{\arabic{table}}

\renewcommand*{\thesection}{Supplementary Note \arabic{section}}
\renewcommand*{\thesubsection}{\arabic{section}.\arabic{subsection}}

% Load supplement content
\section*{Supplementary Notes}
\addcontentsline{toc}{section}{\protect\numberline{}Supplementary Notes}

\subsection*{\centering Contents}
\hrule height 0.4pt
\vspace{0.5em}

\begin{flushleft}
\small
\renewcommand{\arraystretch}{1.4}
\begin{tabularx}{\textwidth}{@{}lXr@{}}
\textbf{Supplementary Note 1} & \hspace{1em}\nameref{suppl:pitfalls} & \textbf{\pageref{suppl:pitfalls}} \\[0.2em]
\textbf{Supplementary Note 2}  & \hspace{1em}\nameref{suppl:consequences-risks} & \textbf{\pageref{suppl:consequences-risks}} \\[0.2em]
\textbf{Supplementary Note 3}  & \hspace{1em}\nameref{suppl-best-practices} & \textbf{\pageref{suppl-best-practices}} \\[0.2em]
\textbf{Supplementary Note 4} & \hspace{1em}\nameref{suppl-screening} & \textbf{\pageref{suppl-screening}} \\[0.2em]
\textbf{Supplementary Note 5} & \hspace{1em}\nameref{suppl:comparison} & \textbf{\pageref{suppl:comparison}} \\
\end{tabularx}
\end{flushleft}

\vspace{0.5em}
\hrule height 0.4pt

%------------------------------------

\newpage
\section{Pitfall descriptions and literature evidence}
\label{suppl:pitfalls}
\begingroup
\let\clearpage\relax
\let\cleardoublepage\relax
To systematically identify common flaws in validating artificial intelligence (AI) for surgical video analysis, we performed a traditional literature review, utilized agentic internet search tools, and conducted a four-stage Delphi process involving surgical data science experts and clinicians. This resulted in a catalog of 18 pitfalls across three pitfall categories. Supplementary Table~\ref{tab:pitfall-descr} provides detailed descriptions of each pitfall, evidence from literature including both papers including these flaws and papers discussing them, as well as references to illustrations in the main manuscript. \\

\begin{footnotesize}
\captionof{table}{Overview of pitfalls related to surgical artificial intelligence (AI) validation, including their categorization, descriptions, supporting evidence, and figure numbers for illustrative examples.}
\label{tab:pitfall-descr}
\vspace{-1.4\baselineskip}
\addtocounter{table}{-1}
\begin{longtable}{ P{0.4cm} P{1.4cm} B{8.3cm} P{1cm} P{1.4cm}  }
 \toprule
 \multicolumn{5}{l}{\textbf{\textit{[P1] Pitfalls related to data}}} \\
\midrule
\textbf{ID} & \textbf{Pitfall} & \textbf{Description} & \textbf{Evidence} & \textbf{Illustration} \\
\midrule
P1.1 & Non-representati\-veness, low quality, and low relevance of data & \noindent The data used for training and/or testing does not accurately reflect the intended real-world use case or lacks clinical relevance. This may, for example, result from geographical imbalance, a lack of diversity in patient demographics, regional differences in surgical standards, or limited or outdated variations in surgical techniques. Additionally, data selection criteria may exclude challenging but clinically relevant cases, such as poor camera quality, low lighting conditions, challenging procedures, or imaging artifacts, which are crucial for assessing model robustness in realistic scenarios. The inclusion of irrelevant data, such as out-of-body frames or unintentionally recorded segments, may also dilute the training signal or introduce misleading patterns. The use of simulated or experimental data may further reduce real-world applicability, while proxy data can misalign validation with clinical objectives. Variability in video quality and preprocessing protocols across institutions may further compromise data fidelity and model reproducibility. In addition, unobserved operator-related heterogeneity, such as multiple surgeons with different experience levels (e.g., trainees and attending surgeons) contributing to the same procedure, particularly in training centers, may introduce additional variance if operator roles are not annotated or reported. & \cite{godau2023deployment, kolbinger2024strategies, koch2024distribution, sellner2023semantic, kilim2022physical, castro2020causality, subasri2025detecting, lavanchy2024challenges, kassem2022federated, mascagni2022multicentric, kong2021accurate, junger2022state, bar2020impact, paranjape2023cross, engelhardt2018improving, sahu2021simulation, huaulme2023peg, kolbinger2023anatomy, kolbinger2024artificial, ding2022carts, childers2021same, jogan2024quality} & Extended Data Fig. 1 \\ \hline

P1.2 & Limited sample size/test cohorts, or data imbalance & \noindent The test data may be limited due to small sample size and/or the absence of validation on independent datasets, including those from different institutions, time periods, or populations (external validation), or from real-world clinical use after deployment (post-deployment validation). Limited data can mislead model development (e.g., algorithm selection), increase the risk of overfitting, or yield unreliable model performance validation, making it difficult to draw meaningful conclusions about the model’s real-world performance. 
These issues are amplified in the presence of class imbalance, where rare classes may appear in only one subset or fold, or be entirely absent, leading to inconsistent model performance depending on the specific validation or test set used. & \cite{bar2020impact, kostiuchik2024surgical, yamlahi2023self, demir2023deep, killeen2023pelphix, maqbool2020m2caiseg, cheng2023deep} & Extended Data Fig. 2 \\ \hline

P1.3 & Existence of spurious correlations within data & \noindent Spurious correlations are statistical patterns in the data that do not reflect robust or generalizable associations with the target task. They can arise from biases in data collection, labeling, or contextual factors unrelated to the underlying clinical objective, and often reflect dataset-specific artifacts that fail to generalize. This may lead to short-cut learning, where a model exploits irrelevant patterns, for example, achieving seemingly good performance by associating certain surgical outcomes with specific recording conditions, such as particular camera types used at different hospitals, or inferring the presence of instruments from irrelevant features such as glove color, if such features are biased within the dataset. & \cite{yuan2024advancing, banerjee2024bias, mahmood2021detecting, geirhos2020shortcut, dammu2023detecting, bykov2023finding} & Extended Data Fig. 3a \\ \hline

P1.4 & Incomplete annotation or missing contextual information & \noindent Annotations of the data are -- intentionally or unintentionally -- incomplete, leading to gaps in the information available for reliable validation. This may include missing metadata, partially annotated video sequences, the omission of specific objects, events, or surgical phases, or insufficient detail for the intended task. Missing metadata, such as age, for example, may lead to undetected confounders and thus biased models. Partially annotated video data may render the application of temporal metrics (e.g., for assessing continuity or stability across frames) infeasible and thus limit the ability to assess performance over time, such as continuity or stability across video frames. While some approaches may use partial annotations intentionally (e.g., in weak supervision), unmanaged or undocumented incompleteness may still compromise reliability and validation validity. The omission of relevant entities can result in inaccurate performance estimates, for example, underestimating sensitivity when relevant instances are not labeled, and may give a misleading impression of model accuracy and clinical utility. & \cite{pangal2021guide, ward2021challenges, tajbakhsh2020embracing, twinanda2016endonet, funke2019video, peng2024reducing, nyangoh2023systematic, murali2023endoscapes} & Extended Data Fig. 3b \\ \hline

P1.5 & Unreliable or inconsistent annotation & \noindent Unreliable or inconsistent annotations can undermine the validity of model validation by introducing ambiguity, bias, or uncontrolled variability. This may result from the lack or insufficiency of a standardized annotation protocol, high inter- or intra-rater variability, or inconsistencies in labeling the same entity across frames. Annotation variability may also arise from annotator fatigue, differences in clinical expertise, or a lack of annotation training and auditing procedures. Additionally, annotation quality may depend on task complexity, which may require multi-rater annotation and inter-/intra-rater agreement analysis. Even if annotations are consistent, they may still be unreliable, for example, if they are based on weak reference sources such as unreliable sensors or flawed clinical definitions. These inconsistencies and sources of unreliability can lead to models learning ambiguous or conflicting patterns, reduce the reliability of reported performance metrics, and make it difficult to interpret model failures or compare results across datasets or studies. & \cite{nwoye2023cholectrack20, pangal2021guide, twinanda2016endonet, sharan2023mvhota, cartucho2024surgt, bano2024placental, mao2024pitsurgrt, radsch2023labelling, luca2022impact, demir2023deep, zisimopoulos2018deepphase, ghamsarian2023cataract} & Fig. 1a \\ \hline

P1.6 & Poor data splitting & \noindent Data are not adequately split into training, validation, and test sets, leading to issues such as overfitting or data leakage. This could, for example, occur due to improper use of test splits for validation or model selection, non-stratified sampling, use of test splits for ablation studies, or the complete lack of a test set. A particularly harmful form of leakage can occur when data from the same patient is included in both training and test sets. This allows the model to leverage patient-specific characteristics, leading to overly optimistic performance estimates and reduced generalizability to new patients. The problem of data leakage is currently becoming even more severe with the emergence of generalist models trained on unknown cohorts, where the lack of transparency in the data composition makes proper separation and assessment especially challenging. & \cite{funke2023metrics, liu2023skit, gao2021trans, wei2022boxpolyp, ahmad2024wavemamba, cervantes2021automatic, he2022empirical, nwoye2022data, kostiuchik2024surgical, jiang2021risk, abdulbaki2018surgical} & Extended Data Fig. 4 \\ \midrule

\pagebreak
\midrule
\multicolumn{5}{l}{\textbf{\textit{[P2] Pitfalls related to metric selection and configuration}}} \\
\midrule
\textbf{ID} & \textbf{Pitfall} & \textbf{Description} & \textbf{Evidence} & \textbf{Illustration} \\
\midrule

P2.1 & Mismatch of metrics and clinical needs & \noindent The used metrics do not reflect the underlying needs of the clinical use case or fail to capture the impact of the model’s performance on clinical decision-making. This often reflects a focus on technical performance rather than clinical relevance. For example, a high frame-wise accuracy in surgical phase recognition tasks may not be clinically meaningful in case the model frequently misidentifies short but clinically relevant phases or critical transitions between surgical phases. Another example includes specific operational requirements such as real-time capabilities, which are often not considered in performance validation, despite being critical for clinical usability. Moreover, clinical needs are shaped not only by the task itself, but also by the intended role of the AI system, for example, whether it supports decision-making or acts autonomously. & \cite{kankanamge2025artificial, hashimoto2024foundation, reinke2024understanding, kolbinger2024strategies, brand2022frame, bar2020impact} & Extended Data Fig. 5 \\ \hline

P2.2 & Lack of metrics that assess temporal aspects & \noindent The used metrics do not assess algorithm properties specific to temporal data. For example, in surgical instrument segmentation, validating performance solely based on frame-wise Dice Similarity Score (DSC) may result in overlooking temporal inconsistencies, such as abrupt appearance or disappearance of instruments between consecutive frames. In addition to temporal coherence of predictions, metrics for evaluating real-time system behavior, such as latency or throughput, are often missing, despite their relevance in time-critical clinical settings. & \cite{dergachyova2016automatic, boers2020improving, zhang2022retrieval, berlet2022surgical, sharan2023mvhota, luiten2021hota, kopf2021robust, funke2023metrics, brand2022frame}& Fig. 1b \\ \hline

P2.3 & Inappropriate metric selection for handling  annotation uncertainties & \noindent Annotation uncertainty is not adequately addressed in metric selection, which can result in misleading performance validation. This uncertainty may arise from inter- and intra-rater variability in subjective tasks, low visibility in surgical videos, or inconsistent labeling of ambiguous regions (e.g., tissue boundaries or transition moments) over time. For example, ambiguity in defining event boundaries - such as the exact moment a surgical phase transition occurs - may introduce inconsistencies in annotations due to inter- or intra-rater variability and missing exact ground truth. & \cite{dergachyova2016automatic, funke2023metrics, berniker2021probabilistic, berlet2022surgical, nwoye2022data} & Extended Data Fig. 6 \\ \hline

P2.4 & Lack of metric robustness across varying real-world conditions & \noindent The selected metrics lack robustness with respect to various real-world conditions such as data quality, acquisition settings, or clinical variability. For example, performance validation may be overly sensitive to changes in frame rate, image resolution, zooming, differences in surgical technique or intraoperative conditions, or annotation granularity. This includes common intraoperative phenomena such as smoke, motion blur, objects temporarily leaving the field of view, or changes in object size due to camera movement or zoom. These factors can distort metric behavior even when model predictions remain stable, and may substantially affect metric stability and reproducibility. & \cite{nwoye2025surgitrack, czempiel2021opera, knoche2021susceptibility} & Extended Data Fig. 7  \\ \hline

P2.5 & Non-standardized configuration and definition of metrics & \noindent Metrics without standardized configuration within a specific use case or definition are used, often without providing details on the concrete formula. For example, hyperparameters such as thresholds for surgical instrument detection, temporal tolerance ranges for transitions in phase recognition (e.g., how many frames of deviation are accepted as correct), or weighting of different error types may not be clearly defined or reported. In addition, even when the same metric is used, differences in spatial or temporal applications, such as calculating accuracy only during annotated segments versus the full video, or validating phase recognition at different frame rates, can lead to non-comparable results and misinterpretation. & \cite{funke2023metrics, gao2021trans, jin2021temporal, jin2017sv} & Extended Data Fig. 8a \\ \hline

P2.6 & Non-suitability of hyperparameters from other domains & \noindent Hyperparameters or default metric configurations adopted from other domains (e.g., general computer vision) may not align with the specific underlying clinical needs. For example, in object detection, a high Intersection over Union (IoU) threshold may be appropriate in general computer vision tasks, but in surgical applications, lower IoU thresholds might be more suitable when only rough localization of objects (e.g., polyps) is needed \cite{tran2023sources}. Similarly, default hyperparameters for event detection in action recognition may not account for the variability in surgical workflows. In marker-less tool tracking (e.g., \cite{hein2025next}), for example, thresholds that define acceptable accuracy in everyday scenarios (e.g., a few millimeters) may be insufficient in surgical contexts, where sub-millimeter precision can be clinically critical. & \cite{tran2023sources, ramesh2023dissecting, hein2025next} & Extended Data Fig. 8b \\ \hline

P2.7 & Intrinsic limitations of individual metrics & \noindent The used metrics harbor pitfalls related to their individual mathematical properties. Even if metrics are well-aligned with the clinical task, they may exhibit problematic behaviors such as sensitivity to class imbalance, non-linearity, or lack of interpretability. The suitability of any given metric should thus be analyzed in light of its known limitations. For example, accuracy may produce misleading values in highly imbalanced data sets. & \cite{reinke2021common, reinke2024understanding} & \\ \midrule

\multicolumn{5}{l}{\textbf{\textit{[P3] Pitfalls related to metric aggregation and reporting}}} \\
\midrule
\textbf{ID} & \textbf{Pitfall} & \textbf{Description} & \textbf{Evidence} & \textbf{Illustration} \\
\midrule

P3.1 & Non-independence within the test set & \noindent Failure to account for non-independence within the test set can yield misleading conclusions. Non-independence can occur when multiple frames from the same surgical video are included, or when data from the same patient, procedure, or institution are used. Aggregation, statistical analysis, and reporting need to account for this lack of independence rather than assuming independent samples. For example, if frames from the same patient are used for validation, performance metric values may appear artificially high due to strong temporal correlation. Other possible consequences include biased statistical estimates, such as underestimated uncertainty or distorted means. & \cite{kostiuchik2024surgical, funke2023metrics, kirtac2022surgical} & Fig. 4 \\ \hline

P3.2 & Clinically uninformative aggregation & \noindent Performance metrics are often simply aggregated, without accounting for the varying importance of different surgical phases or time segments, or for performance differences across clinically relevant conditions. For example, averaging over all frames may obscure poor performance during critical moments, and reporting unstratified metrics may hide failures in challenging scenarios such as for rare phases, low-quality recordings, presence of artifacts, or small anatomical structures, potentially diluting errors during rare but clinically significant events. Aggregating over meaningful temporal units and stratifying results by clinically relevant factors, including those affected by long-tail distributions (e.g., rare events or structures), as well as subgroup characteristics such as disease severity, surgical indication, or operator experience, enables more informative and clinically useful validation. Stratification can also improve transparency and fairness by highlighting differences in performance across subgroups. However, care must be taken in how metrics are applied and aggregated in stratified settings, as some metrics may behave non-intuitively when applied to imbalanced or small subgroups. & \cite{funke2023metrics, kostiuchik2024surgical, zohar2020accurate, czempiel2023symphony, ross2023beyond, nwoye2023cholectrack20, nwoye2025surgitrack, huaulme2024global, hung2019experts} & Extended Data Fig. 9a \\ \hline

P3.3 & Lack of contextualization of performance values & \noindent Performance values are reported without sufficient context, making it difficult to assess their clinical or practical relevance. For example, results may be presented without comparison to human performance, inter-rater agreement, or a meaningful performance threshold that defines an acceptable error rate for the clinical task. Additionally, the significance of observed differences between models may not be validated, resulting in misleading interpretations. Without such contextualization, the clinical utility of a method may remain unclear. Additionally, aspects of how performance information is communicated to end users, including the design of visualizations, thresholds, and labels, can influence clinical perception and decision-making, and should be considered in the contextualization process. & \cite{reinke2024understanding, jogan2024quality, sylolypavan2023impact, ward2021challenges} & Extended Data Fig. 9b \\ \hline

P3.4 & Lack of uncertainty analyses & \noindent Performance values are reported without conveying the uncertainty associated with the results, leading to overconfidence in the model’s performance. For example, confidence intervals, standard errors, or standard deviations may be missing, concealing how much variability is present in the reported metrics, including across time or clinically relevant subgroups. & \cite{christodoulou2024confidence, ban2021aggregating, shi2022attention, hao2023act, wang2023cascade} & Fig. 5 \\ \hline

P3.5 & Insufficient reporting & \noindent Reporting of results lacks sufficient detail and transparency, making it difficult to interpret, compare, and reproduce findings. Critical aspects, such as the strategy and rationale for aggregating performance metrics, details of metric computation, or the limitations of the validation data, are often underreported or unclear. In some cases, reporting may be selective rather than merely incomplete, which can lead to overestimation of performance and misleading impressions of clinical safety or utility. Additionally, established reporting guidelines tailored to specific purposes (e.g., CONSORT-AI for clinical trials, CLAIM for medical imaging) are frequently not followed, leading to incomplete or inconsistent documentation. Another common gap is the lack of documentation regarding data modifications introduced by acquisition hardware or manufacturer-specific processing pipelines (e.g., compression, interlacing, or automatic enhancement), which can affect model performance in subtle and unquantifiable ways. & \cite{koccak2025adherence, loftus2023artificial, navarro2023systematic, marwaha2023appraising, junger2022state, funke2023metrics} & Fig. 6 \\ \bottomrule

\end{longtable}
\end{footnotesize}
\endgroup

\newpage
\section{Descriptions of Pitfall consequences and real-world risks}
\label{suppl:consequences-risks}
Each of the identified pitfalls was mapped to specific potential consequences, such as introduction of biases, unreliable performance assessment, or undetected failure modes, and as to associated real-world risks (e.g., regulatory delay, compromised surgical safety). Supplementary Tables~\ref{tab:consequences} and~\ref{tab:risks} provide detailed descriptions of consequences and risks.
\vspace{0.5em}

\begin{footnotesize}
\captionof{table}{Descriptions of potential consequences of the identified pitfalls.}
\label{tab:consequences}
\vspace{-0.8cm}
\addtocounter{table}{-1}
\begin{longtable}{ P{2.3cm} B{10.7cm}}
    \toprule
      \textbf{Consequence}   &  \noindent\textbf{Description}\\ \hline

        Introduction of biases & \noindent Biases in surgical AI refer to systematic errors arising from flaws in the data used for model development or validation, including imbalanced patient cohorts, underrepresentation of relevant scenarios, sampling artifacts, or spurious correlations. These flaws can lead to issues like shortcut learning, where models exploit statistically predictive but clinically irrelevant patterns. As a result, models may produce non-representative performance estimates, behave unreliably in real-world settings, and fail to generalize across diverse patient populations or surgical workflows. Such biases can not only reduce clinical trust but also risk perpetuating systemic blind spots in algorithmic behavior. \\ \hline
        
        Unreliable performance assessment & \noindent Unreliable performance assessment refers to the generation of performance estimates that are unstable, inconsistent, or sensitive to uncontrolled aspects of the validation setup, or simply uninformative. This may result from non-robust experimental design, poor data splits, or metric configurations that are highly sensitive to implementation details. As a consequence, reported results may fluctuate across settings, making it difficult to draw reliable conclusions, compare methods, or understand true model behavior. \\ \hline
        
        Insufficient transparency and/or reproducibility in performance assessment & \noindent Insufficient transparency and/or reproducibility in performance assessment refers to situations where results cannot be fully interpreted, contextualized, or independently reproduced. This may occur when critical details of the validation process are missing, unclear, or inconsistently applied, including aspects such as aggregation details or data preprocessing steps. As a result, others may be unable to replicate findings, understand the sources of variation in reported results, or assess their applicability to related clinical settings. \\ \hline
        
        Undetected failure modes & \noindent Failure modes are patterns of incorrect or unsafe behavior that an AI model can exhibit under certain conditions. In surgical AI, these often arise from corner/edge cases – rare but clinically important scenarios in surgical video analysis that differ from typical training data – or from inadequate assessment of temporal behavior. Failing to properly validate these cases can lead to unreliable model behavior. When development datasets fail to capture such edge cases, these failure modes remain unnoticed during validation and may only surface in real-world use. Examples include unexpected anatomical variations, poor lighting conditions, occlusions, rare surgical complications, detection delay or instability of predictions across consecutive frames. Overemphasis on common cases leaves model performance in these critical situations unknown. \\ \hline
        
        Leakage and/or model overfitting & \noindent Leakage and model overfitting result from improper use of data during model development and may lead to misleadingly high performance estimates that do not reflect how the model will perform in real-world clinical scenarios, reducing real-world applicability. Data leakage occurs when information from the data used for testing is also present during development e.g., through shared pre-processing, improper data splitting, or data re-use, violating the assumption that test data are independent and unseen. Model overfitting refers to selecting or optimizing models based on patterns that do not generalize beyond the data used during development. It often occurs when no properly untouched test set is reserved, allowing models to perform well only in the corresponding test set rather than showing true generalizability.\\ \hline
        
        Suboptimal resource use & \noindent Suboptimal resource use refers to an inefficient use of time, funding, and (computational) resources in surgical AI development. This may occur when validation does not sufficiently reflect clinical needs, leading researchers to invest in models that address limited or misaligned problems and are unlikely to make a real-world impact. While such models may still contribute to methodological advancement, a lack of alignment with clinical priorities can limit the practical value of the work. Consequently, scientific progress may be slowed. \\ \bottomrule
    
\end{longtable}
\end{footnotesize}

\begingroup
\let\clearpage\relax
\let\cleardoublepage\relax
\vspace{0.6em}
\begin{footnotesize}
\captionof{table}{Descriptions of potential real-world risks of the identified pitfalls.}
\label{tab:risks}
\vspace{-0.5cm}
\addtocounter{table}{-1}
\begin{longtable}{ P{2.3cm} B{10.7cm}}
 \toprule

\textbf{Real-world risk}   & \noindent  \textbf{Description}\\ \hline

        Overestimation of algorithm performance & \noindent Overestimation of algorithm performance refers to a mismatch between how well a surgical AI model is perceived to perform and how it actually behaves in clinical reality. This risk may arise from flawed or incomplete validation, misaligned performance metrics, or misinterpretation of reported results. As a consequence, models may appear more reliable, generalizable, or clinically useful than these truly are. Overestimation can lead to premature deployment, overreliance by clinicians, or inadequate oversight, increasing the risk of downstream errors, inefficiencies, or patient harm.\\ \hline
        
        Unfairness of algorithms& \noindent Unfairness of algorithms refers to systematic performance differences across demographic groups, such as patients of different sexes, races, ethnicities, or socioeconomic backgrounds. This unfairness often arises from imbalanced training data, biased annotations, or model design choices that fail to ensure equitable performance. As a result, some groups may consistently receive more accurate predictions than others, leading to unequal treatment and raising significant ethical and clinical concerns. \\ \hline
       
        Poor generalization to real-world applications & \noindent Poor generalization refers to the inability of an algorithm to maintain consistent performance across diverse real-world settings. This often results from shifts in context, technology, or clinical implementation that are not adequately captured during model development and validation. While a model may perform well in controlled settings, it may struggle when applied to different surgical procedures and teams, clinical workflows and sites, or medical devices (among others). For example, a model trained on data from one hospital may not ne generalized to another due to variations in equipment, imaging quality, or surgeon-specific practices. \\ \hline
        
        Limited clinical utility & \noindent Limited clinical utility refers to the inability of algorithms to provide meaningful benefit in real-world clinical practice. This may occur when models fail to support clinical decision-making, integrate into workflows, or deliver reliable performance in diverse settings. Performance metrics may fail to capture clinically relevant aspects or models’ practical usability in clinical settings, thus concealing their limited clinical utility. As a result, such models may increase operating times, introduce workflow inefficiencies, or place additional cognitive burden on clinicians. \\ \hline
        
        Selection of sub-optimal algorithms & \noindent Selection of sub-optimal algorithms refers to the risk of choosing models based on misleading or incomplete validation rather than true clinical utility. In case performance metrics do not accurately reflect real-world applicability, models may be selected that perform well in development settings but lead to errors, inefficiencies, or reduced quality of care in clinical practice. Conversely, more effective algorithms may be overlooked, limiting clinical benefit and slowing progress in AI-assisted surgery.\\ \hline
        
        Compromised clinical safety & \noindent Compromised clinical safety refers to the potential of surgical AI systems to contribute to medical errors, adverse events, or unsafe clinical decisions. This may occur when models produce inaccurate or misleading outputs, are used beyond their validated scope, are applied in situations in which their behavior is not well understood, or from flawed validation practices, such as inappropriate metric use, data leakage, or poorly designed validation, that fail to reveal limitations prior to clinical use. In such cases, AI use can lead to delays, complications, or inappropriate interventions that put patient safety at risk.\\ \hline
        
        Loss of trust in surgical AI & \noindent Loss of trust in surgical AI refers to clinicians and other stakeholders becoming hesitant to adopt or rely on AI models when these systems fail to perform reliably or align with expectations. This may be caused by repeated failures, biases, poor generalization, unpredictable behavior, or lack of transparency. When models fail to match expectations set during development or produce unreliable predictions, unexplained errors, or inconsistencies across different clinical settings, confidence in their usefulness declines, potentially also reducing clinicians’ willingness to engage with or contribute to future surgical AI research and development. Such loss of trust may also affect regulators, institutional stakeholders, and investors, and can raise concerns around ethical responsibility and accountability in the use of surgical AI. \\ \hline
        
        Stagnation of research progress&\noindent Stagnation of research progress refers to a slowdown in innovation and advancement within surgical AI due to unreliable validation practices or non-comparable benchmarks. When flawed metrics, poor generalization, or inconsistent reporting, among others, obscure the true algorithm performance, it becomes difficult to identify meaningful improvements or reproduce prior findings. As a result, researchers may repeatedly explore already-solved problems, waste resources on uninformative comparisons, or fail to translate experimental insights into clinical progress, ultimately hindering scientific development in the field.\\ \hline
        
        Regulatory rejection or delay&\noindent Regulatory rejection or delay refers to the failure of surgical AI systems to obtain timely approval for clinical use due to insufficient or unconvincing validation or compromised quality control processes. Regulatory pathways are intended to safeguard patient safety and public trust. When AI systems do not adequately demonstrate clinical benefit, fairness, safety, effectiveness, generalizability, or reliability, approval may be withheld – delaying or preventing their availability for real-world use.\\ \hline
        
        Investment risk&\noindent Investment risk refers to the potential for financial, institutional, or strategic resources to be committed to surgical AI systems that ultimately fail to deliver clinical value or broader impact. This may result from flawed validation, misleading performance claims, or inadequate alignment with real-world needs. As a result, organizations may experience financial losses, reputational damage, or delays in innovation.\\ \hline
        
        Lack of clinical/real-world adoption of surgical AI&\noindent Lack of clinical adoption of surgical AI refers to the failure of AI models to be integrated in real-world settings. Challenges such as misleading performance metrics, poor generalization, usability issues, or a mismatch between expected and actual clinical performance can create barriers to the integration and acceptance of surgical AI in the clinical workflow. Even if a model shows strong results in experimental settings, a lack of validation in real-world conditions - such as different surgical teams, workflows, or patient populations - can limit its adoption. If AI tools do not align with clinical needs or fail to provide tangible improvements over existing methods, these may struggle to gain acceptance and practical use in surgical settings.\\ \hline
        
        Reduced or delayed patient/caregiver benefit&\noindent Reduced or delayed patient and caregiver benefit refers to the diminished impact or delayed realization of benefits of surgical AI on patient outcomes and clinical support when models fail to generalize, align with clinical needs, or perform reliably in practice. This may follow from performance expectations that were not met in real-world settings. Instead of enhancing surgical precision, efficiency, or safety, unreliable models may introduce errors, delays, or additional workload for healthcare professionals. In the worst cases, AI-driven mistakes can lead to complications or adverse events, ultimately harming patients rather than improving care.\\ \hline
        
        Negative environmental impact&\noindent Negative environmental impact refers to the unintended ecological consequences of inefficient or unvalidated surgical AI pipelines. Poorly designed benchmarks and redundant training of sub-optimal models can lead to excessive computational resource use, unnecessary data processing, and inflated carbon emissions. Moreover, lack of reproducibility or transparency may cause repeated experiments or model retraining without added value. Together, these factors contribute to an avoidable environmental burden that undermines the sustainability of surgical AI research and deployment.

 \\ \bottomrule

\end{longtable}

\end{footnotesize}

\endgroup

%\include{tables/tab-risks}

%------------------------------------
\section{Practical best practices}
\label{suppl-best-practices}

\begin{tcolorbox}[pitfallbox,title={P1.1 -- Non-representativeness and low relevance of data}]

\textbf{Avoiding non-representativeness and low relevance of data}

\begin{checklist}
\item Identify aspects of variability related to the given clinical use case (e.g., surgical equipment, surgical technique, data quality). \textit{Aspects of variability include:}
\begin{checklist}
    \item Patient-related factors, e.g., age, sex, comorbidities, anatomical variability, disease stage, previous surgeries
    \item Surgeon- and surgical team-related factors, e.g., surgical technique or approach, procedure type and complexity, surgical expertise, intraoperative events or complications
    \item Device- and equipment-related factors, e.g., type and manufacturer of surgical devices or instruments, robotic vs. conventional systems, software versions, and key system parameters (e.g., insufflation settings)
    \item Acquisition-related factors, e.g., image resolution, frame rate, camera angle, zoom level, lighting conditions
    \item Institution- and site-related factors, e.g., hospital type (academic vs. non-academic), geographical region, clinical protocols, infrastructure 
    \item Temporal and workflow-related factors, e.g., surgical phase, duration of procedure, phase transitions, atypical workflows, workflow integration, and digitization level (e.g., differences in how teams interact with digital systems)
\end{checklist}
\item Make sure data are collected in a manner that reflects these sources of variability.
\item Include rare and edge cases (e.g., poor lighting, imaging artifacts, adverse events) based on a predefined coverage plan that specifies which clinically relevant outlier scenarios must be represented.
\item Prioritize data collection from underrepresented locations based on systematic monitoring of geographic and institutional coverage gaps.
\item Prioritize data collection for scenarios identified as high-impact through clinician- or stakeholder-driven needs assessment.
\item Collect and report metadata that enables comparison to the target population (e.g., age, sex, country, comorbidities). If subject-level metadata cannot be shared due to privacy constraints, provide aggregated statistics (e.g., distributions).
\item Use standardized protocols or data descriptors and tools (e.g., data report cards \cite{gebru2021datasheets}) describing scope, inclusion/exclusion criteria, dataset composition, and representational limitations.
\end{checklist}

\textbf{Dealing with non-representativeness and low relevance of data}

\begin{checklist}
\item Define the target population for the dataset (e.g., data from the US only with specific patient characteristics should not be used to generalize to Asia), clearly specifying the clinical, demographic, and procedural characteristics the data are intended to represent.
\item When rare or edge cases are underrepresented, explicitly assess performance on these cases as separate subgroups. If datasets are intentionally enriched to include rare scenarios, report subgroup-specific results separately and avoid interpreting aggregate performance as representative of the general population unless prevalence is accounted for.
\item Quantify and report deviations between dataset characteristics and the intended-use population, including differences in disease prevalence across regions or demographic groups. Avoid claims of global representativeness unless dataset composition reflects known geographic and epidemiological variation.
\end{checklist}

\end{tcolorbox}

\begin{tcolorbox}[pitfallbox,title={P1.2 -- Limited sample size/test cohorts, or data imbalance}]

\textbf{Avoiding limited sample size/test cohorts, or data imbalance}

\begin{checklist}
\item Define planned training, validation, and test cohort size prior to study, informed by the intended clinical use case and anticipated feasibility constraints. Where appropriate, empirically assess whether the available sample size is sufficient by analyzing performance as a function of training set size (e.g., learning curve analysis) to detect instability or saturation effects and predefine the minimum numbers of clinically critical or rare events required for meaningful model development and performance assessment. Make clear which criterion governs decision‐making for model validity in the given clinical use case.
\item When data collection is costly, put high emphasis on data representativeness and coverage of clinically highly relevant cases.
\end{checklist}
\bigskip
\textbf{Dealing with limited sample size/test cohorts, or data imbalance}

\begin{checklist}
\item When using small datasets, clearly state limitations, define the target population, report feasibility constraints, and frame work as proof-of-concept without focusing on generalizability.
\item Quantify performance variability using cross-validation, for example to assess robustness and variability of performance estimates, but do not use cross-validation as a substitute for validation on an independent, unseen test set when assessing generalizability.
\item Complement overall performance metrics by validating performance on rare events or corner cases through targeted error analysis and class- and subgroup-specific analysis.
\item If decision thresholds are optimized for specific clinical scenarios, pre-specify the optimization strategy.
\item In the presence of substantial class imbalance, refrain from using prevalence-dependent metrics (e.g., Accuracy, PPV, F1-Score). Instead, rely on metrics that remain informative under imbalanced conditions (e.g., Balanced Accuracy, Sensitivity, AUROC), report performance at the class level, and follow recommendations from existing frameworks (e.g., Metrics Reloaded \cite{maier2024metrics}).
\item If appropriate, report the expected clinical prevalence of the target condition and quantify the resulting numbers of relevant prediction errors (e.g., false positives and false negatives) at clinically relevant operating points.
\end{checklist}

\end{tcolorbox}

\begin{tcolorbox}[pitfallbox,title={P1.3 -- Existence of spurious correlations within data}]

\textbf{Avoiding and detecting spurious correlations within the data}

\begin{checklist}
\item Use domain knowledge and clinical expertise to identify potential non-causal correlations, for example by flagging variables that are clinically implausible causes of the target outcome but may co-occur with it due to workflow, acquisition, or institutional practices (e.g., specific instruments or devices systematically associated with certain procedures or disease stages).
\item During data collection avoid confounder-label coupling by ensuring confounders vary across cases in every class.
\item Use correlation analyses, subgroup analyses, or systematic data auditing procedures (e.g., structured inspection of metadata-label associations, stratified summaries, or predefined confounder checklists) to identify hidden dependencies and potential confounders (e.g., \cite{pavlak2023data, drenkow2025detecting}).
\item Analyze model behavior under controlled perturbations of potential confounders, including both real-world factors such as acquisition or workflow-related variables and controlled synthetic perturbations used to probe model reliance on spurious cues, to detect whether predictions depend on spurious correlations \cite{veitch2021counterfactual, lapuschkin2019unmasking, mahmood2021detecting}.
\end{checklist}

\bigskip
\textbf{Dealing with spurious correlations within the data}

\begin{checklist}
\item Explicitly analyze and report dataset characteristics and metadata relevant to suspected confounders (e.g., device type, site, surgeon, acquisition protocol) prior to model development, and document how these factors co-occur with target labels (e.g., \cite{lavanchy2024challenges}).
\item Assess, to which extent confounders affect the quality of the model (e.g., by performing stratified validation across clinically or technically relevant subgroups \cite{zech2018variable, oakden2020hidden}).
\end{checklist}

\end{tcolorbox}

\begin{tcolorbox}[pitfallbox,title={P1.4 -- Incomplete annotation or missing contextual information}]

\textbf{Avoiding incomplete annotation or missing contextual information}

\begin{checklist}
\item Define or reuse standardized annotation protocols, including annotator expertise, training background, and annotation materials. Align the annotation protocol with the intended downstream task(s) and performance metric(s) to avoid over-annotating irrelevant details or under-annotating clinically critical information. Where applicable, align label definitions with established clinical ontologies or standards (e.g., OntoSPM \cite{gibaud2018toward}, SNOMED CT \cite{bhattacharyya2015introduction}) to support consistent interpretation. Make protocols publicly accessible and subject to external peer review.
\item Include exemplary images in the annotation protocol, especially for unusual, infrequent, or out-of-distribution cases \cite{radsch2023labelling, mascagni2021surgical}.
\item Include rare, clinically relevant edge cases, including outlier or out-of-distribution cases, in the annotation and annotation protocol and specify which types of events or structures are at highest risk for incomplete labeling.
\item Ensure annotation protocol compliance before data usage and throughout the annotation process ideally by monitoring usability, adherence, and possible annotation drift via periodic audits.
\item Provide practical training and accessible instructions to annotators, especially for non-medical experts, to establish a shared interpretation framework for ambiguous or complex structures.
\item Include metadata required to interpret annotation completeness and context (e.g., timepoints, clinical variables, patient characteristics, clinical information).
\item Use automatic tools to pre-screen for missing annotations or context gaps, but treat automated completion as provisional and have it reviewed by domain experts to avoid introducing annotation biases.
\end{checklist}

\bigskip
\textbf{Dealing with incomplete annotation or missing contextual information}

\begin{checklist}
\item Quantify inter-/intra-rater variability to identify regions/classes/image characteristics associated with potentially high omission risk. 
\item Define completeness criteria, then use them to assess and document the completeness of annotation and metadata completeness. Use statistics to flag likely omissions. Report how incomplete annotations were handled in validation (e.g., exclusion, masking, imputation).
\item Publish annotation characteristics (e.g., label coverage, omission patterns, class prevalences) alongside datasets to support reproducibility and downstream auditing.
\item If resource constraints prevent complete annotation, explicitly document the sampling strategy used to select cases, frames, or regions for annotation, and report coverage metrics (e.g., the proportion of annotatable content that was annotated).
\end{checklist}

\end{tcolorbox}

\begin{tcolorbox}[pitfallbox,title={P1.5 -- Unreliable or inconsistent annotation}]

\textbf{Avoiding unreliable or inconsistent and unreliable annotation}

\begin{checklist}
\item Perform targeted re-annotation of ambiguous cases, such as cases with high inter- or intra-rater variability or systematically inconsistent annotations.
\item Assign a consistent instance identifier (e.g., object or track ID) to the same instance of an object in different frames, especially for temporal tasks, such as tracking.
\item Verify temporal coherence of annotations across frames (e.g., boundary stability over time in the face of object movement).
\item Employ multiple annotators and use inter-/intra-rater variability analysis to identify regions/images of high ambiguity. 
\item Use expert consensus, arbitration, adjudication protocols, or probabilistic labeling strategies where appropriate, and allow fuzzy boundaries when clinically justified.
\item \textit{(see also BP-1.4.1)} Define or reuse standardized annotation protocols, including annotator expertise, training background, and annotation materials. Align the annotation protocol with the intended downstream task(s) and performance metric(s) to avoid over-annotating irrelevant details or under-annotating clinically critical information. Where applicable, align label definitions with established clinical ontologies or standards (e.g., OntoSPM \cite{gibaud2018toward}, SNOMED CT \cite{bhattacharyya2015introduction}) to support consistent interpretation. Make protocols publicly accessible and subject to external peer review.
\item \textit{(see also BP-1.4.2)} Include exemplary images in the annotation protocol, especially for unusual, infrequent, or out-of-distribution cases \cite{radsch2023labelling, mascagni2021surgical}.
\item \textit{(see also BP-1.4.3)} Include rare, clinically relevant edge cases, including outlier or out-of-distribution cases, in the annotation and annotation protocol and specify which types of events or structures are at highest risk for inconsistent or ambiguous labeling.
\item \textit{(see also BP-1.4.4)} Ensure annotation protocol compliance before data usage and throughout the annotation process by monitoring usability, adherence, and possible annotation drift via periodic audits.
\item \textit{(see also BP-1.4.5)} Provide practical training and accessible instructions to annotators, especially for non-medical experts, to establish a shared interpretation framework for ambiguous or complex structures.
\end{checklist}

\bigskip
\textbf{Dealing with unreliable or inconsistent and unreliable annotation}

\begin{checklist}
\item Describe and assess how inconsistent or noisy annotations bias specific metrics.
\item When inter-/intra-rater variability is high, explicitly validate model performance against multiple reference annotations or stratified by annotator, and report variability of performance estimates across observers.
\item Align annotation strategies with task ambiguity and recognize when multiple "correct" interpretations are acceptable, treating such ambiguity as inherent rather than as a quality-control issue. 
\item Identify and report systematic annotation inconsistencies or error patterns (e.g., specific classes, temporal segments, or data modalities) to understand structured sources of annotation unreliability.
\end{checklist}

\end{tcolorbox}

\begin{tcolorbox}[pitfallbox,title={P1.6 -- Poor data splitting}]

\textbf{Avoiding poor data splitting}

\begin{checklist}
\item Identify sources of variability \textit{(see also BP-1.1.1)} and ensure that data splitting explicitly reflects these variability dimensions (e.g., patient, procedure, site, device) and the intended generalization scenario (e.g., stratified vs. leave-one-group-out splits). 
\item Ensure no overlap occurs between training, validation, and test data, such as data derived from the same patient, session, video, or from any temporally adjacent frames that are similar, to prevent data leakage. 
\item Reserve an untouched test set that is only used once for final validation once model development is complete. In cases where data are insufficient to reserve an additional unseen test set, use carefully designed validation schemes (e.g., cross-validation with leave-one-group-out constraints or nested cross-validation), provided that they preserve independence and are not interpreted as evidence of external generalization.
\item If generalization to specific variations is to be tested, designate these variations as unseen test scenarios and document the resulting distributional shift explicitly. Clearly state which dimensions of generalization are validated (e.g., across sites, devices, surgeons, patients, or time). 
\item Use hierarchical splits at higher levels (e.g., hospital, surgeon) if the dataset is large enough.
\item Set up the data splitting strategy a priori along with the indication for use and deployment scenario. Clearly describe and justify the chosen strategy, including rationale, clinical thresholds, and dataset characteristics for each split.
\item Consider temporal splitting for time-sensitive tasks to avoid historical bias and assess whether performance changes when models are validated on chronologically newer data reflecting evolving clinical practice.
\end{checklist}

\bigskip
\textbf{Dealing with poor data splitting}

\begin{checklist}
\item Document the independence of the test set from the training/validation sets. Ensure that the test set  is only used for final validation by tracking and disclosing all instances of test-set access and confirming that no iterative tuning occurred using test data.
\item When comparing models to other studies, report whether those other studies used the same data splits.
\item If post hoc issues in the split are discovered (e.g., leakage), treat the original test set as compromised, re-run the validation with corrected splits, and report how results change. For final claims, reserve a new untouched hold-out (or external dataset) that is not used during any further iteration.
\end{checklist}

\end{tcolorbox}

\begin{tcolorbox}[pitfallbox,title={P2.1 -- Mismatch of metrics and clinical needs}]

\textbf{Avoiding a mismatch of metrics and clinical needs}

\begin{checklist}
\item Involve clinicians early to identify which clinical aspects must be measured (e.g., diagnostic accuracy, robustness, detection speed, sensitivity to high-risk errors, inference latency or real-time capability) and select and justify metrics that appropriately capture these aspects. 
\item Use multiple metrics to capture both technical and clinical performance aspects. Provide a rationale for each metric used, including critical thresholds, clinical implications, and associated risks. 
\item Build on existing metric-selection frameworks (e.g., Metrics Reloaded \cite{maier2024metrics}) to support transparent, context-aware metric selection.
\item Avoid using metrics solely for popularity reasons and justify the exclusion of commonly used metrics when these do not reflect the intended clinical purpose.
\item Assess whether the metrics used appropriately penalize clinically high-risk errors (e.g., missed critical structures, delayed detection) to avoid misleadingly optimistic results.
\item Operationalize clinically meaningful thresholds by tying thresholds to clinical action boundaries, explicitly define what constitutes sufficient performance a priori, and define failure-trigger criteria that, if violated, invalidate deployment regardless of averages.
\end{checklist}

\bigskip
\textbf{Dealing with a mismatch of metrics and clinical needs}

\begin{checklist}
\item Explicitly state clinically relevant aspects that are not reflected by the chosen metrics (e.g., temporal stability or clinician workload implications).
\item Benchmark metric behavior against human perceptual or clinical judgments in challenging scenarios to detect misalignment between metric scores and clinically meaningful assessments of performance.
\end{checklist}

\end{tcolorbox}

\begin{tcolorbox}[pitfallbox,title={P2.2 -- Lack of common metrics that assess temporal aspects}]

\textbf{Avoiding a lack of common metrics that assess temporal aspects}

\begin{checklist}
\item Close the current lack of clinically validated temporal-aware metrics by promoting research that develops and analyzes metrics capturing temporal behavior, transitions, and stability in surgical workflows.
\end{checklist}

\bigskip
\textbf{Dealing with a lack of common metrics that assess temporal aspects}

\begin{checklist}
\item Report both frame-wise and temporal performance and analyze discrepancies between them to reveal inconsistent temporal behavior (e.g., unstable phase transitions) that frame-wise metrics cannot detect.
\item Adopt existing temporal-aware metrics (e.g., time-based F1 \cite{skat2025evaluation}, transitional delay \cite{hisey2021system}, segmental F1 score \cite{farha2019ms}, jitter \cite{zhao2024spatio}) from related domains such as computer vision, provided their underlying assumptions align with surgical workflow dynamics. Validate temporal-aware metrics on real surgical workflows to ensure these capture clinically meaningful temporal patterns, transitions, and delays. Use them to assess temporal consistency, transition stability, and delay tolerance, rather than relying solely on frame-wise metrics.
\item Analyze how errors may propagate across time (e.g., a misclassification early in a sequence leading to cascading downstream errors), particularly in long or complex procedures.
\item Use visualization and interpretable temporal diagnostics (e.g., transition graphs, temporal heatmaps) to contextualize numeric temporal metrics and detect workflow-relevant instabilities.
\item Standardize temporal metric definitions and encourage shared benchmarks to ensure comparability across studies. Standardization should be done use case specific in multidisciplinary consortia.
\end{checklist}

\end{tcolorbox}
\bigskip
\begin{tcolorbox}[pitfallbox,title={P2.3 -- Inappropriate metric selection for handling annotation uncertainties}]

\textbf{Avoiding inappropriate metric selection for handling annotation uncertainties}

\begin{checklist}
\item Establish annotation protocols that explicitly address known ambiguities (e.g., event boundaries, transitions). Characterize these ambiguities by reporting human annotator performance or inter-rater variability as a contextual reference for interpreting model predictions in ambiguous settings.
\item Select metrics whose definition explicitly matches the expected annotation uncertainty structure (e.g., tolerance windows for temporal ambiguity \cite{dergachyova2016automatic, funke2023metrics}) and avoid interpreting deviations within these uncertainty bounds as clinically meaningful errors.
\end{checklist}

\bigskip
\textbf{Dealing with inappropriate metric selection for handling annotation uncertainties}

\begin{checklist}
\item Clearly describe how annotation uncertainty was accounted for in metric computation and performance reporting. Explicitly report limitations, including which types of uncertainty were not incorporated into metric design.
\item Perform sensitivity analyses to assess how annotation uncertainty (e.g., fuzzy event boundaries) impacts metric outcomes.
\end{checklist}

\end{tcolorbox}

\newpage
\begin{tcolorbox}[pitfallbox,title={P2.4 -- Lack of metric robustness across varying real-world conditions}]

\textbf{Avoiding a lack of metric robustness across varying real-world conditions}

\begin{checklist}
\item Assess metric stability using sensitivity analyses or stress tests across varied conditions (e.g., resolution, acquisition settings, clinical workflows, lighting, noise, object sizes, data shuffle) to identify where metrics are fragile or misleading. Quantify the effect of varying conditions on metric values (e.g., percent change, confidence intervals) to clearly communicate robustness limits.
\item Use complementary metrics or joint validation strategies to reduce over-reliance on a single, potentially unstable measure.
\end{checklist}

\bigskip
\textbf{Dealing with a lack of metric robustness across varying real-world conditions}

\begin{checklist}
\item Analyze whether metrics maintain clinical interpretability under realistic intraoperative variability (e.g., changes in object size due to zoom, temporary occlusions, altered illumination). Report conditions under which metrics fail or become misleading.
\item Benchmark metric behavior against human perceptual or clinical judgments in challenging scenarios to detect misalignment between metric scores and clinically meaningful assessments of performance.
\end{checklist}

\end{tcolorbox}
\bigskip
\begin{tcolorbox}[pitfallbox,title={P2.5 -- Non-standardized configuration and definition of metrics}]

\textbf{Avoiding non-standardized configuration and definition of metrics}

\begin{checklist}
\item Adopt and cite established metric definitions and terminology from existing guidelines, such as Metrics Reloaded \cite{maier2024metrics}, or community standards. Where new or complementary metrics are introduced, clearly justify their clinical or technical motivation, fully document the metric definition and implementation (including all parameters and relevant parameter ranges), and discuss how they differ from or extend established measures.
\item Work in large consortia to define new community standards.
\item Pre-define metric definitions and configurations before the analysis is started. Report all metric changes made after initial analysis.
\end{checklist}

\bigskip
\textbf{Dealing with non-standardized configuration and definition of metrics}

\begin{checklist}
\item Explicitly state the intended clinical and technical interpretation of each metric and ensure that the chosen metric definitions and configurations are consistent with this interpretation.
\item Fully report metric configurations, including all hyperparameter choices, thresholds, and aggregation methods.
\item Use or publish standard metric implementations with transparent code and configuration files (e.g., Metrics Reloaded \cite{maier2024metrics}\footnote{\url{https://github.com/Project-MONAI/MetricsReloaded}}). Sanity check metric implementation details.
\item When comparing with prior work, explicitly document and justify differences in metric definitions or configurations that may affect comparability.
\end{checklist}

\end{tcolorbox}

\newpage
\begin{tcolorbox}[pitfallbox,title={P2.6 -- Non-suitability of hyperparameters from unrelated domains}]

\textbf{Dealing with non-suitability of hyperparameters from unrelated domains
}

\begin{checklist}
\item Tune metric-related thresholds (e.g., Intersection over Union (IoU) cutoffs, event timing tolerances) using surgical validation datasets or domain-specific ablation studies, informed by the intended clinical use and clinician input.
\item Do not tune metric hyperparameters directly on the final test set. Reserve separate data (e.g., a validation subset or nested validation scheme) for hyperparameter selection to reduce the risk of overfitting metric thresholds to a specific dataset.
\item Involve clinical experts and relevant clinical literature or guidelines when selecting or interpreting metric hyperparameters tied to task relevance or performance sensitivity.
\item Document and justify all metric hyperparameters, including thresholds values, tolerance windows, and clinically motivated design decisions. 
\item Assess and report how performance metrics and model rankings vary across reasonable ranges of metric hyperparameters (e.g., IoU thresholds or temporal tolerances), rather than relying on a single tuned value, to demonstrate the stability of conclusions. Interpret observed sensitivities in light of their clinical relevance (e.g., effects on critical structure detection).
\end{checklist}

\end{tcolorbox}
\bigskip
\begin{tcolorbox}[pitfallbox,title={P2.7 -- Intrinsic limitations of individual metrics}]

\textbf{Dealing with intrinsic limitations of individual metrics}

\begin{checklist}
\item Critically assess the mathematical properties, assumptions, and inherent limitations of each performance metric in the context of the specific surgical task and data characteristics. Explicitly describe which error types the metric highlights or ignores (e.g., boundary errors vs. temporal delays, frequent minor errors vs. rare critical failures) and assess whether these sensitivities correspond to clinically meaningful failure patterns.
\item Clearly document all known limitations of the chosen metric, including scenarios in which they may produce misleading, unstable, or clinically irrelevant results. Transparently communicate cases where the metric should not be relied upon for model comparison, decision making, or claims of clinical applicability.
\item \textit{(see also BP-2.1.2)} Use multiple metrics to capture both technical and clinical performance aspects. 
\end{checklist}

\end{tcolorbox}

\newpage
\begin{tcolorbox}[pitfallbox,title={P3.1 -- Non-independence within the test set}]

\textbf{Dealing with non-independence within the test set}

\begin{checklist}
\item Clearly describe the full data hierarchy and explicitly define which level is treated as independent for the analysis (e.g., frames of the same video, videos of the same procedure/patient/hospital).
\item When assessing multiple videos (or other sequential data sources, where applicable) of varying lengths, first aggregate performance per video and weight videos appropriately to ensure that video length does not introduce implicit weightings.
\item Use appropriate statistical methods such as hierarchical aggregation, Mixed-Effects Models, Analysis of Variance (ANOVA), or non-parametric techniques (e.g., hierarchical bootstrap). Check if underlying assumptions are met.
\item Report uncertainty exclusively at the level at which cases are treated as independent in the data set (e.g., confidence levels at the patient- or procedure-level). Avoid reporting uncertainty at lower hierarchy levels (e.g.,frame-level confidence intervals) when observations are not independent.
\item Ensure that metric computation aligns with the independence level (e.g., patient-level, procedure-level). Avoid mixing metrics computed at incompatible hierarchical units within the same analysis.
\end{checklist}

\end{tcolorbox}
\bigskip
\begin{tcolorbox}[pitfallbox,title={P3.2 -- Clinically uninformative aggregation}]

\textbf{Avoiding clinically uninformative aggregation}

\begin{checklist}
\item Define the purpose and intended clinical use of the model a priori to ensure that aggregation is justified and clinically meaningful. 
\item Use aggregation schemes that preserve clinically meaningful granularity (e.g., stratified or weighted aggregation with respect to surgical phase, anatomical region). Clearly state the unit of aggregation and justify why this unit reflects the clinical workflow or decision unit relevant for model deployment. Apply weighted aggregation only when the chosen weights can be clinically justified, for example by reflecting how often cases occur in practice or how clinical risk is distributed across cases.
\item Report intermediate results alongside aggregated scores to reveal performance patterns that may be concealed by overall averages. Include reporting results on the worst-performing clinically relevant subgroup.
\item Assess whether aggregation schemes implicitly reweight clinically important error types (e.g., by allowing frequent low-risk errors to dominate aggregate scores, thus concealing rare but clinically unacceptable failures).
\end{checklist}

\end{tcolorbox}
\newpage
\begin{tcolorbox}[pitfallbox,title={P3.3 -- Lack of contextualization of performance values}]

\textbf{Avoiding a lack of contextualization of performance values}

\begin{checklist}
\item Interpret reported performance in the context of the validated experimental or clinical setting (e.g., realistic operating room conditions), including typical hardware, image quality, latency constraints, and workflow integration, to avoid overestimating applicability beyond the validated setting.
\item Specify what level of performance is needed for safe or useful deployment in a clinical setting by consulting clinical experts. Explicitly relate metric values to clinically meaningful consequences (e.g., risk of missed events, delayed intervention, or unnecessary actions), including the identification of clinically unacceptable failure modes or error types that would invalidate certain claims or uses. State which types of claims or conclusions are supported by the reported performance (e.g., feasibility, robustness under specific conditions), and which forms of clinical use or decision-making are not supported.
\item Put performance into clinical perspective by comparing it to relevant references such as human experts, inter-rater variability, existing clinical workflows, or accepted clinical thresholds. Clarify whether differences in metric values (e.g., a 1 percentage point increase in Accuracy) correspond to meaningful changes in clinical decision-making or patient outcomes.
\item Put performance into technical perspective by comparing it to meaningful algorithmic baselines (e.g., prior state-of-the-art methods, ablated variants).
\item Account for task difficulty and expected clinical use when interpreting performance metrics by clarifying which portions of the task space are inherently challenging, subject to ambiguous interpretation, or low clinical impact (e.g., rare transition phases or visually ambiguous events).
\item Present and explain failure cases or performance outliers and map them to their clinical implications to support interpretation and identify model limitations.
\item Provide uncertainty measures (e.g., confidence intervals, variability across conditions) to avoid overinterpreting small differences that fall within measurement noise.
\end{checklist}

\end{tcolorbox}
\bigskip
\begin{tcolorbox}[pitfallbox,title={P3.4 -- Lack of uncertainty reporting}]

\textbf{Avoiding a lack of uncertainty reporting}

\begin{checklist}
\item Report descriptive statistics (e.g., mean/median, interquartile range) alongside measures of model performance uncertainty (e.g., confidence intervals). These can be well complemented by  visualizations that capture performance variability across relevant groups (e.g., patients with specific demographics, specific procedures, surgeons).
\item Incorporate methods to estimate prediction uncertainty (e.g., Monte Carlo dropout, ensembles \cite{lakshminarayanan2017simple}), and interpret their clinical implications \textit{(see also best practices for P3.3)}.
\item Evaluate robustness of model performance by implementing sensitivity analyses (e.g., data perturbations, variable thresholds for threshold-dependent metrics, introduction of randomness factors in training) and repeated experiments (e.g., random seed variation, repeated shuffle splits of data) to quantify how model performance and conclusions vary under context-specific conditions.
\end{checklist}

\end{tcolorbox}
\newpage
\begin{tcolorbox}[pitfallbox,title={P3.5 -- Insufficient reporting}]

\textbf{Avoiding insufficient reporting}

\begin{checklist}
\item Follow AI reporting guidelines that match the study design and output type \cite{kolbinger2024reporting}: 
\begin{checklist}
    \item for image- and video-based algorithm development, use the imaging-focused guideline CLAIM \cite{mongan2020checklist}; 
    \item for diagnostic accuracy designs, use STARD-AI \cite{sounderajah2021developing}; 
    \item when studying the diagnostic accuracy of models based on image- or video-derived features, use TRIPOD+AI \cite{collins2021protocol}; 
    \item and reserve SPIRIT-AI \cite{rivera2020guidelines}, CONSORT-AI \cite{liu2020reporting}, or DECIDE-AI \cite{vasey2022reporting} for prospective clinical trials or early-stage live clinical evaluation of AI decision-support systems. 
    \item Use the checklists provided by the aforementioned guidelines to ensure that all major workflow components (e.g., data sources, inclusion/exclusion criteria, preprocessing, data splitting, validation schemes, metric aggregation procedures, failure cases, metric definitions, and computed statistics) are fully reported, and provide explanatory captions/legends for visualizations.
\end{checklist}
\item Describe the methodology used to identify failure cases (e.g., stratified search, outlier-based analysis, expert review), including explanatory text and visual illustrations rather than only presenting illustrative examples.
\item Make validation code, exact dataset splits, metric implementations, model configurations and version identifiers, and prediction outputs (including configuration files and software versions) openly available to enable full reproduction of the reported results.
\item Publish supplementary materials or intermediate outputs (e.g., per-patient predictions) to allow external metric computation and validation.
\item Track benchmark results across studies using shared datasets, fixed validation protocols, and persistent identifiers (e.g., versioned test sets or leaderboards) to support fair comparison, reuse, and progress monitoring.
\item Use structured documentation formats such as Data Cards \cite{gebru2021datasheets} or AI-specific Report Cards (e.g., \cite{kolossvary2025transparent, mitchell2019model}) to provide standardized, transparent summaries of data characteristics, preprocessing steps, labeling procedures, and known limitations.
\item Report relevant negative or non-improving findings (e.g., comparisons or variants that did not lead to performance gains) to support transparent interpretation of validation results.
\item Establish dataset versioning, including annotations, and update documentation (e.g., versioned releases and change logs) so that validation results remain traceable as surgical practice and technology evolve.
\end{checklist}

\bigskip
\textbf{Dealing with insufficient reporting}

\begin{checklist}
\item Treat incomplete reporting as a methodological limitation that constrains the validity, interpretability, and reuse of the reported results, rather than as a purely stylistic concern.
\end{checklist}

\end{tcolorbox}

%%%%%%%%%%%%

\newpage
\section{Results of the systematic review per pitfall}
\label{suppl-screening}
While pitfalls can theoretically occur in any validation study, their actual prevalence in state-of-the-art surgical AI publications remained unclear. To address this, we conducted a systematic screening of all papers at the 2023 Medical Image Computing and Computer Assisted Intervention (MICCAI) conference that applied deep learning methods to surgical data. In this section, we present the results for each of the identified pitfalls.

\subsection*{[P1] PITFALLS RELATED TO DATA}
\subsubsection*{P1.1: Non-representativeness, low quality, and low relevance of data}\mbox{}

\noindent\textbf{Was the proposed model tested on out-of-distribution data (e.g., data from different centers or different surgeries)?} \\
Yes: 19.6\% \\
No: 54.4\% \\
Unclear: 26.1\% \\
\\
\textbf{Were parts of the data excluded?} (n = 46)\\
Yes: 17.4\% (with clear criteria: 15.2\%; without clear criteria: 2.2\%)\\
No: 41.3\%\\
Unclear: 41.3\% \\
\\
\textbf{Were the datasets public or private?} (n = 46)\\
Public datasets: 41.3\%\\
Private datasets: 32.6\%\\
Combination of private and public datasets: 19.6\%\\
New dataset(s) to be made public: 4.4\%\\
Unclear: 2.2\%\\
\\
\textbf{Which datasets were used?}\\
63\% of tasks specified the exact datasets used\\
A total of 38 datasets were used\\
79\% of datasets were used only once\\
The data sets used the most were EndoVis 2018 (used by 17.2\% of tasks), Cholec80 (13.8\%) and EndoVis 2017 (10.3\%)

\subsubsection*{P1.2: Limited sample size/test cohorts, or data imbalance} \mbox{}

\noindent\textbf{Number of videos}\\

\begin{tabular}{l c c c c}
\hline
 & \textit{Minimum} & \textit{Mean} & \textit{Median} & \textit{Maximum} \\ \hline
\textit{Training} & 6 & 234.2 & 37 & 1,500 \\ 
\textit{Validation} & 2 & 111.4 & 10 & 1,077 \\
\textit{Test} & 2 & 141.6 & 22 & 1,321 \\ \hline
\end{tabular}
\newpage
\noindent\textbf{Were data set sizes reported?} (n = 46)\\ 
Yes: 41.3\%\\ 
Partially: 30.4\%\\ 
No: 21.7\%\\ 
Unclear: 6.5\%\\ 

\noindent\textbf{Was the number of classes within the dataset reported?} (n = 39)\\ 
Yes, they reported in detail for the general data: 48.7\%\\ 
Yes, they reported in detail for each of the data subsets: 2.6\%\\ 
Partially: 10.3\%\\ 
No: 35.9\%\\ 
Unclear: 2.6\%\\ 
\\ 
\textbf{Was reported how classes are distributed in the dataset? (e.g., the percentage of each class in each video)} (n = 39)\\ 
Yes, they reported in detail for the general data: 5.1\%\\ 
Yes, they reported in detail for each of the data subsets: 2.6\%\\ 
Partially: 7.7\%\\ 
No: 84.6\%

\subsubsection*{P1.3: Existence of spurious correlations within data}\mbox{}

\noindent \textbf{Were spurious correlations considered by the authors? (n = 46)} \\
Considered in models: 2.2\% \\
Mentioned by authors, not considered: 6.5\% \\
No: 89.1\% \\
Unclear: 2.2\% 

\subsubsection*{P1.4: Incomplete annotation or missing contextual information} \mbox{}

\noindent \textbf{Were all frames in the test set(s) annotated?} (n = 42) \\
Yes: 48.8\% \\
No: 17.0\% \\
Unclear: 34.2\% 

\subsubsection*{P1.5: Unreliable or inconsistent annotation}\mbox{}

\noindent \textbf{Were object instances tracked (manually) over time for validation?} (n = 33) \\
Yes: 0.0\% \\
No: 42.4\% \\
Unclear: 57.6\% 

\subsubsection*{P1.6: Poor data splitting}\mbox{}

\noindent \textbf{Which principle best described data splitting ?} (n = 46) \\
Train / validation / test: 31.3\% \\
Train / test (single split, no validation): 27.1\% \\
k-fold CV (train / validation) / test: 10.4\% \\
k-fold CV (no test): 6.3\% \\
Train / validation (single split, no test): 2.1\% \\
k-fold CV (no test): 2.1\% \\
Mixture: 2.1\% \\
Other: 2.1\% \\
No splitting: 2.1\% \\
Unclear: 14.6\% 

\noindent\textbf{What were potential sources of data leakage?} (n = 45) \\
None: 26.7\% \\
Hierarchies were not handled properly: 26.7\% \\
Non-independence between training and test samples: 20.0\% \\
Lack of clean separation of training and test dataset: 17.8\% \\
No test set: 17.8\% \\
Unclear: 17.8\% \\
Pre-processing on training and test set: 4.4\% \\
Feature selection on training and test set: 2.2\% \\
Lack of description of data split: 2.2\% \\
Model used features that were not legitimate: 2.2\% \\
No concrete information on data split: 2.2\% \\
Not enough info on which model was trained with which dataset: 2.2\% \\
Temporal leakage: 2.2\% 

\noindent\textbf{Was the test set untouched? (n = 43)} \\
Yes: 46.5\% \\
No: 14.0\% \\
Other: 2.3\% \\
Unclear: 37.2\% 

\subsection*{[P2] PITFALLS RELATED TO METRIC SELECTION AND CONFIGURATION}
\subsubsection*{P2.1: Mismatch of metrics and clinical needs}\mbox{}

\noindent\textbf{Were metric choices justified? }(n = 46)\\
Yes: 30.6\% (by popularity: 20.4\%) \\
No: 69.4\% 

\noindent\textbf{Were clinical needs reflected in the metric choice? }(n = 46)\\
Yes: 6.5\% \\
Partially: 4.4\% \\
No: 8.7\% \\
Unclear: 80.4\% \

\subsubsection*{P2.2: Lack of common metrics that assess temporal aspects}\mbox{}

\noindent\textbf{Were algorithm properties specific to temporal data assessed?} (n = 35)\\
Yes: 8.6\% \\
Partially: 8.6\% \\
No: 77.2\% \\
Unclear: 5.7\% 

\noindent\textit{\textbf{1 paper introduced a temporal consistency metric.}}

\subsubsection*{P2.3: Inappropriate metric selection for handling  annotation uncertainties}\mbox{}

\noindent\textbf{Were event boundaries handled in a specific way?} (only for process-focus tasks; n = 11)\\
Yes: 0.0\% \\
No: 81.8\% \\
Unclear: 18.2\% 

\subsubsection*{P2.4: Lack of metric robustness across varying real-world conditions}\mbox{}

\noindent\textbf{Was the frame rate considered in the validation? }(n = 34)\\
Yes: 17.7\% \\
Partially: 5.9\% \\
No: 58.8\% \\
Unclear: 17.7\% 

\noindent\textbf{Was the image quality considered in the validation (e.g., resolution or complexity)?} (n = 46)\\
Yes: 8.7\% \\
Partially: 6.5\% \\
No: 56.5\% \\
Unclear: 28.3\% 

\subsubsection*{P2.5: Non-standardized configuration and definition of metrics}\mbox{}

\noindent\textbf{Were non-standardized metrics considered for model validation? }(n = 46)\\
Yes: 26.1\% \\
No: 67.4\% \\
Unclear: 6.5\% 

\noindent\textbf{Was explicitly described how the metric was computed (including hyperparameter choice)?} (n = 46)\\
Yes: 8.7\% \\
Partially: 6.5\% \\
No: 82.6\% \\
Unclear: 2.2\% 

\subsubsection*{P2.6: Non-suitability of hyperparameters from unrelated domains}\mbox{}

\noindent\textbf{Did the metrics contain hyperparameters?} (n = 46)\\
Yes: 32.6\% \\
No: 47.8\% \\
Unclear: 19.6\% 

\noindent\textbf{If so, were hyperparameters justified?} (n = 23)\\
Yes: 8.7\% (by popularity: 4.4\%) \\
No: 56.5\% \\
Unclear: 30.4\% 

\newpage
\subsection*{[P3] PITFALLS RELATED TO METRIC AGGREGATION AND REPORTING}
\subsubsection*{P3.1: Non-independence within the test set}\mbox{}

\noindent\textbf{Were test cases independent? }(n = 46)\\
Yes: 10.9\% \\
No: 28.3\% \\
Unclear: 58.7\% \\
Other: 2.2\% 

\noindent\textbf{If hierarchies were present in the data, were they properly addressed?} (n = 39)\\
Yes: 5.1\% \\
Partially: 5.1\% \\
No: 23.1\% \\
Unclear: 66.7\% 

\noindent\textbf{Was the video length considered in the validation?} (n = 34)\\
Yes: 14.7\% \\
Partially: 2.9\% \\
No: 58.8\% \\
Unclear: 23.5\% 

\subsubsection*{P3.2: Clinically uninformative aggregation}\mbox{}

\noindent\textbf{Was relevance considered in aggregation? }(n = 46)\\
Yes: 0.0\% \\
No: 34.8\% \\
Unclear: 65.2\% 

\noindent\textbf{Were results stratified with respect to relevant aspects (e.g., object size / shape, sensor quality, ...)? }(n = 46)\\
Yes: 28.3\% \\
Partially: 2.2\% \\
No: 69.6\% 

\subsubsection*{P3.3: Lack of contextualization of performance values}\mbox{}

\noindent\textbf{Were performance values put into context (e.g., by including inter-rater agreement, by defining of what constitutes a meaningful difference, or by defining of what constitutes a value sufficient to solve the underlying clinical task)? }(n = 46)\\
Yes: 13.0\% \\
Partially: 6.5\% \\
No: 80.4\% 

\subsubsection*{P3.4: Lack of uncertainty reporting}\mbox{}

\noindent\textbf{Manner of reporting variability or uncertainty} (n = 46)\\
None: 39.1\% \\
Standard deviation: 23.9\% \\
Values with +/- (unclear whether it is standard deviation): 21.7\% \\
Interquartile Range (graph): 17.4\% \\
Graphs (other): 4.4\% \\
Confidence intervals: 2.2\% \\
Prediction intervals: 2.2\% \\
Standard deviation in graphs: 2.2\% \\
Standard error: 2.2\% \\
Variability of different runs: 2.2\% \\ 
Statistical tests: 2.2\% 

\noindent\textbf{If standard deviation was reported, how was it calculated? }(n = 21)\\
Not reported: 81.0\% \\
From cross-validation (over non-overlapping samples): 4.8\% \\
From cross-validation (unclear): 4.8\% \\
Over different runs: 4.8\% \\
Other: 4.8\% 

\subsubsection*{P3.5: Insufficient reporting}\mbox{}

\noindent\textbf{Was the reporting comprehensive? }(n = 46)\\
Comprehensive reporting: 2.2\% \\
Only partly described: 97.8\% 

\noindent\textit{\textbf{1 single paper mentioned a reporting guideline but did not fully follow it.}}

\noindent\textbf{Was the aggregation procedure described in detail? }(n = 41)\\
Yes: 4.9\% \\
Partially: 29.3\% \\
No: 56.1\% \\
Unclear: 9.8\% 

\noindent\textbf{Ethical, Legal, Societal Aspects:}\\
\noindent\textit{\textbf{Were ethical aspects reported? }}(n = 46)\\
Yes: 15.2\% \\
Partially: 6.5\% \\
No: 78.3\% 

\noindent\textit{\textbf{Were fairness / bias / equity aspects reported? }}(n = 46)\\
Yes: 4.4\% \\
Partially: 6.5\% \\
No: 89.1\% 

\noindent\textit{\textbf{Were social / legal / governance aspects reported?}} (n = 46)\\
Yes: 2.2\% \\
Partially: 6.5\% \\
No: 91.3\%

\newpage
\section{Comparison of validation practices across MICCAI 2023 and 2025}
\label{suppl:comparison}
To assess whether the widespread validation flaws identified in our primary literature review extend to more recent work, we performed an additional targeted screening of MICCAI 2025 surgical data science papers. This complementary analysis focused on the key pitfalls reported in the main manuscript (see Section 'Validation flaws are widespread in common practice') and was designed to provide an indicative comparison across conference years rather than a full replication of the original screening (as in \ref{suppl-screening}). In this section, we present a side-by-side comparison of validation and reporting practices between MICCAI 2023 and 2025. \\

\begin{small}
\captionof{table}{Comparison of validation practices in MICCAI 2023 and 2025 surgical data science papers. The last column reports the differences between 2023 and 2025 ($\Delta$). Changes towards improved validation practices are highlighted in green, deteriorations in red, and neutral changes in black. Changes to the original phrasing in Section \textit{Validation flaws are widespread in common practice} are added in square brackets to facilitate reading or comparison.}
\label{tab:2023-2025-comp}
\vspace{-0.5cm}
\addtocounter{table}{-1}
\begin{longtable}{ P{2.5cm} B{4.4cm} B{4.4cm} P{1.5cm}}
 \toprule

\textbf{Question} & \noindent  \textbf{MICCAI 2023} & \noindent  \textbf{MICCAI 2025} & \textbf{$\Delta$}\\ \hline
Number of articles & \noindent46 & \noindent89 & \noindent+43\\
& & & \\
Did the data set comprise videos (rather than static frames only)? & \noindent \textbf{74\%} of papers used surgical video data. & \noindent\textbf{87\%} of papers used surgical video data. & \noindent\textcolor{darkgreen}{+13\%} \\ 
& & & \\
Public datasets (if any) & \noindent\textbf{79\%} of datasets were only used once, with \textbf{38} distinct [public] datasets identified across papers. & \noindent\textbf{53\%} of datasets were only used once, with \textbf{86} distinct [public] datasets identified across papers. & \noindent-26\%; +48 \\
& & & \\
Training set size (number of videos) & \noindent The median dataset contained\textbf{ 37} training videos (minimum: \textbf{6} training videos). & \noindent The median dataset contained \textbf{35} training videos (minimum: \textbf{1} training videos). & \noindent \textcolor{darkred}{-2; -5} \\
& & & \\
Validation set size (number of videos) & \noindent The median dataset contained \textbf{10} validation videos (minimum: \textbf{2} validation videos). & \noindent The median dataset contained \textbf{7} validation videos (minimum: \textbf{2} validation videos).
& \noindent \textcolor{darkred}{-3}; 0 \\
& & & \\
Test set size (number of videos)& \noindent The median dataset contained \textbf{22} test videos (minimum: \textbf{2} test videos).& \noindent The median dataset contained \textbf{10} test videos (minimum: \textbf{1} test videos).& \noindent \textcolor{darkred}{-12; -1}\\
& & & \\
Were data set sizes reported?& \noindent \textbf{59\%} of papers did not (fully) report data set sizes.& \noindent \textbf{66\%} of papers did not (fully) report data set sizes.& \noindent \textcolor{darkred}{+7\%}\\
& & & \\
Was the used test set untouched?& \noindent Only \textbf{47\%} explicitly reported an untouched test set, while this was unclear in \textbf{37\%} of papers.& \noindent Only \textbf{22\%} explicitly reported an untouched test set, while this was unclear in \textbf{74\%} of papers.& \noindent \textcolor{darkred}{-25\%; +37\%}\\
& & & \\
Were metric choices justified?& \noindent Only \textbf{30\%} properly justified their metric choice. \textbf{20\%} of those justified by popularity alone.& \noindent Only \textbf{38\%} properly justified their metric choice. \textbf{24\%} of those justified by popularity alone.& \noindent \textcolor{darkgreen}{+8\%};  \textcolor{darkred}{+4\%}\\
& & & \\
Were clinical needs reflected in the metric choice?& \noindent In \textbf{80\%} of papers, it was unclear whether clinical relevance had been considered when selecting metrics.& \noindent In \textbf{75\%} of papers, it was unclear whether clinical relevance had been considered when selecting metrics.& \noindent \textcolor{darkgreen}{-5\%}\\
& & & \\
Were algorithm properties specific to temporal data assessed?& \noindent A total of \textbf{77\%} of papers did not assess properties specific to temporal data.& \noindent A total of\textbf{ 72\%} of papers did not assess properties specific to temporal data.& \noindent \textcolor{darkgreen}{-5\%}\\
& & & \\
Were hierarchies handled in a specific way?& \noindent Among studies involving hierarchical structures, (e.g., patient-level), only \textbf{5\%} explicitly accounted for their dependencies.& \noindent Among studies involving hierarchical structures, (e.g., patient-level), only \textbf{1\%} explicitly accounted for their dependencies.& \noindent \textcolor{darkred}{-4\%}\\
& & & \\
Metrics& \noindent Only a \textbf{single paper [2\%]} used a temporal consistency metric. \textbf{41} metrics were used only once \textbf{[37\%]}. The most commonly used metric was \textbf{Accuracy}.& \noindent Only \textbf{seven papers [8\%]} used a temporal metric. \textbf{93} metrics were used only once \textbf{[72\%]}. The most commonly used metric was \textbf{IoU}.& \noindent \textcolor{darkgreen}{+6\%}; +35; Accuracy $\xrightarrow{}$ IoU\\
& & & \\
Was the aggregation procedure described in detail?& \noindent The aggregation procedure was unclear or not described at all in \textbf{66\%} of papers.& \noindent The aggregation procedure was unclear or not described at all in \textbf{58\%} of papers.& \noindent \textcolor{darkred}{-8\%}\\
& & & \\
Were performance values put into context?& \noindent \textbf{80\%} did not contextualize performance values, for example against human baseline or clinical thresholds.& \noindent \textbf{73\%} did not contextualize performance values, for example against human baseline or clinical thresholds.& \noindent \textcolor{darkgreen}{-7\%}\\
& & & \\
Were confidence or prediction intervals reported?& \noindent Only \textbf{one paper [2\%]} reported confidence intervals, and \textbf{one [2\%]} reported prediction intervals.& \noindent Only \textbf{one paper [1\%]} reported confidence intervals, and \textbf{no paper [0\%]} reported prediction intervals.& \noindent \textcolor{darkred}{-1\%; -2\%}\\
& & & \\
Was inter-rater variability reported?& \noindent \textbf{98\%} of papers did not report inter-rater variability.& \noindent \textbf{95\%} of papers did not report inter-rater variability.& \noindent \textcolor{darkgreen}{-3\%}\\
& & & \\
Did authors report adhering to common reporting guidelines?& \noindent Only \textbf{a single paper [2\%]} reported sufficient detail to enable reproducibility.& \noindent Only \textbf{three papers [3\%]} reported sufficient detail to enable reproducibility.& \noindent \textcolor{darkgreen}{+1\%}\\
& & & \\
Were ethical aspects reported?& \noindent \textbf{78\%} [of papers] lacked ethical reporting.& \noindent \textbf{84\%} [of papers] lacked ethical reporting.& \noindent \textcolor{darkred}{+6\%}\\
& & & \\
Were fairness-/ bias-/ equity-related aspects reported?& \noindent \textbf{89\%} [of papers] ignored fairness or biases.& \noindent \textbf{94\%} [of papers] ignored fairness or biases.& \noindent \textcolor{darkred}{+5\%}\\
& & & \\
Were societal-/ legal-/ governance-related aspects reported?& \noindent \textbf{91\%} [of papers] omitted social, legal, or governance considerations.& \noindent \textbf{89\%} [of papers] omitted social, legal, or governance considerations.& \noindent \textcolor{darkgreen}{-2\%}

 \\ \bottomrule

\end{longtable}

\end{small}

%%
%% The next two lines define the bibliography style to be used, and
%% the bibliography file.
%\bibliographystyle{ACM-Reference-Format}

% \section{Full Author Affiliations}
% \label{app:authors}
% \printauthors

% \input{Appendix.tex}

\end{document}